\documentclass[prc,preprint,showpacs,tightenlines,%
                floatfix,amsfonts,nofootinbib]{revtex4}

\usepackage{graphicx}
\usepackage{bm}
\usepackage{amssymb}
\usepackage{latexsym}
\usepackage{amsmath}

\begin{document}

%
%
%
%
\newcommand{\covdev}{{\widetilde \partial}}
\newcommand{\finalnewpage}{\newpage}
\newcommand{\lagrang}{{\cal L}}
\newcommand{\newg}{{\skew2\overline g}_\rho}
\newcommand{\Nbar}{\skew3\overline N \mkern2mu}
\newcommand{\stroke}[1]{\mbox{{$#1$}{$\!\!\!\slash \,$}}}

\newcommand{\dgamma}[1]{\gamma_{#1}}
\newcommand{\ugamma}[1]{\gamma^{#1}}
\newcommand{\dsigma}[1]{\sigma_{#1}}
\newcommand{\usigma}[1]{\sigma^{#1}}
\newcommand{\ugammafive}{\gamma^{5}}
\newcommand{\dgammafive}{\gamma_{5}}
\newcommand{\ket}[1]{\vert#1\rangle}
\newcommand{\bra}[1]{\langle#1\vert}
\newcommand{\inprod}[2]{\langle#1\vert#2\rangle}
\newcommand{\psibar}[1]{\overline{#1}}
\newcommand{\half}{\ensuremath{\frac{1}{2}}}
\newcommand{\threehalf}{\ensuremath{\frac{3}{2}}}
\newcommand{\slashed}[1]{\not\!#1}
\newcommand{\lag}{\mathcal{L}}
\newcommand{\uhalftau}[1]{\frac{\tau^{#1}}{2}}
\newcommand{\dhalftau}[1]{\frac{\tau_{#1}}{2}}
\newcommand{\ucpartial}[1]{\widetilde{\partial}^{#1}}
\newcommand{\dcpartial}[1]{\widetilde{\partial}_{#1}}
\newcommand{\mn}{\mu\nu}
\newcommand{\ab}{\alpha\beta}
\newcommand{\Tdagger}[2]{T^{\dagger \,#1}_{#2}}
\newcommand{\T}[2]{T^{#1}_{\,\,#2}}
\newcommand{\vbg}{\mathsf{v}}
\newcommand{\abg}{\mathsf{a}}
\newcommand{\sbg}{\mathsf{s}}
\newcommand{\pbg}{\mathsf{p}}
\newcommand{\modular}[2]{\vert \vec{#1}_{#2}\vert}
\def\Tr{\mathop{\rm Tr}\nolimits}

%
%
\def\dthree#1{\intback{\rm d}^3\intback #1}
\def\dthreex{\dthree{x}}
\def\fpi{f_{\pi}}
\def\gammafive{\gamma^5}
\def\gammafivel{\gamma_5}
\def\gammamu{\gamma^{\mu}}
\def\gammamul{\gamma_{\mu}}
\def\intback{\kern-.1em}
\def\kfermi{k_{\sssize {\rm F}}}    
\def\mn{{\mu\nu}}
\def\sigmamunu{\sigma^\mn}
\def\sigmamunul{\sigma_\mn}
\def\Tr{\mathop{\rm Tr}\nolimits}

\let\dsize=\displaystyle
\let\tsize=\textstyle
\let\ssize=\scriptstyle
\let\sssize\scriptscriptstyle
%
%
%


\newcommand{\beq}{\begin{equation}}
\newcommand{\eeq}{\end{equation}}
\newcommand{\beqa}{\begin{eqnarray}}
\newcommand{\eeqa}{\end{eqnarray}}

\def\Inthelimit#1{\lower1.9ex\vbox{\hbox{$\
   \buildrel{\hbox{\Large \rightarrowfill}}\over{\scriptstyle{#1}}\ $}}}

\title{Weak Pion and Photon Production off Nucleons in a \\
       Chiral Effective Field Theory}

\author{Brian D. Serot}\email{serot@indiana.edu}
\author{Xilin Zhang}\email{xilzhang@indiana.edu}
\affiliation{Department of Physics and Center for Exploration of
             Energy and Matter\\
             Indiana University, Bloomington, IN\ \ 47405}

%
\author{\null}
\noaffiliation

%
\date{\today\\[20pt]}

\begin{abstract}
Neutrino-induced pion and photon production from nucleons and nuclei
are important for the interpretation of neutrino-oscillation
experiments, and these processes are potential backgrounds in the
MiniBooNE experiment [A. A. Aquilar-Arevalo \textit{et al.}
(MiniBooNE Collaboration), Phys.\ Rev.\ Lett.\ {\bf 100}, 032301
(2008)]. Pion and photon production are investigated at intermediate
energies, where the $\Delta$ resonance becomes important. The
Lorentz-covariant effective field theory contains nucleons, pions,
Deltas, isoscalar scalar ($\sigma$) and vector ($\omega$) fields,
and isovector vector ($\rho$) fields.  The lagrangian exhibits a
nonlinear realization of (approximate) $SU(2)_L \otimes SU(2)_R$
chiral symmetry and incorporates vector meson dominance. Power
counting for vertices and Feynman diagrams involving the $\Delta$ is
explained. Because of the built-in symmetries, the vector currents
are automatically conserved, and the axial-vector currents satisfy
PCAC. The irrelevance of so-called off-shell $\Delta$ couplings and
the structure of the dressed $\Delta$ propagator, which has a pole
only in the spin-3/2 channel, are discussed. To calibrate the
axial-vector transition current $(N\! \leftrightarrow \Delta)$, pion
production from the nucleon is used as a benchmark and compared to
bubble-chamber data from Argonne and Brookhaven National
Laboratories. At low energies, the convergence of our power-counting
scheme is investigated, and next-to-leading-order tree-level
corrections are found to be very small.
\end{abstract}

\smallskip

\maketitle

\section{Introduction}
\label{sec:intro}

Weak pion production from nucleons and nuclei plays an important
role in the interpretation of neutrino-oscillation experiments, such
as MiniBooNE  \cite{MINIBOONE} and K2K \cite{K2K}. Pion absorption
after production will lead to events that mimic quasielastic
scattering. Moreover, neutral current (NC) $\pi^0$ and photon
production produce detector signals that resemble those of the
desired $e^\pm$ signals.  Finally, NC $\pi^0$ and photon production
might explain the excess events seen at low energies in MiniBooNE.

Ultimately, the calculations must be done on nuclei, which are the
primary detector materials in oscillation experiments. To separate
the many-body effects from the reaction mechanism and to calibrate
the elementary amplitude, we will study charged current (CC) and NC
pion and photon production from free nucleons. We will apply our
full lagrangian to the many-body problem in a forthcoming paper.

In this work, we use a recently proposed Lorentz-covariant
meson--baryon effective field theory (EFT) that was originally
motivated by the nuclear many-body problem
\cite{SW86,SW97,FST97,FSp00,FSL00,EvRev00,LNP641,EMQHD07}. (This
formalism is often called \emph{quantum hadrodynamics} or QHD.) This
QHD EFT includes all the relevant symmetries of the underlying QCD;
in particular, the approximate, spontaneously broken $SU(2)_L
\otimes SU(2)_R$ chiral symmetry is realized nonlinearly. The
motivation for this EFT and some calculated results are discussed in
Refs.~\cite{SW97,FST97,HUERTAS02,HUERTASwk,HUERTAS04,MCINTIRE04,%
MCINTIRE05,JDW04,MCINTIRE07,HU07,MCINTIRE08,BDS10}.

Here we consistently incorporate the $\Delta (1232)$ resonance as an
explicit degree of freedom in this EFT, while respecting the
underlying symmetries of QCD noted earlier. We are concerned with
the intermediate-energy region $(E^{\mathrm{Lab}}_\nu <
1\,\mathrm{GeV})$, where the resonant behavior of the $\Delta$ is
important. Couplings to electroweak fields are included using the
external field procedure \cite{Gasser84}, which allows us to deduce
the electroweak currents. Because of the approximate symmetries
contained in the lagrangian, the vector currents are explicitly
conserved, and the axial-vector currents satisfy PCAC. Form factors
are generated within the theory by vector meson dominance (VMD),
which avoids introducing phenomenological form factors and makes
current conservation manifest.  We discuss the power counting of
both vertices and diagrams on and off resonance and consistently
keep all tree-level diagrams through next-to-leading order.

The goal of this work is to use CC and NC pion production from
nucleons to serve as a benchmark calculation. In a future paper, we
will include the electroweak response of the nuclear many-body
system to discuss pion production from nuclei.

There have been numerous earlier studies of weak pion production off
nucleons in the energy regime where the $\Delta$ is important
\cite{ADLER68,LS72,SCHREINER73,REIN81,AR99,SATO03,OL05,OL06,%
HERNANDEZ07,GRACZYK08,GRACZYK09,OL10}. It is typically assumed that
the vector part of the $N \to \Delta$ transition current is well
constrained by electromagnetic interactions \cite{OL06,GRACZYK08}.
The uncertainty is in the axial-vector part of the current, which is
determined by fitting to ANL \cite{RADECKY82} and BNL
\cite{KITAGAKI86} bubble-chamber data. The data has large error
bars, which leads to significant model dependence in the fitted
results \cite{HERNANDEZ07,GRACZYK09,PRAET09}.  In this work, we
choose one recently fitted parametrization \cite{GRACZYK09} and use
it to determine the momentum dependence of the transition current
vertices.  We also use it to determine the constants of our VMD
parametrization.  We then compare calculations with both sets of
vertices to the data at low and intermediate neutrino energies.

This paper is organized as follows: in Sec.~\ref{sec:form}, we
introduce our EFT lagrangian and calculate several current matrix
elements that will be useful for the subsequent Feynman diagram
calculations. The theory involving the $\Delta$ is emphasized, and
the pathologies of introducing the $\Delta$ in quantum field theory
are clarified, which is the basis of the lagrangian construction.
Then the transition current basis and form factors are discussed
carefully. In Sec.~\ref{sec:diag}, we show the detailed calculations
for CC and NC pion production and for the NC photon production cross
sections. We initially insist on approaching this problem within
EFT, and hence consider only the low-energy region:
$E^{\mathrm{Lab}}_{\nu} \leqslant 0.5 \, \mathrm{GeV}$. After that,
we show our results in Sec.~\ref{sec:res}. Whenever possible, we
compare our results with available data and present our analysis.
Finally, our conclusions are summarized in Sec.~\ref{sec:sum}.

In the Appendixes, we discuss isospin conventions; $C$, $P$, and $T$
properties of various fields; form factors; the $\Delta$ propagator;
and kinematics.

\section{formalism} \label{sec:form}
\subsection{Notation} \label{subsec:not}
In this calculation, we use the metric
$g_{\mn}=\mathrm{diag}(1,-1,-1,-1)_{\mn}$, and the convention for
the Levi-Civita symbol $\epsilon^{\mu\nu\alpha\beta}$ is
$\epsilon^{0123}=1$. The Dirac matrices are represented as (here
$\sigma^{i}$ is a Pauli matrix)
\beq
\gamma^0=
\begin{pmatrix}
\mathbf{1}& \,\, 0\\
0\,\,&\mathbf{-1}
\end{pmatrix} \ ,\quad
\gamma^1=
\begin{pmatrix}
0&\sigma^x\\
-\sigma^x&0
\end{pmatrix} \ ,\quad
\gamma^2=
\begin{pmatrix}
0&\sigma^y\\
-\sigma^y&0
\end{pmatrix} \ ,\quad
\gamma^3=
\begin{pmatrix}
0&\sigma^z\\
-\sigma^z&0
\end{pmatrix} \ ,
\eeq
and $\gamma^5=i\gamma^0\gamma^1\gamma^2\gamma^3$.

Since we are going to include the $\Delta$, which is the lowest $N$
resonance, and whose isospin is $I=3/2$, we will define the
conventions for isospin indices. We will work with spherical vector
components for the pion field, which requires some care with signs.
Begin with
\begin{eqnarray}
\Delta^{\ast a}&\equiv& T^{a}_{\,\,iA} \Delta^{\ast iA} \ ,
\end{eqnarray}
Here $a=\pm 3/2, \pm 1/2$, $i=\pm 1, 0$, and $A=\pm 1/2$. The upper
components labeled `$a$', `$i$', and `$A$' furnish
$\mathcal{D}^{(3/2)}$, $\mathcal{D}^{(1)}$, and
$\mathcal{D}^{(1/2)}$ representations of the isospin $SU(2)$ group.
We can immediately see that $T^{a}_{\,\,iA}=\langle 1,\frac{1}{2}
;i,A \vert \frac{3}{2}; a\rangle$, which are CG coefficients. It is
well known that the conjugate representation of $SU(2)$ is
equivalent to the representation itself, so we introduce a metric
linking the two representations to raise or lower the indices $a,i$,
and $A$. For example, $\Delta_{a}\equiv(\Delta^{\ast
a})^{\ast}=T^{\dagger \,\,iA}_{a} \Delta_{iA}$, where $T^{\dagger
\,\,iA}_{a} = \langle \frac{3}{2}; a \vert 1,\frac{1}{2};i,A
\rangle$, can be written as
\begin{eqnarray}
\Delta_{a}=T^{iA}_{a} \Delta_{iA} \equiv T^{b}_{jB}\,
\widetilde\delta_{ba}\,\widetilde\delta^{ji}\,\widetilde\delta^{BA}\,\Delta_{iA}
 \ . 
\end{eqnarray}
Here, $\widetilde\delta$ denotes a metric for one of the three
representations. So in this convention, $T^{\dagger
\,\,iA}_{a}=T^{iA}_{a}$, which is straightforward to prove. Details
about the conventions are given in Appendix~\ref{app:indices}.

\subsection{Lagrangian without $\Delta$(1232)}

It is widely accepted that the approximate global chiral symmetry
$SU(2)_L \otimes SU(2)_R \otimes U(1)_B$ in two-flavor QCD is
spontaneously broken to $SU(2)_V \otimes U(1)_B$, while also being
manifestly broken due to the small quark masses. To implement such
broken global symmetry in the effective lagrangian using hadronic
degrees of freedom, it was found that there exists a general
nonlinear realization of such symmetry
\cite{Weinb68,Coleman69A,Coleman69B}. Sometime later, in
Ref.~\cite{Weinb79}, the concept of phenomenological lagrangians was
revisited, leading to the wide use of effective field theory, in
which chiral symmetry is realized nonlinearly.

Here we will make use of the background field method to construct a
low-energy effective field theory. In this method, we elevate the
global symmetry $SU(2)_L \otimes SU(2)_R \otimes U(1)_B$ to a local
symmetry \cite{Gasser84,Gasser85,schererprimer}.

First, we will only briefly discuss how the elevated local symmetry
is realized in two-flavor QCD, since this material can be easily
found elsewhere (see Ref.~\cite{schererprimer}, for example), and
then we show how this symmetry is realized nonlinearly in QHD. This
theory is well developed in Refs.~\cite{FST97,SW97,EMQHD07}. Since
we take a different approach, we will detail the chiral symmetry
realization in this section. Finally, we will talk about the
electroweak interactions of hadrons.

\subsubsection{Chiral symmetry realization} \label{subsubsec:chisym}

The charge algebra of $SU(2)_L \otimes SU(2)_R \otimes U(1)_B$ is
(we will henceforth ignore the \emph{tilde} on $\widetilde{\delta}$
and $\widetilde\epsilon\,$)
\begin{eqnarray}
\left[Q_{L}^{i} \, , \, Q_{L}^{j}\right] &=& i \epsilon^{ijk} Q_{Lk} \ , \notag \\
\left[Q_{R}^{i} \, , \, Q_{R}^{j}\right] &=& i \epsilon^{ijk} Q_{Rk} \ , \notag \\
\left[Q_{L}^{i} \, , \, Q_{R}^{j}\right] &=& 0 \ , \notag \\
\left[Q_{B}  \, , \, Q_{L,R}^{i}\right] &=& 0 \ ,  \quad
\text{where} \quad i, j, k=+1,0,-1\ . 
\end{eqnarray}
We can immediately see that massless, two-flavor QCD has this
symmetry, with background fields including $\vbg^{\mu}\equiv
\vbg^{i\mu} \tau_{i}/{2}$ \ (isovector vector), $\vbg_{(s)}^{\mu}$ \
(isoscalar vector), $\abg^{\mu}\equiv \abg^{i\mu} \tau_{i}/{2}$ \
(isovector axial-vector), $\sbg \equiv \sbg^{i} \tau_{i}/{2}$ \
(isovector scalar), and $\pbg\equiv \pbg^{i} \tau_{i}/{2}$ \
(isovector pseudoscalar), where $i=x,y,z \ \text{or} \ +1,0,-1$:
\begin{eqnarray}
\lag &=& \lag_{QCD}+ \psibar{q} \dgamma{\mu} (\vbg^{\mu}+ B
       \vbg_{(s)}^{\mu}
        + \dgammafive \abg^{\mu}) q - \psibar{q} (\sbg-i \dgammafive \pbg) q \notag \\
     &=& \lag_{QCD}+ \psibar{q}_{L} \dgamma{\mu} (l^{\mu}+ B \vbg_{(s)}^{\mu}) q_{L}
        + \psibar{q}_{R} \dgamma{\mu} (r^{\mu}+ B \vbg_{(s)}^{\mu} )q_{R} \notag \\
     &&\quad {}-\psibar{q}_{L} (\sbg-i\pbg) q_{R}-\psibar{q}_{R} (\sbg+i  \pbg) q_{L} \ .
     \label{eqn:qcdL}
\end{eqnarray}
Here, $r^{\mu}= \vbg^{\mu}+\abg^{\mu}$,
$l^{\mu}=\vbg^{\mu}-\abg^{\mu}$, $q_{L}=\frac{1}{2}(1-\ugammafive)\,
q$, $q_{R}=\frac{1}{2}(1+\ugammafive)\, q $, and $B = 1/3$ is the
baryon number. The symmetry transformation rules are
\begin{eqnarray}
q_{LA}&\to& \exp \left[-i\frac{\theta(x)}{3} \right]
      \left( \exp \left[ -i\theta_{Li}(x)\,\frac{\tau^{i}}{2} \right] \right)_{A}^{\;B}
      q_{LB} \equiv \exp \left[-i\frac{\theta(x)}{3} \right]
      (L)_{A}^{\;B} q_{LB} \ ,\notag \\[5pt]
q_{R} &\to& \exp \left[ -i\frac{\theta(x)}{3} \right]
      \exp \left[ -i\theta_{Ri}(x)\,\frac{\tau^{i}}{2} \right] q_{R}
      \equiv \exp \left[ -i\frac{\theta(x)}{3} \right] \, R q_{R} \ , \notag
      \\[5pt]
l^{\mu} &\to& L \, l^{\mu} L^{\dagger} + i L \, \partial^{\mu} L^{\dagger} \ , \notag
      \\[5pt]
r^{\mu} &\to& R \, r^{\mu} R^{\dagger} + i R \, \partial^{\mu} R^{\dagger} \ , \notag
      \\[5pt]
\vbg_{(s)}^{\mu} &\to& \vbg_{(s)}^{\mu} - \partial^{\mu} \theta \ ,
\notag
      \\[5pt]
\sbg+i\pbg &\to& R(\sbg +i\pbg )L^{\dagger} \ , \qquad \sbg-i\pbg
      \to L(\sbg -i\pbg)R^{\dagger}
      \ . 
\end{eqnarray}
We can also construct field strength tensors that transform
homogeneously:
\begin{eqnarray}
f_{L\mu\nu}&\equiv& \partial_{\mu}l_{\nu}-\partial_{\nu}l_{\mu}
       -i \left[l_{\mu} \, , \, l_{\nu}\right] \to L f_{L\mu\nu}L^{\dagger} \ , \notag \\
f_{R\mu\nu}&\equiv& \partial_{\mu}r_{\nu}-\partial_{\nu}r_{\mu}
       -i \left[r_{\mu} \, , \, r_{\nu}\right] \to R f_{R\mu\nu}R^{\dagger} \ , \notag \\
f_{s\mu\nu}&\equiv&
       \partial_{\mu}\vbg_{(s)\nu}-\partial_{\nu}\vbg_{(s)\mu} \to
       f_{s\mu\nu} \ .
\end{eqnarray}
Meanwhile, to conserve $C, P$, and $T$ symmetry, we have the
corresponding transformation rules shown in
Appendix~\ref{app:cptsymQCD}.

Now we proceed to discuss low-energy nuclear theory involving
$\pi^{i}$, $\rho^{i}_{\mu}$, $N^{A}$, and the chiral singlets
$V_{\mu}$ and $\phi$ \cite{FST97}. As noted earlier, chiral symmetry
is spontaneously broken at low energy in the chiral limit, and the
symmetry is realized nonlinearly:
\begin{eqnarray}
U&\equiv&\exp \left[ 2i\frac{\pi_{i}(x)}{f_{\pi}}\, t^{i} \right]
       \to  LUR^{\dagger} \ , \notag \\[5pt]
\xi&\equiv& \sqrt{U}= \exp \left[ i\frac{\pi_{i}}{f_{\pi}}\, t^{i}
       \right]
       \to L\xi h^{\dagger}=h \,\xi R^{\dagger} \ , \notag \\[5pt]
\widetilde v_{\mu}&\equiv& \frac{-i}{2} [\xi^{\dagger}(\partial_{\mu}
       -il_{\mu})\xi+\xi(\partial_{\mu}-ir_{\mu})\xi^{\dagger}]
       \equiv \widetilde v_{i\mu}t^{i}
       \to h \, \widetilde v_{\mu} h^{\dagger} -ih \, \partial_{\mu}h^{\dagger} \ ,
       \notag \\[5pt]
\widetilde a_{\mu}&\equiv& \frac{-i}{2} [\xi^{\dagger}(\partial_{\mu}
       -il_{\mu})\xi-\xi(\partial_{\mu}-ir_{\mu})\xi^{\dagger}]
       \equiv \widetilde a_{i\mu}t^{i} \to h \, \widetilde a_{\mu} h^{\dagger} \ ,
       \notag \\[5pt]
\dcpartial{\mu}U&\equiv& \partial_{\mu} U -i l_{\mu} U +i U r_{\mu}
       \to L \, \dcpartial{\mu}UR^{\dagger} \ , \notag \\[5pt]
(\dcpartial{\mu}\psi)_{\alpha}&\equiv& (\partial_{\mu}+i\,\widetilde v_{\mu}
       -i\vbg_{(s)\mu}B)_{\alpha}^{\;\beta} \psi_{\beta}
       \to \exp \left[ -i\theta(x)B \right]
       h_{\alpha}^{\;\beta} (\dcpartial{\mu}\psi)_{\beta} \ , \notag
       \\[5pt]
\widetilde{v}_{\mu\nu}&\equiv& -i [\widetilde{a}_{\mu} \, , \,
       \widetilde{a}_{\nu}]
       \to h \, \widetilde{v}_{\mu\nu} h^{\dagger} \ ,  \notag
       \\[5pt]
F^{(+)}_{\mu\nu}&\equiv&\xi^{\dagger}f_{L\mu\nu}\,\xi
       + \xi f_{R\mu\nu}\, \xi^{\dagger}
       \to hF^{(+)}_{\mu\nu}h^{\dagger} \ , \notag \\[5pt]
F^{(-)}_{\mu\nu}&\equiv&\xi^{\dagger}f_{L\mu\nu}\,\xi
       - \xi f_{R\mu\nu}\, \xi^{\dagger}
       \to hF^{(-)}_{\mu\nu}h^{\dagger} \ , \notag \\[5pt]
\dcpartial{\lambda} F^{(\pm)}_{\mu\nu} &\equiv&
       \partial_{\lambda}F^{(\pm)}_{\mu\nu} + i [\widetilde{v}_{\lambda}
       \, , \, F^{(\pm)}_{\mu\nu}] \to h\,\dcpartial{\lambda}
F^{(\pm)}_{\mu\nu}h^{\dagger} \ . \label{eqn:allchiral} 
\end{eqnarray}
In the preceding equations, $t^{i}$ are the generators of reducible
representations of $SU(2)$. Specifically, they could be generators
of $\mathcal{D}^{(1/2)}_{N} \oplus \mathcal{D}^{(1)}_{\rho} \oplus
\mathcal{D}^{(3/2)}_{\Delta}$, which operate on non-Goldstone
isospin multiplets including the nucleon, $\rho$ meson, and
$\Delta$. We will generically label these fields by
$\psi_{\alpha}=\left( N_{A}, \rho_{i}, \Delta_{a} \right)_{\alpha}$.
Most of the time, the choice of $t^{i}$ is clear from the context.
$B$ is the baryon number of the particle.  The transformations of
the isospin and chiral singlets $V_{\mu}$ and $\phi$ are trivial. We
will also make use of the dual field tensors, for example,
$\psibar{F}^{\,(\pm)\,\mu\nu} \equiv \epsilon^{\mu\nu\alpha\beta}
F^{(\pm)}_{\alpha\beta}$, which have the same chiral transformations
as the ordinary field tensors. Here we do not include the background
fields $\sbg$ and $\pbg$ mentioned in Eq.~(\ref{eqn:qcdL}), which
are the source of manifest chiral-symmetry breaking in the Standard
Model.

The realizations of $C$, $P$, and $T$ symmetries are given in
Appendix~\ref{app:cptsymQHD}.

\subsubsection{Power counting and the lagrangian}
\label{subsubsec:lagN}

Based on the transformation rules of the building blocks listed
above, we can begin to construct the low-energy EFT lagrangian. The
organization of interaction terms in this lagrangian is based on
power counting \cite{FST97}. The power counting in EFT essentially
assumes that when the interaction structure becomes more
complicated, i.e., more fields and more derivatives are introduced,
its contribution to physical observables becomes less important.
Similarly, loop contributions will be suppressed more when the
number of loops gets bigger. The validity of this assumption is
connected with Naive Dimensional Analysis (NDA)
\cite{GEORGI84,GEORGI93} which assumes the strength (coupling) of
the interaction is of order unity when the appropriate dimensional
scale factors have been included. This ``naturalness'' can be
checked only after the calculations are finished.

To make the power counting transparent, we can associate with each
interaction term an index
\begin{equation}
\hat{\nu} \equiv d+ \frac{n}{2} + b \ . 
\end{equation}
Here $d$ is the number of derivatives (small momentum transfer) in
the interaction, $n$ is the number of fermion fields, and $b$ is the
number of heavy meson fields. Since the lagrangian is well developed
in Refs.~\cite{tang98,EMQHD07}, we just outline the lagrangian here.
We begin with
\begin{eqnarray}
\lag_{N (\hat{\nu} \leqslant 3)}&=&
        \psibar{N}(i\ugamma{\mu}[\dcpartial{\mu}
        +ig_{\rho}\rho_{\mu}+ig_{v}V_{\mu}]+g_{A}\ugamma{\mu}\ugammafive\,
        \widetilde{a}_{\mu}-M+g_{s}\phi)N  \notag \\[5pt]
&& {}-\frac{f_{\rho}g_{\rho}}{4M}\, \psibar{N}\rho_{\mu\nu}
        \usigma{\mu\nu}N
        -\frac{f_{v}g_{v}}{4M}\, \psibar{N}V_{\mu\nu} \usigma{\mu\nu}N
        -\frac{\kappa_{\pi}}{M}\, \psibar{N}\,\widetilde{v}_{\mu\nu}
        \usigma{\mu\nu}N   \notag \\[5pt]
&& {}+\frac{4\beta_{\pi}}{M}\, \psibar{N}N \Tr(\widetilde{a}_{\mu}\widetilde{a}^{\mu})
        +\frac{i\kappa_{1}}{2M^{2}}\, \psibar{N}
        \dgamma{\mu}\overset{\leftrightarrow}{\dcpartial{\nu}} N
        \Tr\left(\widetilde{a}^{\mu}\widetilde{a}^{\nu}\right)    \notag
        \\[5pt]
&& {}+\frac{1}{4M}\, \psibar{N} \usigma{\mu\nu}( 2
        \lambda^{(0)}f_{s\mu\nu}+\lambda^{(1)}F^{(+)}_{\mu\nu} ) N \ , 
\end{eqnarray}
where $\dcpartial{\mu}$ is defined in Eq.~(\ref{eqn:allchiral}),
$\overset{\leftrightarrow}{\dcpartial{\nu}} \equiv \dcpartial{\nu} -
(\overset{\leftarrow}{\partial_{\nu}} - i\widetilde v_{\nu}+
i\vbg_{(s)\nu})$, and the new field tensors are $V_{\mu\nu} \equiv
\partial_{\mu} V_{\nu} - \partial_{\nu} V_{\mu}$ and
\begin{equation}
\rho_{\mn} \equiv \partial_{[\mu}\rho_{\nu
        ]}+i\overline{g}_{\rho}[\rho_{\mu}\,
        , \, \rho_{\nu}]
        + i ([\widetilde{v}_{\mu}\, , \, \rho_{\nu}] - \mu \leftrightarrow
        \nu) \to h \, \rho_{\mn} h^{\dagger}\ .
\end{equation}
The superscripts ${}^{(0)}$ and ${}^{(1)}$ denote the isospin.

Next is a purely mesonic term:
\begin{eqnarray}
\lag_{\mathrm{meson} (\hat{\nu} \leqslant4)} &=& \half \,
       \partial_{\mu}\phi\,\partial^{\mu}\phi
       + \frac{1}{4} f^{2}_{\pi} \Tr[\dcpartial{\mu}U(\ucpartial{\mu}U)^{\dagger}]
       +\frac{1}{4} f^{2}_{\pi}\, m^{2}_{\pi}\Tr(U+U^{\dagger}-2) \notag
       \\[5pt]
&& {}-\half \Tr(\rho_{\mu\nu}\rho^{\mu\nu}) -\frac{1}{4} \, V^{\mu\nu}V_{\mu\nu}
       \notag \\[5pt]
&& {}+\half \left(1+\eta_{1}\frac{g_{s}\phi}{M}
       +\frac{\eta_{2}}{2}\frac{g^{2}_{s}\phi^{2}}{M^{2}}\right)m^{2}_{v}\, V_{\mu}V^{\mu}
       + \frac{1}{4!}\, \zeta_{0} \, g^{2}_{v}(V_{\mu}V^{\mu})^{2} \notag
       \\[5pt]
&& {}+\left(1+\eta_{\rho}\frac{g_{s}\phi}{M}\right) m^{2}_{\rho} \Tr(\rho_{\mu}\rho^{\mu})
       -\left(\half+\frac{\kappa_{3}}{3!}\frac{g_{s}\phi}{M}
       +\frac{\kappa_{4}}{4!}\frac{g_{s}^{2}\phi^{2}}{M^{2}}\right)m_{s}^{2}\phi^{2}
       \notag \\[5pt]
&& {}+\frac{1}{2g_{\gamma}} \left(
       \Tr(F^{(+)\mu\nu}\rho_{\mu\nu})+\frac{1}{3}\, f_{s}^{\mu\nu}V_{\mu\nu}
       \right) \ . 
\end{eqnarray}
The $\nu=3$ and $\nu=4$ terms in $\lag_{\mathrm{meson} (\hat{\nu}
\leqslant4)}$ are important for describing the bulk properties of
nuclear many-body systems \cite{FST97,FTS95,FST96}. The only
manifest chiral-symmetry breaking is through the nonzero pion mass.

Finally, we have
\begin{eqnarray}
\lag_{N,\pi (\hat{\nu} =4)}&=&\frac{1}{2M^{2}}\,
       \psibar{N}\dgamma{\mu}(2\beta^{(0)}
       \partial_{\nu}f_{s}^{\mu\nu}+\beta^{(1)}\dcpartial{\nu}F^{(+)\mu\nu}
       +\beta_{A}^{(1)}\ugammafive \dcpartial{\nu}F^{(-)\mu\nu})N   \notag
       \\[5pt]
&& {}-\omega_{1}\Tr(F^{(+)}_{\mu\nu}\, \widetilde{v}^{\mu\nu})
       +\omega_{2} \Tr(\widetilde{a}_{\mu}\dcpartial{\nu}F^{(-)\mu\nu})
       +\omega_{3} \Tr \left( \widetilde{a}_{\mu} i
       \left[\widetilde{a}_{\nu} \, , \, F^{(+)\mu\nu} \right]\right)  \notag
       \\[5pt]
&& {}-g_{\rho\pi\pi}\frac{2f^{2}_{\pi}}{m^{2}_{\rho}}
       \Tr(\rho_{\mu\nu}\widetilde{v}^{\mu\nu})  \notag  \\[5pt]
&& {}+\frac{c_{1}}{M^{2}}\, \psibar{N}\ugamma{\mu} N
       \Tr \left(\widetilde{a}^{\nu}\, \psibar{F}^{(+)}_{\mu\nu}\right)
       +\frac{e_{1}}{M^{2}}\, \psibar{N}\ugamma{\mu}\,
       \widetilde{a}^{\nu}N \,
       \psibar{f}_{s\mu\nu} \notag \\[5pt]
&& {}+\frac{c_{1\rho}g_{\rho}}{M^{2}}\, \psibar{N}\ugamma{\mu} N\Tr
       \left(\widetilde{a}^{\nu}\, \psibar{\rho}_{\mu\nu}\right)
       +\frac{e_{1v}g_{v}}{M^{2}}\, \psibar{N}\ugamma{\mu}\,
       \widetilde{a}^{\nu}N \, \psibar{V}_{\mu\nu}\ . 
\end{eqnarray}
Note that $\lag_{N,\pi (\hat{\nu} =4)}$ is not a complete list of
all possible $\hat{\nu}=4$ interaction terms. However, $\beta^{(0)}$
and $\beta^{(1)}$ will be used in the form factors of the nucleon's
vector current, $\omega_{1,2,3}$ will contribute to the form factor
of the pion's vector current, and $g_{\rho\pi\pi}$ will be used in
the form factors that incorporate vector meson dominance. The
constants $c_{1}, e_{1}, c_{1\rho}$, and $e_{1\rho}$ will be
explained later when we discuss photon production.

\subsubsection{ Contributions to current matrix elements from
irreducible diagrams}

By comparing Eq.~(\ref{eqn:qcdL}) with the electroweak interactions
of quarks in the Standard Model \cite{IZ80,DGH92}, we can determine
the form of the background fields in terms of the vector bosons
$W^{\pm}_{\mu}$, $Z_{\mu}$, and $A_{\mu}$:
\begin{eqnarray}
l_{\mu}&=&-e\, \frac{\tau^{0}}{2}\, A_{\mu}
       +\frac{g}{\cos\theta_{w}}\sin^{2}\theta_{w}\, \frac{\tau^{0}}{2}\, Z_{\mu} \notag
       \label{eqn.lmubackground} \\[5pt]
&& {}-\frac{g}{\cos\theta_{w}}\frac{\tau^{0}}{2}\, Z_{\mu}
       -g V_{ud}\, \left( W^{+1}_{\mu}\, \frac{\tau_{+1}}{2}
       +W^{-1}_{\mu}\frac{\tau_{-1}}{2} \right) 
       \ , \\[5pt]
r_{\mu}&=&-e\, \frac{\tau^{0}}{2}\, A_{\mu}
       +\frac{g}{\cos\theta_{w}}\sin^{2}\theta_{w}\, \frac{\tau^{0}}{2}\, Z_{\mu} 
       \ , \label{eqn.rmubackground} \\[5pt]
\vbg_{(s)\mu}&=&-e\, \half\,
       A_{\mu}+\frac{g}{\cos\theta_{w}}\sin^{2}\theta_{w}\, \half\,
       Z_{\mu} \ , \label{eqn.vsmubackground}
\end{eqnarray}
where $g$ is the $SU(2)$ charge, and $\theta_{w}$ is the weak mixing
angle. Furthermore:
\begin{eqnarray}
f_{L\mu\nu}&=&-e\, \uhalftau{0} A_{[\nu , \mu]}
       + \frac{g}{\cos\theta_{w}}\sin^{2}\theta_{w}\, \uhalftau{0}\, Z_{[\nu ,\mu]}
       -\frac{g}{\cos\theta_{w}}\uhalftau{0}\, Z_{[\nu , \mu]} \notag
       \\[5pt]
&& {}-gV_{ud}\, \uhalftau{+1}\, W_{+1[\nu , \mu]}
       -gV_{ud}\, \uhalftau{-1}\, W_{-1[\nu , \mu]}  \notag \\[5pt]
&& {}+ \text{interference \ terms \ including } (WZ) , (WA) , (WW),
       \text{\ but \ no }\
       (ZA) \ , 
       \\[5pt]
f_{R\mu\nu}&=&-e\, \uhalftau{0}\, A_{[\nu , \mu]}
       + \frac{g}{\cos\theta_{w}}\sin^{2}\theta_{w}\, \uhalftau{0}\, Z_{[\nu , \mu]}
       \quad\text{\ (no \ interference \ terms)} \ , 
       \\[5pt]
f_{s\mu\nu}&=& -e\, \half\, A_{[\nu , \mu]}+
       \frac{g}{\cos\theta_{w}}\sin^{2}\theta_{w}\, \half\, Z_{[\nu , \mu]} \ . 
\end{eqnarray}
If we define [see Eq.~(\ref{eqn:qcdL})]
\begin{eqnarray}
\lag_{\mathrm{ext}}&\equiv&\vbg_{i\mu}V^{i\mu}-\abg_{i\mu}A^{i\mu}+\vbg_{(s)\mu}J^{B\mu}
       \notag \\[5pt]
&=& J^{L}_{i\mu}\, l^{i\mu}+J^{R}_{i\mu}\, r^{i\mu}+\vbg_{(s)\mu}J^{B\mu} \ , 
       \\[5pt]
\lag_{I}&=& -eJ^{EM}_{\mu}A^{\mu}
       -\frac{g}{\cos\theta_{w}}J^{NC}_{\mu}Z^{\mu}-gV_{ud}\,
       J^{L}_{+1 \mu}W^{+1\mu}-gV_{ud}\, J^{L}_{ -1 \mu}W^{-1\mu}\ ,  
\end{eqnarray}
and use Eqs.~(\ref{eqn.lmubackground}) to
(\ref{eqn.vsmubackground}), we can easily discover
\begin{eqnarray}
J^{L}_{i \mu} &\equiv& \half\, (V_{i\mu}+A_{i\mu}) \ , 
       \\[5pt]
J^{R}_{i \mu} &\equiv& \half\, (V_{i\mu}-A_{i\mu}) \ , 
       \\[5pt]
J^{EM}_{\mu}&=&V^{0}_{\mu}+\half\, J^{B}_{\mu} \ , 
       \\[5pt]
J^{NC}_{\mu}&=& J^{L0}_{\mu}-\sin^{2}\theta_{w}\, J^{EM}_{\mu} \ .
       \label{eqn:ncdef}
\end{eqnarray}
Here, $J^{B}_{\mu}$ is the baryon current, defined to be coupled to
$\vbg_{(s)}^{\mu}$. These relations are consistent with the charge
algebra $Q=T^{0}+B/2$. ($B$ is the baryon number.) $V^{i\mu}$ and
$A^{i\mu}$ are the isovector vector current and the  isovector
axial-vector current, respectively. We do not discuss ``seagull''
terms of higher order in the couplings because they do not enter in
our calculations \cite{EMQHD07,XZThesis}.

Based on the equations given above and the EFT lagrangian, we can
calculate the matrix elements $\bra{N} V^{i}_{\mu},A^{i}_{\mu},
J^{B}_{\mu} \ket{N}$ and $\bra{N; \pi} V^{i}_{\mu}, A^{i}_{\mu},
J^{B}_{\mu} \ket{N}$ at tree level; loops are not
included.\footnote{%
The expressions for the currents listed below differ from those in
Refs.~\protect{\cite{EMQHD07,AXC02}} because contributions from
non-minimal and vector meson dominance terms are included here.}
Since non-Goldstone vector bosons are included here, then by vector
meson dominance (VMD), we can extrapolate the current away from
$Q^{2}=0$ to some extent \cite{EMQHD07,BDS10}. The results are given
below, and the explicit calculations are shown in
Appendix~\ref{app:ff}. Note that $q^{\mu}$ is defined as the
\emph{incoming} momentum transfer at the vertex; in terms of initial
and final nucleon momenta, $q^{\mu} \equiv p^{\mu}_{nf} -
p^{\mu}_{ni}$. Similarly, $q^{\mu} + p^{\mu}_{ni} = p^{\mu}_{nf} +
k^{\mu}_{\pi}$ for pion production.

\begin{eqnarray}
\bra{N, B} V^{i}_{\mu} \ket{N, A}&=& \bra{B} \uhalftau{i}\ket{A}\,
          \psibar{u}_{f}
          \left(\dgamma{\mu}
          +2\delta F_{1}^{V,md}\,\frac{q^{2}\dgamma{\mu}-\slashed{q} q_{\mu}}{q^{2}}
          +2F_{2}^{V,md}\,\frac{\dsigma{\mn}iq^{\nu}}{2M}\right)u_{i} \notag \\
&&\null   \\[5pt]
&\equiv& \bra{B} \uhalftau{i}\ket{A}\, \psibar{u}_{f}
          \Gamma_{V\mu}(q) u_{i}
          \label{eqn:NNvectorcurrentwithff} \\[5pt]
&\overset{\mathrm{on\ shell}}{\equiv}&  \bra{B}\uhalftau{i}\ket{A}\,
          \psibar{u}_{f} \left( 2F_{1}^{V,md}\dgamma{\mu}
          +2F_{2}^{V,md}\,\frac{\dsigma{\mn}iq^{\nu}}{2M} \right)u_{i}\ , 
\end{eqnarray}

\begin{eqnarray}
\bra{N, B} J^{B}_{\mu} \ket{N, A}&=&  \delta_{B}^{A}\,
          \psibar{u}_{f} \left( \dgamma{\mu}
          +2\delta F_{1}^{S,md}\, \frac{q^{2}\dgamma{\mu}-\slashed{q} q_{\mu}}{q^{2}}
          +2F_{2}^{S,md}\frac{\dsigma{\mn}iq^{\nu}}{2M} \right) u_{i} 
          \\[5pt]
&\equiv& \delta_{B}^{A}\, \psibar{u}_{f} \Gamma_{B\mu}(q) u_{i}
          \label{eqn:NNbaryoncurrentwithff} \\[5pt]
&\overset{\mathrm{on\ shell}}{\equiv}&  \delta_{B}^{A} \,
          \psibar{u}_{f}
          \left(2F_{1}^{S,md}\,\dgamma{\mu}+2F_{2}^{S,md}\,
          \frac{\dsigma{\mn}iq^{\nu}}{2M} \right) u_{i} \ ,
\end{eqnarray}

\begin{eqnarray}
%
\bra{N, B; \pi,j,k_{\pi}} A^{i}_{\mu} \ket{N, A} &=&
-\frac{\epsilon^{i}_{\,jk}}{f_{\pi}}\,
          \bra{B}\uhalftau{k}\ket{A}\,
          \psibar{u}_{f}\ugamma{\nu}u_{i} \notag \\[5pt]
&&{}\times \left[g_{\mn}+2\delta F_{1}^{V,md}((q-k_{\pi})^{2})
          \frac{ q\cdot(q-k_{\pi})g_{\mn}-(q-k_{\pi})_{\mu}q_{\nu}}{(q-k_{\pi})^{2}}
          \right] \notag \\[5pt]
%
%
&&{}-\frac{\epsilon^{i}_{\,jk}}{f_{\pi}}\, \bra {B}
\uhalftau{k}\ket{A}\,
          \psibar{u}_{f}\frac{\dsigma{\mn}iq^{\nu}}{2M}\, u_{i}
          \bigg[ 2\lambda^{(1)}  \notag \\[5pt]
&& \qquad\qquad {}\left. +2\delta F^{V,md}_{2}((q-k_{\pi})^{2})\,
          \frac{q\cdot(q-k_{\pi})}{(q-k_{\pi})^{2}} \right] 
          \\[5pt]
&\equiv&  \frac{\epsilon^{i}_{\,jk}}{f_{\pi}}\, \bra {B}
          \uhalftau{k}\ket{A}\, \psibar{u}_{f} \Gamma_{A\pi \mu}(q,k_{\pi})
          u_{i} \ . \label{eqn:NNpionaxialcurrentwithff}
\end{eqnarray}

Here ($m_{\rho}=0.776\ \mathrm{GeV}$, $m_{v}=0.783 \ \mathrm{GeV}$):
\begin{eqnarray}
F_{1}^{V,md}&=&\half \left(1+ \frac{\beta^{(1)}}{M^{2}}\,  q^{2}
          -\frac{g_{\rho}}{g_{\gamma}} \frac{q^{2}}{q^{2}-m^{2}_{\rho}} \right),
          \ \beta^{(1)}=-1.35, \ \frac{g_{\rho}}{g_{\gamma}}=2.48  \label{eqn:F1vmd}
          \ , \\[5pt]
F_{2}^{V,md}&=&\half
          \left(2\lambda^{(1)}-\frac{f_{\rho}g_{\rho}}{g_{\gamma}}
          \frac{q^{2}}{q^{2}-m^{2}_{\rho}} \right),
          \ \lambda^{(1)}=1.85, \, f_{\rho}=3.04 \label{eqn:F2vmd}
          \ , \\[5pt]
F_{1}^{S,md}&=&\half \left(1+ \frac{\beta^{(0)}}{M^{2}}\, q^{2}
          -\frac{2 g_{v}}{3 g_{\gamma}} \frac{q^{2}}{q^{2}-m^{2}_{v}} \right),
          \ \beta^{(0)}=-1.40, \ \frac{g_{v}}{ g_{\gamma}}=3.95 \label{eqn:F1smd}
          \ , \\[5pt]
F_{2}^{S,md}&=&\half \left(2\lambda^{(0)}-\frac{2
          f_{v}g_{v}}{3g_{\gamma}}
          \frac{q^{2}}{q^{2}-m^{2}_{v}} \right),
          \ \lambda^{(0)}=-0.06, \ f_{v}=-0.19 \label{eqn:F2smd}
          \ , \\[5pt]
\delta F_{1}^{V/S,md}(q^{2}) &\equiv & F_{1}^{V/S,md}(q^{2})- F_{1}^{V/S,md}(0)
          \ , \\[5pt]
\delta F_{2}^{V/S,md}(q^{2}) &\equiv &F_{2}^{V/S,md}(q^{2})-
          F_{2}^{V/S,md}(0) \ . 
\end{eqnarray}

If we follow a procedure similar to that used in calculating
$\bra{N} V^{i}_{\mu} \ket{N}$ and $\bra{N; \pi} A^{i}_{\mu}\ket{N}$,
we can expand the axial-vector current in powers of $q^2$ using the
lagrangian constants $g_{A}$ and $\beta_{A}^{(1)}$. In fact, we can
improve on this by including the axial-vector meson ($a_{1\mu}$)
contribution to the matrix elements, which would arise from the
interactions: $g_{a_{1}}\psibar{N}\ugamma{\mu}\ugammafive a_{1\mu}
N$ and $c_{a_{1}}\Tr \left(F^{(-)\mn} a_{1\mn} \right)$. Here
$a_{1\mu}=a_{1i\mu}\tau^{i}/2$ and $a_{1\mn}\equiv\dcpartial{\mu}
a_{1\nu}-\dcpartial{\nu} a_{1\mu}$, where $a_{1i\mu}$ are the fields
of the $a_{1}$ meson (whose mass is denoted as $m_{a_{1}}=1.26 \
\mathrm{GeV}$). Then we obtain (details can be found in
Appendix~\ref{app:ff})
\begin{eqnarray}
\bra{N, B} A^{i}_{\mu} \ket{N, A} &=& -G_{A}^{md}(q^{2})\,
          \bra{B} \uhalftau{i}\ket{A}\, \psibar{u}_{f} \left(
          \dgamma{\mu}-\frac{q_{\mu}\slashed{q}}{q^{2}-m_{\pi}^{2}}\right)
          \ugammafive u_{i} \notag \\[5pt]%
&\equiv& \bra{B} \uhalftau{i}\ket{A}\, \psibar{u}_{f}\Gamma_{A \mu}
(q) u_{i}\ ,
          \label{eqn:NNaxialcurrentff}  \\[5pt]%
\bra{N, B,\pi, j}  V^{i}_{\mu} \ket{N, A}&=&
\frac{\epsilon^{i}_{\,jk}}{f_{\pi}}\,
          \bra {B} \uhalftau{k}\ket{A}\, \psibar{u}_{f}\bigg( G_{A}^{md}(0)
          \dgamma{\mu}\ugammafive \notag \\[5pt]
&& {}+\delta G_{A}^{md}((q-k_{\pi})^{2})\,
          \frac{q\cdot (q-k_{\pi})g_{\mn}-(q-k_{\pi})_{\mu}
          q_{\nu}}{(q-k_{\pi})^{2}}\,
          \ugamma{\nu}\ugammafive \bigg) u_{i} \ , 
          \\[5pt]
&\equiv& \frac{\epsilon^{i}_{\,jk}}{f_{\pi}}\,
          \bra {B} \uhalftau{k}\ket{A}\, \psibar{u}_{f} \Gamma_{V\pi \mu}(q,k_{\pi})u_{i}
          \ . \label{eqn:NNpionvectorcurrentff}
\end{eqnarray}

Here, the definitions of $G_{A}^{md}(q^{2})$ and $\delta
G_{A}^{md}(q^{2})$ are
\begin{eqnarray}
G_{A}^{md}(q^{2})&\equiv&
          g_{A}-\beta_{A}^{(1)}\,\frac{q^{2}}{M^{2}}
          - \frac{2c_{a_{1}}g_{a_{1}}q^{2}}{q^{2}-m_{a_{1}}^{2}}\ , \notag \\[5pt]
g_{A}&=&1.26, \ \beta_{A}^{(1)}=2.27, \ c_{a_{1}}g_{a_{1}}=3.85\ ,
          \label{eqn:defofGA1} \\[5pt]
\delta G_{A}^{md}(q^{2}) &\equiv&
G_{A}^{md}(q^{2})-G_{A}^{md}(0)=G_{A}^{md}(q^{2})-g_{A} \ . \label{eqn:defofGA2} 
\end{eqnarray}%

For the pion's vector current form factor \cite{FST97}:
\begin{eqnarray}
\bra{\pi,k,k_{\pi}} V^{i}_{\mu}\ket{\pi,j,k_{\pi}-q}
&=&i\epsilon^{ij}_{\;\; k}
          (2k_{\pi}-q)_{\mu}
          \left( 1-\frac{g_{\rho\pi\pi}}{g_{\gamma}}\frac{q^{2}}{q^{2}-m^{2}_{\rho}}
          \right) \quad\text{pion on shell}
          \notag \\[5pt]
&\equiv&  i \epsilon^{ij}_{\;\; k}\, (2k_{\pi}-q)_{\mu}
          F_{\pi}^{md}(q^{2}), \quad
          \frac{g_{\rho\pi\pi}}{g_{\gamma}}=1.20
          \label{eqn:Fpivmd}
          \\
\Longrightarrow  &&  \text{(pion off shell)} \notag \\
\bra{\pi,k,k_{\pi}} V^{i}_{\mu}\ket{\pi,j,k_{\pi}-q} &=&
i\epsilon^{ij}_{\;\; k}\,
          \left[(2k_{\pi}-q)_{\mu} +2 \delta F_{\pi}^{md}(q^{2}) \left(k_{\pi \mu}
          -\frac{q\cdot k_{\pi}}{q^{2}}\, q_{\mu}\right)\right] \notag
          \\[5pt]
&\equiv&  i\epsilon^{ij}_{\;\; k}\, P_{V \mu}(q,k_{\pi}) \ ,
          \label{eqn:pionvectorcurrentff}          \\[5pt]
\text{with} \qquad  \delta F_{\pi}^{md}(q^{2})
          &=&F_{\pi}^{md}(q^{2}) -F_{\pi}^{md}(0) \ . 
\end{eqnarray}

To determine the couplings in Eqs.~(\ref{eqn:F1vmd}),
(\ref{eqn:F2vmd}), (\ref{eqn:F1smd}), (\ref{eqn:F2smd}),
(\ref{eqn:defofGA1}) and (\ref{eqn:Fpivmd}), we compare our results
with the conventional experimentally fitted form factors
\cite{FST97,kelly04}. We make the behavior of the form factors near
$q^{2}=0$ as close to the conventional form factors as possible. In
this calculation, we fit our `md' form factors to those in
\cite{kelly04} for the nucleon's vector and baryon current. The
conventional nucleon's axial-vector current used to fit our
$G_{A}^{md}$ is parameterized the in literature \cite{EW88} as
$G_{A}(q^{2})=g_{A}/(1-q^{2}/M_{A}^{2})^{2},$ with $g_{A}=1.26$ and
$M_{A}=1.05 \, \mathrm{GeV}$. As shown in Ref.~\cite{BDS10}, the
form factors due to vector meson dominance become inadequate at
$Q^{2} \approx 0.3 \, \mathrm{GeV}^{2}$. This is also true of the
axial parametrization. This indicates that the EFT lagrangian is
only applicable for $E_{l} \leqslant 0.5 \, \mathrm{GeV}$ in
lepton--nucleon interactions, above which $Q^{2}$ exceeds the limit.
This will be clarified in the kinematical analysis of
Sec.~\ref{subsec:kinematics}.

\subsection{Lagrangian involving $\Delta (1232)$}

\subsubsection{Chiral symmetry and power counting}
The $\Delta^{\ast a}$ belong to an $I=3/2$ multiplet as
non-Goldstone particles in the low-energy theory. The chiral
symmetry realization involving non-Goldstone particles,
$\psi_{\alpha}=\left( N_{A}, \rho_{i}, \Delta_{a} \right)_{\alpha}$,
has been given generally in Sec.~\ref{subsubsec:chisym}. Moreover,
in the power counting of vertices, the $\Delta$ is counted the same
way as the nucleon.

\subsubsection{Spin-$3/2$ particles as manifest degrees of freedom in field theory
 \cite{RaritaSchwinger}}

We briefly discuss the well-known pathologies of high-spin particles
in field theory
\cite{JohnsonSudarshan,CRHagen,Capri1980,VeloZwanzinger,LPSSingh73}.
It was discovered that with strong couplings, or strong fields, or
large field variations, a field theory involving high-spin fields
cannot be self-consistent. These investigations were carried out in
the lagrangian formalism with a finite number of interaction terms.
The reason for the pathology is that the unphysical degrees of
freedom of the $\Delta$ may contribute when the constraints on the
$\Delta$ fields are lost after adding interactions. However, it has
more recently been realized that the number of degrees of freedom is
correct in low-energy effective field theory, which has, in
principle, an infinite number of interaction terms, as long as we
work in the limit of low-energy and weak (boson) fields
\cite{Pascalutsa98,Pascalutsa01,Krebs09}. As we know, in the
low-energy effective field theory, the interactions and boson fields
are scaled by $1/M \approx 1/(1 \,\mathrm{GeV})$. The essence of the
argument is that in the perturbative picture, the spin-1/2
components of the off-shell $\Delta$ (treating the $\Delta$ as a
stable particle) will behave as \emph{local contact interactions}
without pole structure, which should be calibrated together with the
complete set of contact terms in the effective lagrangian.

Another issue is about the so-called off-shell couplings, which have
the form $\gamma_{\mu} \psi^{\mu}$, $\partial_{\mu} \psi^{\mu},
\overline{\psi}^{\mkern3mu\mu} \gamma_{\mu},$ and $\partial_{\mu}
\overline{\psi}^{\mkern3mu\mu}$, and which had also been discussed
together with pathologies. From the modern effective field theory
viewpoint, it has been concluded \cite{tang96,Krebs10} that these
off-shell couplings are redundant. The physical picture of this will
be clarified after introducing the lagrangian in
Sec.~\ref{subsubsec:deltaprop}.

\subsubsection{Lagrangian and $\Delta$ propagator renormalization}
\label{subsubsec:deltaprop}

Consider first $\lag_{\Delta; \pi,\rho,V,\phi}\ (\hat{\nu}\leqslant
3)$, which is essentially a copy of the corresponding lagrangian for
nucleons:
\begin{eqnarray}
\lag_{\Delta; \pi,\rho,V,\phi}&=&\frac{-i}{2}\,
          \psibar{\Delta}^{\mkern4mu a}_{\mu}\{\usigma{\mn}\, , \, (i\slashed{\ucpartial{}}
          -h_{\rho}\slashed{\rho}-h_{v}\slashed{V}-m
          +h_{s}\phi)\}_{a}^{\mkern3mu b}\, \Delta_{b\nu}
          + \widetilde{h}_{A}\psibar{\Delta}^{\mkern4mu a}_{\mu}
          \slashed{\widetilde{a}}_{a}^{\mkern3mu b} \ugammafive \Delta^{\mu}_{b}
          \notag \\[5pt]
&&{}-\frac{\widetilde{f}_{\rho}h_{\rho}}{4m}\,
          \psibar{\Delta}_{\lambda}\,
          \rho_{\mn}\usigma{\mn}\Delta^{\lambda}
          -\frac{\widetilde{f}_{v}h_{v}}{4m}\,
          \psibar{\Delta}_{\lambda} V_{\mn}\usigma{\mn}\Delta^{\lambda} \notag
          \\[5pt]
&&{}-\frac{\widetilde{\kappa}_{\pi}}{m}\,
          \psibar{\Delta}_{\lambda}\widetilde{v}_{\mn}\usigma{\mn}\Delta^{\lambda}
          +\frac{4\widetilde{\beta}_{\pi}}{m}\,
          \psibar{\Delta}_{\lambda}\Delta^{\lambda}\Tr(\widetilde{a}^{\mu}\,
          \widetilde{a}_{\mu}) \ .
\end{eqnarray}
Here the sub- and superscripts $a, b = (\pm 3/2, \pm 1/2)$, so the
isovector mesons use the isospin-3/2 representation. A few words on
pathologies and off-shell couplings are in order. Based on
$\lag_{\Delta; \pi,\rho,V,\phi}$, we can find the free propagator
for the $\Delta$. Details are shown in
Appendix~\ref{app:deltapropagator}, and here we ignore $\mn$ indices
and isospin indices:
\begin{eqnarray}
S_{F}^{0}(p)&\equiv& S_{F}^{0 (\threehalf)}(p)
          + S_{F}^{0(\threehalf \perp)}(p) \notag
          \\[5pt]
&=&P^{(\frac{3}{2})} \frac{-1}{\slashed{p}-m+i \epsilon} P^{(\frac{3}{2})} \notag
          \\[5pt]
&&{}+ P^{(\threehalf \perp)}
          \left[-\frac{1}{\sqrt{3}m}P^{(\half)}_{12}-\frac{1}{\sqrt{3}m}P^{(\half)
          }_{21}+P^{(\half)}_{22}\frac{2}{3m^{2}}(\slashed{p}+m)
          P^{(\half)}_{22} \right] P^{(\threehalf \perp)} \ . 
\end{eqnarray}
The operator $P^{(\frac{3}{2})}$ projects the general
Rarita--Scwhinger field $\psi_{\mu}$ to spin-3/2 objects, while
$P^{(\frac{3}{2}\perp)} \equiv {\bf{1}}-P^{(\frac{3}{2})}$ is the
orthogonal projection operator. In
Appendix~\ref{app:deltapropagator}, we also show that the
self-energy $\Sigma_{\mn} =\Sigma^{\Delta} g_{\mn}+\delta
\Sigma_{\mn}$ can be written as
\begin{eqnarray}
\Sigma &=&  P^{(\frac{3}{2})}\Sigma^{\Delta} P^{(\frac{3}{2})}
          + P^{(\threehalf \perp)} \Sigma  P^{(\threehalf \perp)} \notag
          \\[3pt]
&\equiv& \Sigma^{(\frac{3}{2})} + \Sigma^{(\frac{3}{2} \perp)} \ .
\end{eqnarray}
We can then renormalize the $\Delta$ propagator as follows:
\begin{eqnarray}
S_{F}&=& (S_{F}^{0 (\threehalf)} + S_{F}^{0 (\threehalf \perp)})
          +(S_{F}^{0 (\threehalf)} + S_{F}^{0 (\threehalf \perp)})
          (\Sigma^{(\frac{3}{2})} + \Sigma^{(\frac{3}{2} \perp)})
          (S_{F}^{0 (\threehalf)} + S_{F}^{0 (\threehalf \perp)})+ \dotsb \notag
          \\[5pt]
&=& S_{F}^{0 (\threehalf)}+ S_{F}^{0 (\threehalf)} \Sigma^{(\frac{3}{2})}
          S_{F}^{0 (\threehalf)}+ \dotsb \notag \\[3pt]
&&{}+S_{F}^{0 (\threehalf \perp)}+S_{F}^{0 (\threehalf \perp)}
          \Sigma^{(\frac{3}{2} \perp)} S_{F}^{0 (\threehalf \perp)}+
          \dotsb
\end{eqnarray}

From this we can conclude that the renormalized propagator
$S_{F}\equiv S_{F}^{(\threehalf)}+S_{F}^{(\threehalf \perp)}$. The
resonant contribution is
$S_{F}^{(\threehalf)}=S_{F}^{0(\threehalf)}+S_{F}^{0(\threehalf)}
\Sigma^{(\frac{3}{2})} S_{F}^{(\threehalf)}$. The  background
contribution is $S_{F}^{(\threehalf \perp)}=S_{F}^{0(\threehalf
\perp)}+S_{F}^{0(\threehalf \perp)} \Sigma^{(\frac{3}{2} \perp)}
S_{F}^{(\threehalf \perp)}$. We can see that renormalization will
shift the pole position of the resonant part. Moreover, as long as
power counting is valid, i.e., $O(\Sigma /M) \ll 1$, we will always
be far away from the unphysical pole in the renormalized
non-resonant part proportional to $\{ 1/[1 - O(\Sigma /M)]\}^{-1}$.
This also suggests that we will not see an unphysical pole in the
renormalized propagator, when we work in the low-energy perturbative
region, where power counting makes sense. So we indeed have the
right degrees of freedom in our low-energy theory, namely, a single
pole at the resonance. Thus perturbative unitarity is not obviously
violated in this theory with high-spin fields.

Another issue is the $1/p^{2}$ singularity in the projection
operators. In principle, when we are in the low-energy region,
$p^{2}=(p_{ni}+k)^{2}$ is always positive (timelike) in all the
channels. Here $k$ is some general small momentum compared to
$p_{ni}$, which is the nucleon's momentum that is almost on shell.

The preceding discussion also helps to clarify the redundancy of the
off-shell couplings. The self-energy due to these couplings will not
contribute in the renormalizaton of $S_{F}^{(\threehalf)}$, but it
will indeed change the non-resonant part. However, the effects due
to these couplings can be expanded in powers of the momenta. So they
will essentially look like \emph{higher-order contact terms} without
the $\Delta$. This justifies the redundancy of these couplings. To
ignore these couplings in a way that does not break term-by-term
chiral symmetry, we can always combine $\partial^{\mu}$ with pion
fields so that it becomes $\ucpartial{\mu}$. This indicates that
those couplings with $\ucpartial{\mu}$ or $\ugamma{\mu}$ contracted
with $\Delta_{\mu}$ can be ignored without breaking manifest chiral
symmetry \cite{Krebs09,Krebs10}. For a concrete example of these
results, see Appendix~\ref{app:deltalagrangianexample}.

To produce the $N \leftrightarrow \Delta$ transition currents, we
construct the following lagrangians ($\hat{\nu}\leqslant4$):
\begin{eqnarray}
\lag_{\Delta;N;\pi}&=&h_{A}\psibar{\Delta}^{\mkern4mu a
          \mu}\, \Tdagger{iA}{a}\, \widetilde{a}_{i\mu}N_{A} +C.C. \ , 
\end{eqnarray}
\begin{eqnarray}
\lag_{\Delta;N;\mathrm{background}}&=& \frac{ic_{1\Delta}}{M}\,
          \psibar{\Delta}^{\mkern4mu a}_{\mu}\dgamma{\nu}\ugammafive\,
          \Tdagger{iA}{a}F_{i}^{(+)\mn}N_{A}
          +\frac{ic_{3\Delta}}{M^{2}}\,
          \psibar{\Delta}^{\mkern4mu a}_{\mu}\, i\ugammafive\, \Tdagger{iA}{a}
          (\dcpartial{\nu}F^{(+)\mn})_{i} N_{A}  \notag \\[5pt]
&&{}+\frac{c_{6\Delta}}{M^{2}}\, \psibar{\Delta}^{\mkern4mu
          a}_{\lambda}
          \dsigma{\mu\nu} \Tdagger{iA}{a}
          (\ucpartial{\lambda}\psibar{F}^{(+)\mn})_{i} N_{A} \notag \\[5pt]
&&{}-\frac{d_{2\Delta}}{M^{2}}\, \psibar{\Delta}^{\mkern4mu
          a}_{\mu}\,
          \Tdagger{iA}{a} (\dcpartial{\nu}F^{(-)\mn})_{i}N_{A}
          -\frac{id_{4\Delta}}{M}\, \psibar{\Delta}^{\mkern4mu a}_{\mu} \dgamma{\nu}\,
          \Tdagger{iA}{a} F_{i}^{(-)\mn}N_{A} \notag \\[5pt]
&&{}-\frac{id_{7\Delta}}{M^{2}}\, \psibar{\Delta}^{\mkern4mu
          a}_{\lambda}
          \dsigma{\mn}\Tdagger{iA}{a}(\ucpartial{\lambda}F^{(-)\mn})_{i}N_{A}
%
          + C.C. \ , \label{eqn:truelagdeltaNbackground}
\end{eqnarray}
\begin{eqnarray}
\lag_{\Delta;N;\rho}&=&\frac{ic_{1\Delta\rho}}{M}\,
          \psibar{\Delta}^{\mkern4mu a}_{\mu}\,\dgamma{\nu}\ugammafive\,
          \Tdagger{iA}{a}\rho_{i}^{\mn}N_{A}
          +\frac{ic_{3\Delta\rho}}{M^{2}}\, \psibar{\Delta}^{\mkern4mu a}_{\mu}\,
          i\ugammafive\, \Tdagger{iA}{a} (\dcpartial{\nu}\rho^{\mn})_{i} N_{A}
          \notag \\[5pt]
&&{} +\frac{c_{6\Delta\rho}}{M^{2}}\, \psibar{\Delta}^{\mkern4mu
          a}_{\lambda} \dsigma{\mu\nu}\,
          \Tdagger{iA}{a} (\ucpartial{\lambda}\, \psibar{\rho}^{\mkern2mu\mn})_{i} N_{A}
          + C.C.  \label{eqn:truelagdeltaNrho}
\end{eqnarray}
Terms omitted from these lagrangians are either redundant or are not
relevant to our calculations \cite{XZThesis}.

\subsubsection{Transition currents}

It is easy to expect the validity of the following definitions:
\begin{eqnarray}
\bra{\Delta, a,p_{\Delta}} V^{i\mu} \ket{N, A,p_{N}} &\equiv&
          \Tdagger{iA}{a}\, \psibar{u}_{\Delta\alpha}(p_{\Delta})\,
          \Gamma_{V}^{\alpha\mu}(q)\, u_{N}(p_{N})\ , 
          \label{eqn:transitioncurrentVvertex} \\[5pt]
\bra{\Delta, a,p_{\Delta}} A^{i\mu} \ket{N, A,p_{N}} &\equiv&
          \Tdagger{iA}{a}\, \psibar{u}_{\Delta\alpha}(p_{\Delta})\,
          \Gamma_{A}^{\alpha\mu}(q)\, u_{N}(p_{N})\ .
\label{eqn:transitioncurrentvertex}
\end{eqnarray}
Based on the lagrangians given previously, we find (note that
$\sigma_{\mn} \epsilon^{\mu\nu\alpha\beta} \propto i
\sigma^{\alpha\beta} \ugammafive$)
\begin{eqnarray}
\Gamma_{V}^{\alpha\mu}&=&\frac{2c_{1\Delta}(q^{2})}{M}\,
          (q^{\alpha}\ugamma{\mu}-\slashed{q}g^{\alpha\mu}) \ugammafive
          +\frac{2c_{3\Delta}(q^{2})}{M^{2}}\, (q^{\alpha}q^{\mu}-g^{\alpha\mu}q^{2})
          \ugammafive  \notag \\[5pt]
&&{}-\frac{8c_{6\Delta}(q^{2})}{M^{2}}\,
          q^{\alpha}\usigma{\mn}iq_{\nu}\ugammafive \ ,
          \\[5pt]
\Gamma_{A}^{\alpha\mu}&=&-h_{A} \left(
          g^{\alpha\mu}-\frac{q^{\alpha}q^{\mu}}{q^{2}-m^{2}_{\pi}} \right)
          +\frac{2d_{2\Delta}}{M^{2}}\, (q^{\alpha}q^{\mu}-g^{\alpha\mu}q^{2})
          -\frac{2d_{4\Delta}}{M}\, (q^{\alpha}\ugamma{\mu}-g^{\alpha\mu}\slashed{q})
          \notag \\[5pt]
&& {}-\frac{4d_{7\Delta}}{M^{2}}\, q^{\alpha}\usigma{\mn}iq_{\nu}\ ,
          \\[5pt]
c_{i\Delta}(q^{2}) &\equiv& c_{i\Delta}
          +\frac{c_{i\Delta\rho}}{2g_{\gamma}}\, \frac{q^{2}}{q^{2}-m^{2}_{\rho}}\ ,
          \qquad i=1, 3, 6, \\[5pt]
c_{1\Delta}&=&1.21, \qquad c_{3\Delta}=-0.61, \qquad c_{6\Delta}=-0.078, \notag \\
\frac{c_{1\Delta\rho}}{g_{\gamma}}&=&-4.58, \quad
\frac{c_{3\Delta\rho}}{g_{\gamma}}=2.32, \qquad
\frac{c_{6\Delta\rho}}{g_{\gamma}}=0.30. \label{eqn:cimd}
\end{eqnarray}

Quite similar to the $c_{i \Delta}(q^{2})$, we can also introduce
axial-vector meson exchange into the axial transition current, which
leads to a structure for the $d_{i \Delta}(q^{2})$ that is similar
to the vector transition current form factors. There is one subtlety
associated with the realization of $h_{A}(q^{2})$, which is the same
as the one detailed in Appendix~\ref{app:ff} for
$G_{A}^{md}(q^{2})$: with our lagrangian, we have a pion-pole
contribution associated only with the $h_{A}$ coupling, and all the
higher-order terms contained in $\delta h_{A}(q^{2})\equiv
h_{A}(q^{2})-h_{A}$ conserve the axial transition current. With the
limited information about manifest chiral-symmetry breaking, we will
ignore this subtlety and still use the form similar to the
$c_{1\Delta}(q^{2})$ to parameterize $h_{A}(q^{2})$. The
axial-vector meson couplings $h_{\Delta a_{1}}$ and $d_{i \Delta
a_{1}}$ will be combinations of $g_{a_{1}}$ and the coupling
strength of $\Delta a_{1} N$ interactions. So we have
\begin{eqnarray}
h_{A}(q^{2}) &\equiv& h_{A}+h_{\Delta a_{1}} \,  \frac{q^{2}}{q^{2}-m^{2}_{a_{1}}}\ ,
          \\[5pt]
d_{i\Delta}(q^{2}) &\equiv& d_{i\Delta}
          + d_{i \Delta a_{1}}\, \frac{q^{2}}{q^{2}-m^{2}_{a_{1}}}\ ,
          \qquad i=2, 4, 7,  \\[5pt]
h_{A}&=&1.40, \qquad   d_{2\Delta}=-0.087, \qquad d_{4\Delta}=0.20,
          \qquad d_{7\Delta}=-0.04,   \notag \\
h_{\Delta a_{1}}&=&-3.98, \quad  d_{2 \Delta a_{1}}=0.25, \qquad
          d_{4 \Delta a_{1}}= -0.58, \qquad d_{7 \Delta a_{1}}=0.12.
          \label{eqn:dimd}
\end{eqnarray}

To determine the coefficients in the transition form factors shown
in Eqs.~(\ref{eqn:cimd}) and (\ref{eqn:dimd}), we will compare ours
with the conventional ones used in the literature. In
Refs.~\cite{HERNANDEZ07, GRACZYK09} for example, the definition is
\begin{eqnarray}
\bra{\Delta,\half} j_{cc+}^{\mu} \ket{N,-\half} &\equiv&
          \psibar{u}_{\alpha}(p_{\Delta}) \left\{ \left[
          \frac{C_{3}^{V}}{M}\, (g^{\alpha\mu} \slashed{q}-q^{\alpha}
          \ugamma{\mu}) +\frac{C_{4}^{V}}{M^{2}}\, (q\cdot
          p_{\Delta}\, g^{\alpha\mu}-q^{\alpha} p_{\Delta}^{\mu}) \right. \right. \notag
          \\[5pt]
&&{}+ \left. \frac{C_{5}^{V}}{M^{2}}\, (q\cdot p_{N}\,
          g^{\alpha\mu}-q^{\alpha}
          p_{N}^{\mu}) \right] \ugammafive  \notag
          \\[5pt]
&&{}+\left[ \frac{C_{3}^{A}}{M}\,
          (g^{\alpha\mu}\slashed{q}-q^{\alpha}\ugamma{\mu})
          +\frac{C_{4}^{A}}{M^{2}}\, (q\cdot p_{\Delta}\, g^{\alpha\mu}-q^{\alpha}
          p_{\Delta}^{\mu}) \right. \notag
          \\[5pt]
&&{}+\left. \left. C_{5}^{A}g^{\alpha\mu}+\frac{C_{6}^{A}}{M^{2}}\,
          q^{\mu}q^{\alpha} \right] \right\} u(p_{N}) 
          \\[5pt]
&\equiv& -\sqrt{\frac{2}{3}}\
          \psibar{u}_{\alpha}(p_{\Delta})\left(\Gamma_{V}^{\alpha\mu}
          +\Gamma_{A}^{\alpha\mu}\right)
          u(p_{N}) \ ,  
\end{eqnarray}
where $\Gamma_{V}^{\alpha\mu}$ and $\Gamma_{A}^{\alpha\mu}$ are
defined in Eqs.~(\ref{eqn:transitioncurrentVvertex}) and
(\ref{eqn:transitioncurrentvertex}). The basis given above is known
to be complete for the transition matrix element. The
phenomenological form factors are listed below \cite{GRACZYK09}:
\begin{eqnarray}
C_{3}^{V}(q^{2})&=& \frac{2.13}{1- (q^{2}/4M_{V}^{2})}\,
          G_{D}(q^{2}) \ , 
          \\[5pt]
C_{4}^{V}(q^{2})&=&
\frac{-1.51}{1- (q^{2}/4M_{V}^{2})}\, G_{D}(q^{2})\ , 
          \\[5pt]
C_{5}^{V}(q^{2})&=& \frac{0.48}{1- (q^{2}/ 0.776 M_{V}^{2})}\,
          G_{D}(q^{2}) \ . 
          \\[5pt]
\text{Here} \qquad
          G_{D}(q^{2})&\equiv& \frac{1}{[1- (q^{2}/M_{V}^{2})]^{2}}\
          ,
          \qquad \text{and} \ M_{V}=0.84 \, \mathrm{GeV.}  
          \\[5pt]
C_{3}^{A}(q^{2})&=& 0 \ , 
          \\[5pt]
C_{4}^{A}(q^{2})&=&-\frac{1}{4}\, C_{5}^{A}(q^{2}) \ , 
          \\[5pt]
C_{6}^{A}(q^{2})&=&C_{5}^{A}(q^{2})\, \frac{M^{2}}{m_{\pi}^{2}-q^{2}}\ , 
          \\[5pt]
C_{5}^{A}(q^{2})&=&1.14 \left(1 + \frac{1.21
          q^{2}}{2\,\mathrm{GeV}^2 -q^{2}} \right)
          \frac{1}{[1 - (q^{2}/M_{A}^{2})]^{2}} \ ,  \quad \text{where}\
          M_{A}=1.29 \, \mathrm{GeV.}
          \notag \\
&\null&
\end{eqnarray}

To equate the two different representations of the transition
currents when $q^{2}=0$ and the baryons are on shell, we have:
\begin{eqnarray}
c_{1\Delta}&=&\sqrt{\frac{3}{2}}\,\left[\frac{C_{3}^{V}}{2}+\frac{m-M}{2M}\frac{(C_{4}^{V}
          +C_{5}^{V})}{2}\right] \ ,
\label{eqn:axialtransitionff1}
          \\[5pt]
c_{3\Delta}&=&\sqrt{\frac{3}{2}}\ \frac{(C_{4}^{V}-C_{5}^{V})}{4} \ , 
\label{eqn:axialtransitionff2}
          \\[5pt]
c_{6\Delta}&=&\sqrt{\frac{3}{2}}\ \frac{(C_{4}^{V}+C_{5}^{V})}{16} \ . 
\label{eqn:axialtransitionff3}
          \\[5pt]
h_{A}&=&\sqrt{\frac{3}{2}}\ C_{5}^{A} \ ,
\label{eqn:axialtransitionff4}
          \\[5pt]
d_{2\Delta}&=&\sqrt{\frac{3}{2}}\ \frac{C_{4}^{A}}{4}  , 
\label{eqn:axialtransitionff5}
          \\[5pt]
d_{4\Delta}&=&-\sqrt{\frac{3}{2}} \left(\frac{C_{3}^{A}}{2}
          +\frac{m+M}{2M} \frac{C_{4}^{A}}{2}\right) \ , 
\label{eqn:axialtransitionff6}
          \\[5pt]
d_{7\Delta}&=&\sqrt{\frac{3}{2}}\ \frac{C_{4}^{A}}{8} \ .
\label{eqn:axialtransitionff}
\end{eqnarray}

To determine our own form factors, we assume that the relations
above hold not only when $q^{2}=0$, but also in kinematic regimes
with finite $q^{2}$. It can be shown that at \emph{low energy}, the
differences in observables due to using the two bases, with these
relations applied, are negligible. This shows that our basis of
invariants is also complete and partially justifies extending the
preceding linear relations to kinematics with finite $q^{2}$.
Alternatively, all of the $q^{2}$ dependence of these $c_{i\Delta}$
and $d_{i\Delta}$ form factors can be realized in terms of meson
dominance. We then require that the meson dominance form factors be
as close as possible to the ones produced by the conventional form
factors in Eqs.~(\ref{eqn:axialtransitionff1}) to
(\ref{eqn:axialtransitionff}). However, when we compare our meson
dominance form factors with those in literature, we clearly see the
inadequacy of the leading-order meson dominance expressions above
$Q^{2} \approx 0.3 \, \mathrm{GeV}^{2}$.

\section{Feynman diagrams}
\label{sec:diag}

\begin{figure}[ht]
\begin{center}
\includegraphics[scale=0.5]{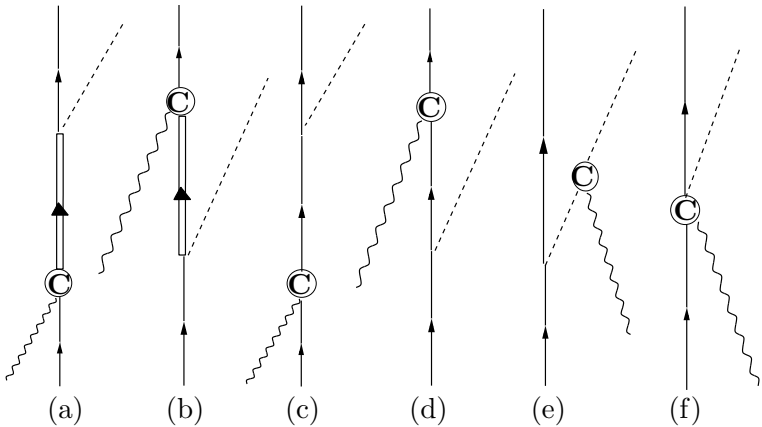}
\put(-270,37){$\bf{C}$} \put(-225,106){$\bf{C}$}
\put(-179,37){$\bf{C}$} \put(-130,104){$\bf{C}$}
\put(-72,78){$\bf{C}$} \put(-33,65){$\bf{C}$} \put(-270,-10){(a)}
\put(-225,-10){(b)} \put(-180,-10){(c)} \put(-133,-10){(d)}
\put(-87,-10){(e)} \put(-35,-10){(f)} \caption{Feynman diagrams for
pion production. Here, $\bf{C}$ stands for various types of currents
including vector, axial-vector, and baryon currents. Some diagrams
may be zero for some specific type of current. For example, diagrams
(a) and (b) will not contribute for the (isoscalar) baryon current.
Diagram (e) will be zero for the axial-vector current. The pion-pole
contributions are included in the vertex functions of the currents.}
\label{fig:feynmanpionproduction}
\end{center}
\end{figure}

Tree-level Feynman diagrams for pion production due to the vector
current, axial-vector current, and baryon current are shown in
Fig.~\ref{fig:feynmanpionproduction}. In this section, we will begin
to calculate different matrix elements for pion production and
photon production. The Feynman diagrams for photon production can be
viewed as diagrams in Fig.~\ref{fig:feynmanpionproduction} with an
outgoing $\pi$ line changed to a $\gamma$ line. It is easy to see
that diagram (e) in Fig.~\ref{fig:feynmanpionproduction} will not
contribute to photon production, since there is no vertex connecting
a pion and a photon.

\subsection{Renormalized $\Delta$ propagator}

Since the one-pion-loop self-energy has already been calculated in
Ref.~\cite{tang98}, we simply quote the result here:
\begin{eqnarray}
S_{F}^{\mn}(p) &\equiv&
          -\frac{\slashed{p}+m}{p^{2}-m^{2}-\Pi(\eta_{p})+im\Gamma(p^{2})}
          P^{(\frac{3}{2}) \mn}-\frac{1}{\sqrt{3}m}P^{(\half)
          \mn}_{12}-\frac{1}{\sqrt{3}m}P^{(\half) \mn}_{21}
          \notag \\[5pt]
&&{}+\frac{2}{3m^{2}}\,(\slashed{p}+m) P^{(\half) \mn}_{22}
          + O(\Gamma /m) \times \text{non-pole terms,} 
          \\[5pt]
\Gamma(p^{2}) &=& \frac{\pi}{12mp^{4}}\, \frac{h_{A}^{2}}{(4\pi
          f_{\pi})^{2}}\, (p^{2}+M^{2}+2Mm) \notag
          \\[5pt]
&&{}\times\left[(p^{2}-M^{2})^{2}-(p^{2}+3M^{2})m_{\pi}^{2}\right]
          \sqrt{(p^{2}-M^{2})^{2}-4p^{2}m_{\pi}^{2}}\ .
          \label{eqn:tang98width} 
\end{eqnarray}
However, to make the calculation simpler, we will set $\Pi=0$. We
take $m=1232$ MeV as the Breit--Wigner mass \cite{pdg2008}. Note
that $\Gamma$ is implicitly associated with a factor of
$\Theta[p^{2}-(M+m_{\pi})^{2}]$.

A few words on the $1/p^{2}$ singularities in the projection
operators are in order here. (See Appendix
\ref{app:deltapropagator}.) When the $\Delta$ is in the $s$ channel,
we never see these singularities, since $p^2$ is timelike. When the
$\Delta $ is in other channels, for $p^{2} \geqslant
(m_{\pi}+M)^{2}$, we never see the singularities. For $p^{2}
\leqslant (m_{\pi}+M)^{2}$ and hence $\Gamma(p^{2})=0$, the apparent
singularities are actually canceled out in our approximation scheme.
It can be easily checked that if we set $\Pi=0$, then when
$\Gamma=0$, $S_{F}^{\mn}(p) \longrightarrow S_{F}^{0 \mn}(p)$, so no
singularities will appear. This is the advantage of setting $\Pi=0$.

We have to remember, however, that the whole calculation is valid
only in the low-energy limit, and in this limit, we would not see a
$\Delta$ in the $u$ channel that is far off shell, because both $q$
and $k_{\pi}$ are tiny. Hence the $1/p^{2}$ singularities should not
be a problem in low-energy effective theory from a very general
perspective.

\subsection{Power counting of Feynman diagrams ($\nu$)\ \cite{Pascalutsa08}}
\label{subsec:powercounting}

First let's outline the calculation of the interaction amplitude
$M$. Consider CC pion production (in the one-weak-boson-exchange
approximation):
\begin{eqnarray}
M&=&4\sqrt{2}\, G_{F} V_{ud}\, \langle J_{L i \mu}^{(lep)} \rangle
\langle J_{L}^{(had) i \mu} \rangle_{\pi}\ ,
\label{eqn:amplitudedefccpion}
\end{eqnarray}
where $i=+1,-1$. In Eq.~(\ref{eqn:amplitudedefccpion}), $G_{F}$ is
the Fermi constant, $V_{ud}$ is the CKM matrix element corresponding
to $u$ and $d$ quark mixing, $ \langle J_{L i \mu}^{(lep)}\rangle
\equiv \bra{l(\bar{l})} J_{L i \mu} \ket{\nu_{l} (\bar{\nu_{l}})}$,
and $ \langle J_{L}^{(had) i \mu} \rangle_{\pi} \equiv \bra{N B, \pi
j } J_{L}^{i \mu }\ket{N A}$. $ \langle J_{L i \mu}^{(lep)}\rangle$
is well known, so in the following, we focus on calculating $\langle
J_{L}^{(had) i \mu} \rangle_{\pi}$.

For NC pion production:
\begin{eqnarray}
M&=&4\sqrt{2}\, G_{F} \langle J_{NC  \mu}^{(lep)} \rangle \langle
J_{NC}^{(had)  \mu} \rangle_{\pi}\ ,
\label{eqn:amplitudedefncpion}
\end{eqnarray}
where $ \langle J_{NC \mu}^{(lep)} \rangle$ is the well-known
leptonic neutral current matrix element, and  $\langle J_{NC}^{(had)
\mu} \rangle_{\pi} \equiv \bra{N B, \pi j } J_{NC}^{ \mu }\ket{N
A}$. For NC photon production, we have a similar expression:
\begin{eqnarray}
M&=&4\sqrt{2}\, G_{F} \langle J_{NC  \mu}^{(lep)} \rangle \langle
J_{NC}^{(had)  \mu} \rangle_{\gamma} \ ,
\label{eqn:amplitudedefncphoton}
\end{eqnarray}
where $ \langle J_{NC}^{(had)  \mu} \rangle_{\gamma} \equiv \bra{N
B, \gamma } J_{NC}^{ \mu }\ket{N A}$.

Now consider the power counting for $ \langle J^{(had) \mu}
\rangle_{\pi(\gamma)}$ in
Eqs.~(\ref{eqn:amplitudedefccpion}),~(\ref{eqn:amplitudedefncpion}),
and~(\ref{eqn:amplitudedefncphoton}). The order of the diagram
$(\nu)$ is given by normal power counting \cite{tang98}: $\nu=2L+2-
\frac{1}{2} \, E_{n} +\sum_{i} \#_{i}(\hat{\nu}_{i}-2)$, where $L$
is the number of loops, $E_{n}$ is the number of external baryon
lines, $\hat{\nu}_{i}\equiv d_{i}+\frac{1}{2}\, n_{i}+b_{i}$ is the
order of the vertex $(\hat{\nu})$ defined in
Sec.~\ref{subsubsec:lagN}, and $\#_{i}$ is the number of times that
particular vertex appears.

As pointed out in Ref.~\cite{Pascalutsa08}, by including $\Delta$
resonances in calculations, we have a new mass scale $\delta \equiv
m-M \approx 300$ MeV. We must also consider the order of the
$\Delta$ width $\Gamma$. Formally, it is counted as $
O(Q^{3}/M^{2})$; however, numerical calculations with
Eq.~(\ref{eqn:tang98width}) indicate that it should be counted as
$O(Q^{3}/M^{2} \times 10)$. Because of these two issues, we have to
rethink the power counting of diagrams involving $\delta$ in two
energy regimes. One is near the resonance, while the other is at
lower energies, away from the resonance. In the resonance region,
the $\Delta$ propagator scales like
\begin{eqnarray}
S_{F} \sim \frac{1}{i \Gamma} + O \left(\frac{1}{M} \right) \approx
\frac{1}{10 i\, O(Q^{3}/M^{2})} \approx \frac{1}{i\, O(Q^{2}/M)}
\sim \frac{1}{O(Q)} \frac{M}{i\, O(Q)}\ ,
\end{eqnarray}
where the $O({1}/{M})$ comes from non-pole terms. In the
lower-energy region,
\begin{eqnarray}
S_{F} &\sim & \frac{1}{2[\delta-O(Q)] - 10 i\, O(Q^{3}/M^{2})} +O
\left(\frac{1}{M} \right)\sim \frac{1}{O(Q)} \frac{O(Q)}{\delta} +O
\left(\frac{1}{M} \right)\approx \frac{1}{O(Q)} \frac{O(Q)}{M}\ .
\notag \\
&\null &
\end{eqnarray}
So compared to the normal power counting mentioned above, in which
the nucleon propagator scales as ${1}/{O(Q)}$, for diagrams
involving one $\Delta$ in the $s$ channel, we take $\nu \to \nu -1$
in the resonance regime and $\nu \to \nu+1 $ away from the
resonance. This partially justifies the strategy of incorporating
non-resonant diagrams at low energies, while ignoring them in the
resonance region when fitting the form factors in this region
\cite{GRACZYK08,GRACZYK09}.

\subsection{CVC and PCAC}

We will calculate matrix elements of currents and test the
conservation of the vector current and the baryon current, and also
partial conservation of the axial-vector current.

\subsubsection{Diagram (a)}
Diagram (a) in Fig.~\ref{fig:feynmanpionproduction} leads to a
vector current ($k_{\pi}$ is the \emph{outgoing} pion's momentum)
\begin{eqnarray}
\langle V^{i \mu}\rangle_{\pi} &=& -\frac{ih_{A}}{f_{\pi}}\,
\T{a}{Bj}\, \Tdagger{iA}{a}\, \psibar{u}_{f} k_{\pi}^{\lambda}\,
S_{F\lambda \alpha}(p)\, \Gamma_{V}^{ \alpha \mu}(p;q,p_{i}) u_{i}\
. \label{eqn:vcdelta}
\end{eqnarray}
Here $\Gamma_{V}^{ \alpha \mu}(p;q,p_{i})$ is defined in
Eq.~(\ref{eqn:transitioncurrentVvertex}). Momentum conservation
gives $p=q+p_{i}$, and $\T{a}{jB}\, \Tdagger{iA}{a}
=\delta_{j}^{i}\delta_{B}^{A}-\frac{1}{3}(\tau_{j}\tau^{i})_{B}^{\;A}
=\frac{2}{3}\delta_{j}^{i}\delta_{B}^{A}-\frac{i}{3}\epsilon_{j}^{\,
ik}\,
(\tau_{k})_{B}^{A}=\frac{2}{3}\delta_{j}^{i}\delta_{B}^{A}+\frac{i}{3}\epsilon^{i}_{\,
jk}\, (\tau^{k})_{B}^{A}$, where the subscript $j$ denotes the
isospin of the outgoing pion. Vector current conservation is
obvious, and $\nu_{nr} \geqslant 3$ in the lower-energy region,
while $\nu_{r} \geqslant 1$ in the resonance region. Here the
higher-order terms in $\nu$ come from including form factors at the
vertices.

The axial-vector current matrix element is
\begin{eqnarray}
\langle A^{i \mu}\rangle_{\pi} &=& -\frac{ih_{A}}{f_{\pi}}\,
\T{a}{Bj}\, \Tdagger{iA}{a}\, \psibar{u}_{f} k_{\pi}^{\lambda}\,
S_{F\lambda \alpha}(p)\, \Gamma_{A}^{ \alpha \mu}(p;q,p_{i}) u_{i}\
. \label{eqn:acdelta}
\end{eqnarray}
Here $\Gamma_{A}^{ \alpha \mu}(p;q,p_{i})$ is defined in
Eq.~(\ref{eqn:transitioncurrentvertex}). PCAC is also obvious, if we
check the structure of $\Gamma_{A}^{\alpha \mu}$, and $\nu_{nr}
\geqslant 2, \nu_{r} \geqslant 0$.

The baryon current matrix element is
\begin{eqnarray}
\langle J_{B}^{\mu}\rangle_{\pi}=0\ .
\label{eqn:bcdelta}
\end{eqnarray}

Now we examine the NC matrix element $\langle J_{NC}^{(had)  \mu}
\rangle_{\gamma}$. First, based on the relations given in
Eq.~(\ref{eqn:ncdef}), we define
\begin{eqnarray}
\Gamma_{N}^{\alpha\mu}(p;q,p_{i}) &\equiv& (\half-\sin^{2}\theta_{w})
          \Gamma_{V}^{\alpha\mu}(p;q,p_{i})+\half\, \Gamma_{A}^{\alpha\mu}(p;q,p_{i})\ ,  
          \\[5pt]
\psibar{\Gamma}_{N}^{ \mu \alpha}(p_{f};q,p) &\equiv &\ugamma{0}
          \Gamma_{N}^{\dagger \alpha \mu}(p;-q,p_{f})\ugamma{0}\ , 
          \\[5pt]
\bra{\Delta, a, p } J^{\mu}_{NC}(q) \ket{N,A , p_{i}} &\equiv
          &\Tdagger{0A}{a}\, \psibar{u}_{\alpha}(p)\Gamma_{N}^{\alpha\mu}(p;q,p_{i})
          u(p_{i})\ , 
          \\[5pt]
\bra{N , A, p_{f}} J^{\mu}_{NC}(q) \ket{\Delta,a, p } &\equiv
          &\T{a}{0A}\, \psibar{u}(p_{f}) \psibar{\Gamma}_{N}^{\mu
          \alpha}(p_{f};q,p)u_{\alpha}(p)\ . 
\end{eqnarray}
Then we find [$k$ is the outgoing photon's momentum and
$\epsilon_{\lambda}^{\ast}(k)$ is its polarization]
\begin{eqnarray}
\langle J_{NC}^{\mu}\rangle _{\gamma}&=& e \T{a}{0B}\,
          \Tdagger{0A}{a}\, \psibar{u}_{f}\, \epsilon_{\lambda}^{\ast}(k)
          \psibar{\Gamma}_{V}^{\lambda \alpha}(p_{f};-k,p) S_{F\alpha\beta}(p)
          \Gamma_{N}^{\beta \mu}(p;q,p_{i}) u_{i}\ .
          \label{eqn:ncdelta}
\end{eqnarray}
CVC and PCAC are straightforward to verify here. For the vector
current, $\nu_{nr} \geqslant 4$, $\nu_{r} \geqslant 2$, while for
the axial-vector current, $\nu_{nr} \geqslant 3$, $\nu_{r} \geqslant
1$.

\subsubsection{Diagram (b)}

Diagram (b) in Fig.~\ref{fig:feynmanpionproduction} leads to the
vector current
\begin{eqnarray}
\langle V^{i \mu}\rangle_{\pi} &=& -\frac{ih_{A}}{f_{\pi}}\,
          \T{ai}{B}\, \Tdagger{A}{ja}\, \psibar{u}_{f} \psibar{\Gamma}_{V}^{\mu
          \alpha }(p_{f};q,p) S_{F\alpha \lambda }(p)\,
          k_{\pi}^{\lambda}\, u_{i}\ .
          \label{eqn:vcdeltacross}
\end{eqnarray}
Here $\psibar{\Gamma}_{V}^{ \mu \alpha}(p_{f};q,p) \equiv \ugamma{0}
\Gamma_{V}^{\dagger \alpha \mu}(p;-q,p_{f})\ugamma{0}$,
$p=-q+p_{f}$, and $\T{ai}{B}\, \Tdagger{A}{aj}
=\delta^{i}_{j}\delta_{B}^{A}-\frac{1}{3}(\tau^{i}
\tau_{j})_{B}^{\;A}=\frac{2}{3}\, \delta^{i}_{j}\delta_{B}^{A}
-\frac{i}{3}\, \epsilon^{i}_{\,jk} (\tau^{k})_{B}^{A}$. The
conservation of the vector current is obvious, and $\nu_{nr}
\geqslant 3$.

The axial-vector current matrix element is
\begin{eqnarray}
\langle A^{i \mu}\rangle_{\pi} &=& -\frac{ih_{A}}{f_{\pi}}\,
          \T{ai}{B}\, \Tdagger{A}{ja}\, \psibar{u}_{f} \psibar{\Gamma}_{A}^{\mu
          \alpha }(p_{f};q,p) S_{F\alpha \lambda }(p)\,
          k_{\pi}^{\lambda}\, u_{i}\ .
          \label{eqn:acdeltacross}
\end{eqnarray}
Here $\psibar{\Gamma}_{A}^{ \mu \alpha}(p_{f};q,p) \equiv \ugamma{0}
\Gamma_{A}^{\dagger \alpha \mu}(p;-q,p_{f})\ugamma{0}$,
$p=-q+p_{f}$, PCAC is again obvious, and $\nu_{nr}  \geqslant 2$.

The baryon current matrix element is zero $( \langle
J_{B}^{\mu}\rangle_{\pi}=0 )$,
and the NC current matrix element for photon production is
\begin{eqnarray}
\langle J_{NC}^{\mu}\rangle _{\gamma}&=& e \T{a0}{B}\,
          \Tdagger{A}{a0}\, \psibar{u}_{f} \psibar{\Gamma}_{N}^{\mu \alpha
          }(p_{f};q,p) S_{F\alpha\beta}(p)  \Gamma_{V}^{\beta \lambda
          }(p;-k,p_{i})\, \epsilon_{\lambda}^{\ast}(k)\,  u_{i}\ .
\label{eqn:ncdeltacross}
\end{eqnarray}
Both CVC and PCAC are obvious here. For the vector current,
$\nu_{nr} \geqslant 4$, while for the axial-vector current,
$\nu_{nr} \geqslant 3$.

\subsubsection{Diagrams (c) and (d)}

These two diagrams lead to a vector current
\begin{eqnarray}
\langle V^{i \mu}\rangle_{\pi} &=&-\frac{ig_{A}}{f_{\pi}}\,
          \bra{B} \dhalftau{j}\uhalftau{i}\ket{A}\, \psibar{u}_{f}
          \slashed{k_{\pi}}\ugammafive S_{F}(p) \Gamma_{V}^{\mu}(q) u_{i} \notag
          \\[5pt]
&&{}-\frac{ig_{A}}{f_{\pi}}\, \bra{B}
          \uhalftau{i}\dhalftau{j}\ket{A}\, \psibar{u}_{f}\Gamma_{V}^{\mu}(q)
          S_{F}(p) \slashed{k_{\pi}}\ugammafive u_{i}\ .
          \label{eqn:vcn}
\end{eqnarray}
Here $S_{F}(p)=S_{F}(q+p_{i})$ is the nucleon propagator,
$\Gamma_{V}^{\mu}(q)$ has been defined in
Eq.~(\ref{eqn:NNvectorcurrentwithff}), and $\nu \geqslant 1$. To
prove CVC, one must consider diagrams (c), (d), (e), and (f)
together.

For the axial-vector current, we find
\begin{eqnarray}
\langle A^{i \mu}\rangle_{\pi} &=&-\frac{ig_{A}}{f_{\pi}}\, \bra{B}
          \dhalftau{j}\uhalftau{i}\ket{A}\, \psibar{u}_{f}\slashed{k_{\pi}}\ugammafive S_{F}(p)
          \Gamma_{A}^{\mu}(q)\, u_{i} \notag
          \\[5pt]
&&{}-\frac{ig_{A}}{f_{\pi}}\, \bra{B}
          \uhalftau{i}\dhalftau{j}\ket{A}\, \psibar{u}_{f}\Gamma_{A}^{\mu}(q)
          S_{F}(p) \slashed{k_{\pi}}\ugammafive \, u_{i}\ .
          \label{eqn:acn}
\end{eqnarray}
Here, $\Gamma_{A}^{\mu}(q)$ has been defined in
Eq.~(\ref{eqn:NNaxialcurrentff}), PCAC is obvious, and $\nu\geqslant
1$.

For the baryon current we have
\begin{eqnarray}
\langle J_{B}^{ \mu}\rangle_{\pi} &=&-\frac{ig_{A}}{f_{\pi}}\, \bra{B} \dhalftau{j}\ket{A}\,
          \psibar{u}_{f}\slashed{k_{\pi}}\ugammafive S_{F}(p) \Gamma_{B}^{\mu}(q)\, u_{i}  \notag
          \\[5pt]
&&{}-\frac{ig_{A}}{f_{\pi}}\, \bra{B} \dhalftau{j}\ket{A}\,
          \psibar{u}_{f}\Gamma_{B}^{\mu}(q) S_{F}(p)
          \slashed{k_{\pi}}\ugammafive\, u_{i}\ .
          \label{eqn:bcn}
\end{eqnarray}
Here, $\Gamma_{B}^{\mu}(q)$ has been defined in
Eq.~(\ref{eqn:NNbaryoncurrentwithff}). It is easy to see that the
baryon current is conserved and that $\nu\geqslant 1$.

Finally, for NC photon production, we get
\begin{eqnarray}
\langle J_{NC}^{\mu}\rangle _{\gamma}&=& e\, \psibar{u}_{f}\,
          \epsilon^{\ast}_{\lambda}(k) \left( (\uhalftau{0})_{B}^{\;C}\,
          \Gamma_{V }^{\lambda}(-k)
          + \frac{\delta_{B}^{\;C}}{2}\, \Gamma_{B }^{\lambda}(-k) \right) S_{F}(p)  \notag
          \\[5pt]
&&{}\times \left((\uhalftau{0})_{C}^{\; A}\left[(\half - \sin^{2}{\theta_{w}})
          \Gamma_{V}^{\mu}(q)+\half\Gamma_{A}^{\mu}(q)\right]
          - \frac{\delta_{C}^{\; A}}{2}\sin^{2}{\theta_{w}}\, \Gamma_{B}^{\mu}(q)
          \right) u_{i} \notag
          \\[5pt]
&&{}+ e \, \psibar{u}_{f} \left((\uhalftau{0})_{B}^{\; C}\left[(\half - \sin^{2}{\theta_{w}})
          \Gamma_{V}^{\mu}(q)+\half\Gamma_{A}^{\mu}(q)\right]
          - \frac{\delta_{B}^{\; C}}{2}\sin^{2}{\theta_{w}} \Gamma_{B}^{\mu}(q) \right) \notag
          \\[5pt]
&&{}\times S_{F}(p)\,  \epsilon^{\ast}_{\lambda}(k) \left(
          (\uhalftau{0})_{C}^{\; A}\, \Gamma_{V }^{\lambda}(-k) +
          \frac{\delta_{C}^{\; A}}{2}\, \Gamma_{B }^{\lambda}(-k) \right)
          u_{i}\ ,
\label{eqn:ncn}
\end{eqnarray}
where we use the shorthand
\begin{equation}
(\uhalftau{0})_{B}^{\; A} = \bra{B} \uhalftau{0}\ket{A}\ .
\end{equation}
One can verify the conservation of the vector current and the baryon
current, as well as the partial conservation of the axial-vector
current. For all three currents, power counting gives $\nu \geqslant
1$. However, this naive power counting does not give an accurate
comparison between the $\Delta$ contributions and the $N$
contributions at low energies, as we discuss below.

\subsubsection{Diagrams (e) and (f)}
\label{subsubsec:diagramef}

The two diagrams lead to a vector current
\begin{eqnarray}
\langle V^{i \mu}\rangle_{\pi} &=& \frac{g_{A}}{2f_{\pi}}\,
          \epsilon^{i}_{\;jk} (\tau^{k})_{B}^{A}\,
          \frac{P_{V}^{\mu}(q,k_{\pi})}{(q-k_{\pi})^{2}-m_{\pi}^{2}}\,
          \psibar{u}_{f} (\slashed{q}-\slashed{k_{\pi}}) \ugammafive \, u_{i} 
          \\[5pt]
&&{}{}+\frac{\epsilon^{i}_{\,jk}}{f_{\pi}}\, \bra {B}
          \uhalftau{k}\ket{A}\, \psibar{u}_{f} \Gamma_{V\pi
          }^{\mu}(q,k_{\pi})\, u_{i} \ .
          \label{eqn:vcpfcontact}
\end{eqnarray}
Here, $P_{V}^{\mu}(q,k_{\pi})$ is defined in
Eq.~(\ref{eqn:pionvectorcurrentff}), $\Gamma_{V\pi
}^{\mu}(q,k_{\pi})$ is defined in
Eq.~(\ref{eqn:NNpionvectorcurrentff}), and $\nu \geqslant 1 $.
Finally, we can combine diagrams (c), (d), (e), and (f) to get
vector current conservation.

For the axial-vector current, diagram (e) does not contribute, and
we find
\begin{eqnarray}
\langle A^{i \mu}\rangle_{\pi}
          &=&\frac{\epsilon^{i}_{\,jk}}{f_{\pi}}\, \bra {B}
          \uhalftau{k}\ket{A}\, \psibar{u}_{f} \Gamma_{A\pi}^{\mu}(q,k_{\pi})\, u_{i}
          +\frac{\epsilon^{i}_{\;jk}}{f_{\pi}}\,
          \bra{B}\uhalftau{k}\ket{A}\,
          \frac{q^{\mu}}{q^{2}-m^{2}_{\pi}}\, \psibar{u}_{f}\,
          \frac{(\slashed{q}+\slashed{k_{\pi}})}{2} \, u_{i}  
          \\[5pt]
&&{}+\frac{\epsilon^{i}_{\,jk}}{f_{\pi}}\, \bra {B} \uhalftau{k}\ket{A}\,
          4\kappa_{\pi}\,\psibar{u}_{f} \left(\frac{\usigma{\mn}ik_{\pi \nu}}{2M}
          +\frac{q^{\mu}}{q^{2}-m^{2}_{\pi}}\,
          \frac{\usigma{\ab}i k_{\pi\alpha}q_{\beta}}{2M} \right)u_{i} 
          \\[5pt]
&&{}+\frac{\delta_{j}^{\,i}}{f_{\pi}}\,
          \delta_{B}^{\,A}\, 4\beta_{\pi}
          \frac{1}{M}\left(- ik_{\pi}^{\mu}+\frac{i q\cdot k_{\pi}\, q^{\mu}}{q^{2}-m_{\pi}^{2}}\right)
          \psibar{u}_{f}u_{i} 
          \\[5pt]
&&{}+\frac{\delta_{j}^{\,i}}{f_{\pi}}\, \delta_{B}^{\,A}\,
          \frac{-i\kappa_{1}}{4}\, \frac{1}{M^{2}}\,
          \psibar{u}_{f}\left(q_{\nu}(p_{f}+p_{i})^{\{\nu}\ugamma{\mu\}}
          -\frac{q \cdot (p_{f}+p_{i})\,
          q^{\mu}}{q^{2}-m_{\pi}^{2}}\,
          (\slashed{q}+\slashed{k_{\pi}})\right) u_{i} \, .
          \label{eqn:accontact}
\end{eqnarray}
Here, $\Gamma_{A\pi }^{\mu}(q,k_{\pi})$ is given in
Eq.~(\ref{eqn:NNpionaxialcurrentwithff}) and leads to a
$\nu\geqslant1$ contribution. The contributions due to
$\kappa_{\pi}$, $\beta_{\pi}$, and $\kappa_{1}$ are at $\nu=2$. It
is easy to check that PCAC holds.

For the baryon current, diagrams (e) and (f) do not contribute at
order $\nu=1$: $\langle J_{B}^{\mu}\rangle_{\pi}=0.$

For the NC photon production matrix element we find
\begin{eqnarray}
\langle J_{NC}^{\mu} \rangle_{\gamma} &=&
          \delta_{B}^{A}\, \frac{-iec_{1}}{M^{2}}\,
          \epsilon^{\mn\alpha\beta}\,
          \psibar{u}_{f}\dgamma{\nu}k_{\alpha}\epsilon^{\ast}_{\beta}(k) u_{i} \notag
          \\[5pt]
&&{} + \delta_{B}^{A}\, \frac{-iec_{1}
          q^{\mu}}{M^{2}(q^{2}-m^{2}_{\pi})}\, \epsilon^{\lambda \nu
          \alpha\beta}\, \psibar{u}_{f}\dgamma{\lambda} q_{\nu} k_{\alpha}
          \epsilon^{\ast}_{\beta}(k) u_{i} \notag
          \\[5pt]
&&{}+\left(\uhalftau{0}\right)_{B}^{A}\, \frac{-ie e_{1}}{2 M^{2}}\,
          \epsilon^{\mn\alpha\beta}\, \psibar{u}_{f}\dgamma{\nu}k_{\alpha}\epsilon^{\ast}_{\beta}(k)
          u_{i} \notag
          \\[5pt]
&&{}+ \left(\uhalftau{0}\right)_{B}^{A}\, \frac{-ie e_{1} q^{\mu}}{2
          M^{2}(q^{2}-m^{2}_{\pi})}\, \epsilon^{\lambda \nu
          \alpha\beta}\,
          \psibar{u}_{f}\dgamma{\lambda} q_{\nu} k_{\alpha}
          \epsilon^{\ast}_{\beta}(k) u_{i} \ .
          \label{eqn:nccontact}
\end{eqnarray}
It is straightforward to see that PCAC is satisfied. Here $\nu=3$;
for $\nu < 3$, there are no contact vertices contributing to the NC
photon production channel. By power counting, we expect that at low
energy, these terms can be neglected compared to the $\nu=1$ terms.
However, as claimed in Ref.~\cite{RHill09}, these contact vertices
have possible high-energy extrapolations due to the anomalous decays
of the $\omega$ and $\rho$. According to Ref.~\cite{RHill09}, these
terms may play an important role in coherent photon production.
However, we must realize that the constants $c_{1}$ and $e_{1}$ can
only be fixed by experiment, and it is not clear that only anomalous
meson decay will contribute to these operators at low energy. As
shown in the lagrangian, we can construct meson dominance by
coupling mesons instead of photons to the vertex. Moreover, these
terms are also the same as operators induced by the off-shell
parameters in the $\Delta$ lagrangian.


\section{Results}
\label{sec:res}

In this section, after introducing the kinematics, we will discuss
our results for CC and NC pion production, and also NC photon
production, and compare them with available data whenever possible.
Aiming at the excessive events in the MiniBooNE experiment, we will
focus on the scattering of $\nu_{\mu}$ and $\bar{\nu}_{\mu}$ off
nucleons with $E_{\nu, \bar{\nu}} \leqslant 0.5 \ \mathrm{GeV}$.

\subsection{Kinematics}
\label{subsec:kinematics}

\begin{figure}[!ht]
 \begin{center}
 \includegraphics[scale=0.6]{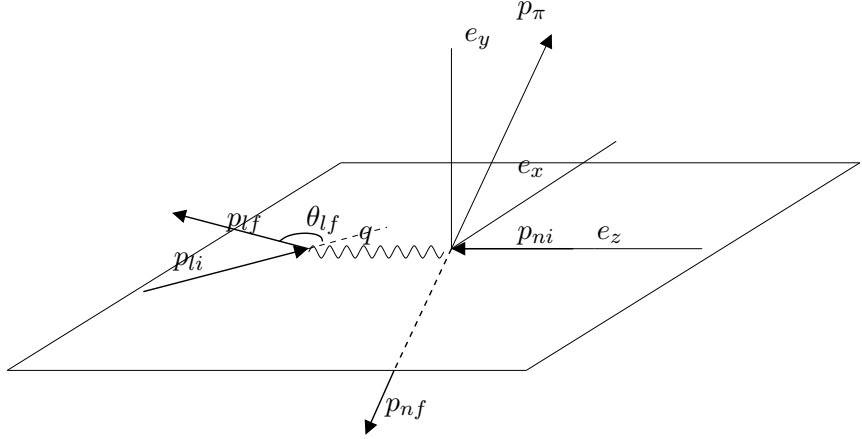}
 \put(-260,65){$p_{li}$}
 \put(-240,80){$p_{lf}$}
 \put(-190,75){$q$}
 \put(-130,75){$p_{ni}$}
 \put(-130,160){$p_{\pi}$}
 \put(-180,10){$p_{nf}$}
  \put(-210,80){$\theta_{lf}$}
  \put(-100,75){$e_{z}$}
   \put(-130,100){$e_{x}$}
   \put(-150,150){$e_{y}$}
 \end{center}
 \caption{The configuration in the isobaric frame.}
 \label{fig:isobaricconfig}
\end{figure}

Figure~\ref{fig:isobaricconfig} shows the configuration in the
isobaric frame, i.e., the cm frame of the final nucleon and pion.
The momenta are measured in this frame, except those labeled as
$p^{L}$, which denotes a momentum measured in the Lab frame with the
initial nucleon stationary. Detailed analysis of the kinematics is
given in Appendix~\ref{app:kinmatics}. The expression for the total
cross section for this process is ($\overline{\vert M
\vert}^{\mkern3mu 2}$ is the averaged total interaction amplitude
squared.)
\begin{eqnarray}
\sigma &=& \int \frac{\overline{\vert M \vert}^{\mkern3mu 2}}{4\vert
          p_{li}^{L} \cdot p_{ni}^{L}\vert}\, (2\pi)^{4}
          \delta^{(4)} \big( \sum_{i}p_{i} \big)  \frac{d^{3}
          \vec{p}_{lf}^{\;L}}{(2\pi)^{3}2E_{lf}^{L}}\,
          \frac{d^{3} \vec{p}_{\pi}^{\;L}}{(2\pi)^{3}2E_{\pi}^{L}}\,
          \frac{d^{3} \vec{p}_{nf}^{\;L}}{(2\pi)^{3}2E_{nf}^{L}} \notag
          \\[5pt]
&=&\int \frac{\overline{\vert M \vert}^{\mkern3mu 2}}{32 M_{n}}\,
          \frac{1}{(2\pi)^{5}}\,  \frac{\modular{p}{\pi}}{E_{\pi}+E_{nf}}
          \frac{\vert \vec{p}_{lf}^{\;L}\vert }{\vert \vec{p}_{li}^{\;L}\vert}\,
          d\Omega_{\pi} dE_{lf}^{L} d\Omega_{lf}^{L}  \notag
          \\[5pt]
&=&\int \frac{\overline{\vert M \vert}^{\mkern3mu 2}}{64M_{n}^{2}}\,
          \frac{1}{(2\pi)^{5}}\,
          \frac{\modular{p}{\pi}}{E_{\pi}+E_{nf}}\,
          \frac{\pi}{\vert \vec{p}_{li}^{\;L}\vert E_{li}^{L}}\, d\Omega_{\pi}
          dM_{\pi n}^{2} d Q^{2} \ . 
\end{eqnarray}
Based on the equations in Appendix~\ref{app:kinmatics}, we can make
the following estimates:

For CC pion production:
\begin{itemize}
\item When $E_{\nu}^{L}=0.4 \, \mathrm{GeV}$, $(M_{\pi n})_{max}
      \approxeq 1.17 \, \mathrm{GeV}, Q^{2}_{max} \approxeq 0.2 \,
      \mathrm{GeV}^{2}$.
\item When $E_{\nu}^{L}=0.5 \, \mathrm{GeV} $,
      $(M_{\pi n})_{max} \approxeq 1.24 \, \mathrm{GeV}, Q^{2}_{max}
      \approxeq 0.3 \, \mathrm{GeV}^{2}$.
\end{itemize}
We can see that above $E_{\nu}^{L}=0.4 \,\mathrm{GeV}$, the
interaction begins to be dominated by the $\Delta$ resonance.
However, when $E_{\nu}^{L}=0.75 \, \mathrm{GeV}$, $(M_{\pi n})_{max}
\approxeq 1.4 \, \mathrm{GeV}$, and higher resonances, for example
$P_{11}(1440)$, may play a significant role. The exception is that
for $\nu_{\mu} + p \longrightarrow \mu^{-}+ p + \pi^{+}$, only $I=
3/2$ can contribute, and the next resonance in this channel is the
$\Delta(1600)$, which is accessible only when $E_{\nu}^{L} \geqslant
1.8\, \mathrm{GeV}$.

For NC pion production:
\begin{itemize}
\item When $E_{\nu}^{L}=0.3\, \mathrm{GeV}$, $(M_{\pi n})_{max}
      \approxeq 1.2 \, \mathrm{GeV}, Q^{2}_{max} \approxeq 0.1 \,
      \mathrm{GeV}^{2}$.
\item When $E_{\nu}^{L}=0.5 \, \mathrm{GeV}$, $(M_{\pi n})_{max}
      \approxeq 1.35 \, \mathrm{GeV}, Q^{2}_{max} \approxeq 0.3 \,
      \mathrm{GeV}^{2}$.
\end{itemize}
Here, we can see that above $E_{\nu}^{L}=0.3\, \mathrm{GeV}$, the
interaction begins to be dominated by the $\Delta$. However, when
$E_{\nu}^{L}=0.6\, \mathrm{GeV}$, $(M_{\pi n})_{max} \approxeq 1.4
\, \mathrm{GeV}$, and higher resonances may play a significant role.

For NC photon production ($E_{\gamma}^{L} \geqslant 0.2\,
\mathrm{GeV} $):
\begin{itemize}
\item When $E_{\nu}^{L}=0.3 \, \mathrm{GeV}$, $(M_{\gamma n})_{ max}
      \approxeq 1.2 \, \mathrm{GeV}, Q^{2}_{max} \approxeq 0.1 \,
      \mathrm{GeV}^{2}$.
\item When $E_{\nu}^{L}=0.5 \, \mathrm{GeV} $, $(M_{\gamma n})_{ max}
      \approxeq 1.35 \, \mathrm{GeV}, Q^{2}_{max} \approxeq 0.3 \,
      \mathrm{GeV}^{2}$.
\end{itemize}
Here, we expect the $\Delta$ to dominate when $E_{\nu }^{L}
\geqslant 0.3\, \mathrm{GeV} $. But, similar to the case of NC pion
production, higher resonances may need to be considered when
$E_{\nu}^{L} \geqslant 0.6\, \mathrm{GeV}$.

From the analysis outlined above, we can expect our EFT to be valid
at $E_{\nu}^{L}\leqslant 0.5 \, \mathrm{GeV}$, since only the
$\Delta$ resonance can be excited, and $Q^{2} \leqslant 0.3 \,
\mathrm{GeV}^{2}$, so that meson dominance works for various
currents' form factors \cite{BDS10}. To go beyond this energy regime
when we show our results, we will require $M_{\pi n} \leqslant 1.4\,
\mathrm{GeV}$ and will use standard phenomenological form factors
that work when $Q^{2} \geqslant 0.3 \, \mathrm{GeV}^{2}$.

\subsection{CC pion production}

\begin{figure}[!ht]
\centering
\includegraphics[scale=0.7,angle=-90]{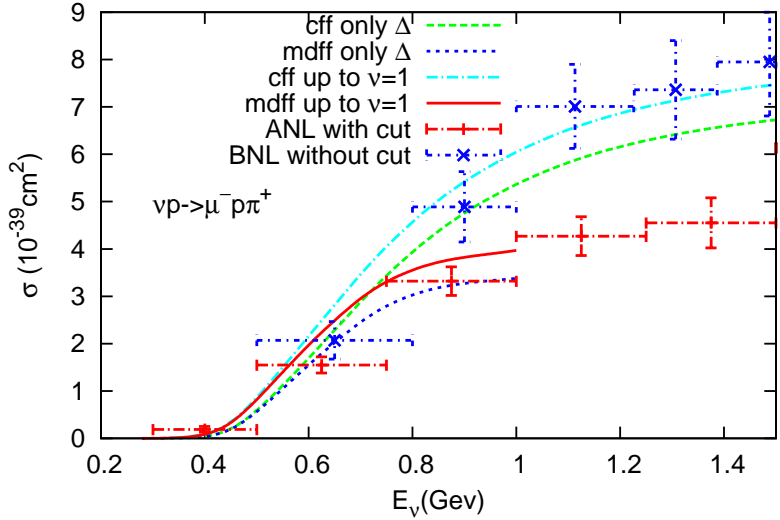}
\caption{Total cross section for $\nu_{\mu}+p\longrightarrow
\mu^{-}+p+\pi^{+}$. `Only $\Delta$' indicates that only diagrams
with $\Delta$ (both $s$ and $u$ channels) are included. `Up to
$\nu=1$' includes all the diagrams at leading order. The code `cff'
indicates that the calculations are done with conventional form
factors, while `mdff' indicates the calculations are based on the
EFT lagrangian with meson dominance. In the ANL data, $M_{\pi n}
\leqslant$ 1.4 $\mathrm{GeV}$ is applied, while no such cut is
applied in the BNL data. For all calculations, $ M_{\pi n}
\leqslant$ 1.4 $\mathrm{GeV}$ is applied. } \label{Fig:pppion+}
\end{figure}

\begin{figure}[!ht]
\centering
\includegraphics[scale=0.7,angle=-90]{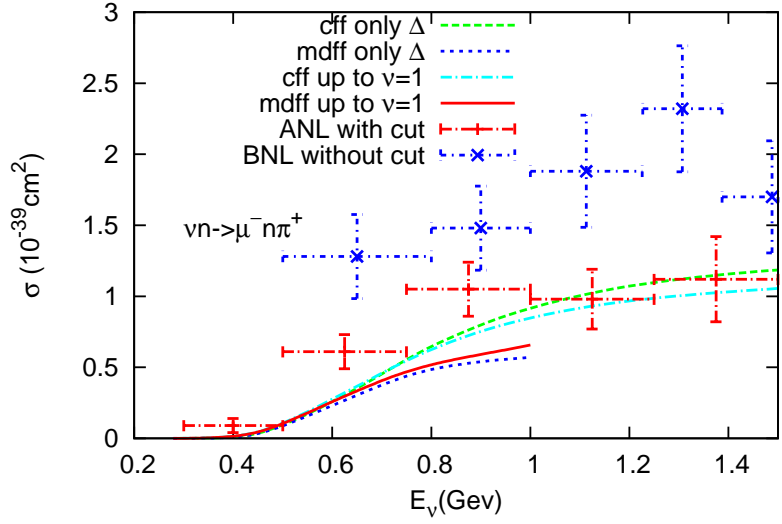}
\caption{Total cross section for $\nu_{\mu}+n\longrightarrow
\mu^{-}+n+\pi^{+}$. The curves are defined as in
Fig.~\ref{Fig:pppion+}.} \label{Fig:nnpion+}
\end{figure}

\begin{figure}[!ht]
\centering
\includegraphics[scale=0.7,angle=-90]{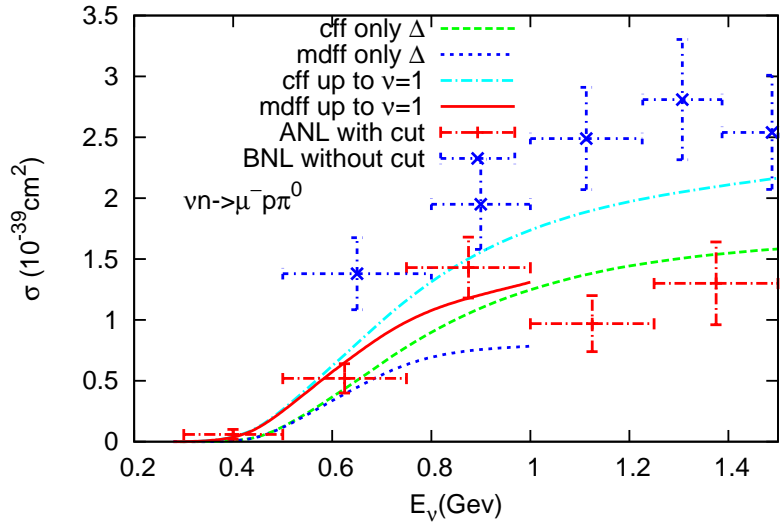}
\caption{Total cross section for $\nu_{\mu}+n\longrightarrow
\mu^{-}+p+\pi^{0}$. The curves are defined as in
Fig.~\ref{Fig:pppion+}.} \label{Fig:nppion0}
\end{figure}

In this section, we will compare CC pion neutrinoproduction results
with ANL~\cite{RADECKY82} and BNL~\cite{KITAGAKI86} measurements. In
both experiments, the targets are hydrogen and deuterium. (All the
other experiments use much heavier nuclear targets in (anti)neutrino
scattering, and to explain this, we must examine many-body effects.)
The beam is $\nu_{\mu}$, the average energy of which is $1\,
\mathrm{GeV}$ and $1.6\, \mathrm{GeV}$ for ANL and BNL,
respectively. In the ANL data, there is a cut on the invariant mass
of the pion and final nucleon system: $M_{\pi n} \leqslant 1.4\,
\mathrm{GeV}$. In the BNL data, there is no such cut. Based on the
phase-space analysis discussed above, this cut clearly reduces the
number of events when $E_{\nu}$ is above $0.5\thicksim0.6\,
\mathrm{GeV}$. Since the data stretch above this limit, in the first
three figures:~\ref{Fig:pppion+}, \ref{Fig:nnpion+}, and
\ref{Fig:nppion0}, we show our conventional form factor (`cff')
calculations with the $M_{\pi n}$ constraint. That is, for $F^{md}$,
$G^{md}$, $c_{\Delta}$, and $d_{\Delta}$ we substitute the
conventional form factors used in the literature \cite{GRACZYK09}.
Then we apply our lagrangian to the meson dominance form factor
(`mdff') calculations. As we have already concluded that this `mdff'
approach is inadequate above $E_{\nu}=0.5\, \mathrm{GeV}$, in the
following figures:~\ref{Fig:pppion+low}, \ref{Fig:nnpion+low},
\ref{Fig:nppion0low}, \ref{Fig:pppion-low}, \ref{Fig:nnpion-low},
and \ref{Fig:pnpion0low}, we show the `mdff' results with $E_{\nu}
\leqslant 0.5 \thicksim 0.6\, \mathrm{GeV}$, for which $M_{\pi n}
\leqslant 1.4\, \mathrm{GeV}$ holds automatically. Since we believe
the EFT is applicable in this low-energy regime, in these figures,
we show results including Feynman diagrams up to order $\nu=1$ and
$\nu=2$.

In Fig.~\ref{Fig:pppion+}, we show the data and calculations for
$\nu_{\mu}+p\longrightarrow \mu^{-}+p+\pi^{+}$. The ANL data is
systemically smaller than the BNL data, due to enforcing the $M_{\pi
n}$ constraint at ANL. As mentioned above, we make use of the
conventional form factors and include in the `cff only $\Delta$'
calculation the Feynman diagrams with the $\Delta$ in both $s$ and
$u$ channels and in the `cff up to $\nu=1$' all the Feynman diagrams
up to leading order. These two calculations are quite similar to
those done in Ref.~\cite{HERNANDEZ07}. Indeed, our results are
consistent with theirs for the conventional value of $C_{5}^{A}$.
(In Ref.~\cite{HERNANDEZ07}, only the $s$ channel contribution is
included in the calculation with `only $\Delta$'.) Next, we apply
our lagrangian in the `mdff' calculations, in which form factors are
realized in terms of meson dominance. In Fig.~\ref{Fig:pppion+}, we
show both the result with only $\Delta$ diagrams and the result with
all the leading-order diagrams in the `mdff' calculations, so that
we can compare the `mdff' approach with the `cff' approach.

First, we can see that both `cff' and `mdff' with only $\Delta$
diagrams are consistent with the data at $E_{\nu} \leqslant 0.5\,
\mathrm{GeV}$. Introducing other diagrams up to order $\nu=1$ is
still allowed by the data at low energy, although they indeed
increase the cross section noticeably. Second, the two approaches
with the same diagrams begin to differ from each other beyond
$E_{\nu}=0.5 \, \mathrm{GeV}$, which is also consistent with the
analysis of phase space and the discussion of the validity of meson
dominance. In Ref.~\cite{HERNANDEZ07}, a reduced $C_{5}^{A}(0)$ is
introduced, primarily to reduce the calculated cross sections above
$E_{\nu}=1\, \mathrm{GeV}$. However, since we are only concerned
with the $E_{\nu} \leqslant 0.5\, \mathrm{GeV}$ region, in which we
see satisfactory agreement between our calculations and the data, we
will stick to the $C_{5}^{A}(0)$ fitted from the $\Delta$ free width
in the framework of our EFT lagrangian. Furthermore, in the original
spectrum-averaged $d\sigma/dQ^{2}$ data of ANL~\cite{RADECKY82}, the
contributions from $E_{\nu} \leqslant 0.5\, \mathrm{GeV}$ neutrinos
are excluded, so comparing calculations with data at low energy is
not feasible at this stage, and we will not show our
$d\sigma/dQ^{2}$ here.

In Figs.~\ref{Fig:nnpion+} and~\ref{Fig:nppion0}, we show the data
and calculations for $\nu_{\mu}+n\longrightarrow \mu^{-}+n+\pi^{+}$
and $\nu_{\mu}+n\longrightarrow \mu^{-}+p+\pi^{0}$. We can see that
the situations in these two processes are quite similar to the one
in Fig.~\ref{Fig:pppion+}: the results of the `cff' and `mdff'
approaches are consistent with the data at low energy. Again the
differences between the two approaches with the same diagrams begin
to show up when the neutrino energy goes beyond $0.5\,
\mathrm{GeV}$. Although the pion production is still dominated by
the $\Delta$, other diagrams introduce significant contributions,
which violate the naive estimate of the ratio of the three channels'
cross sections based on isospin symmetry and $\Delta$ dominance.

\begin{figure}[!ht]
\centering
\includegraphics[scale=0.7,angle=-90]{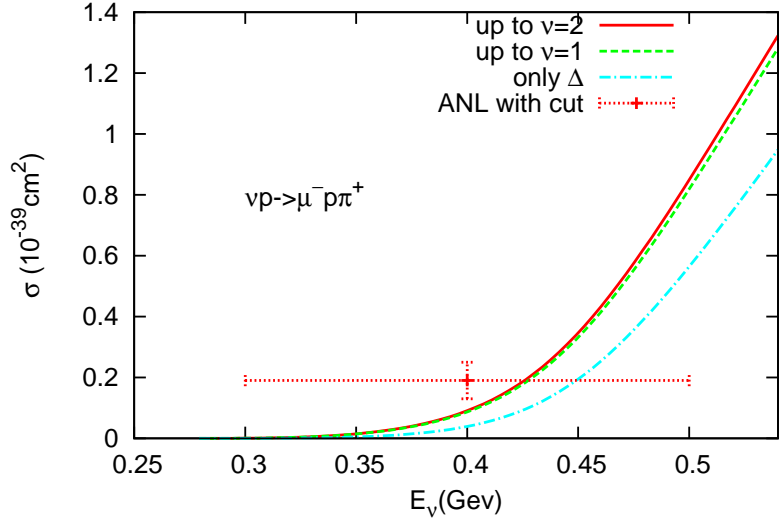}
\caption{Total cross section for $\nu_{\mu}+p\longrightarrow
\mu^{-}+p+\pi^{+}$. `Only $\Delta$' indicates that only diagrams
with $\Delta$ (both $s$ and $u$ channels) are included. `Up to
$\nu=1$' includes all the diagrams at leading order. `Up to $\nu=2$'
includes higher-order contact terms, whose couplings are from
\protect{Ref.~\cite{tang97simple}}. In the ANL data, $M_{\pi n}
\leqslant 1.4 \ \mathrm{Gev}$. For calculations, $ M_{\pi n}
\leqslant 1.4 \ \mathrm{Gev}$ is applied. } \label{Fig:pppion+low}
\end{figure}

\begin{figure}[!ht]
\centering
\includegraphics[scale=0.7,angle=-90]{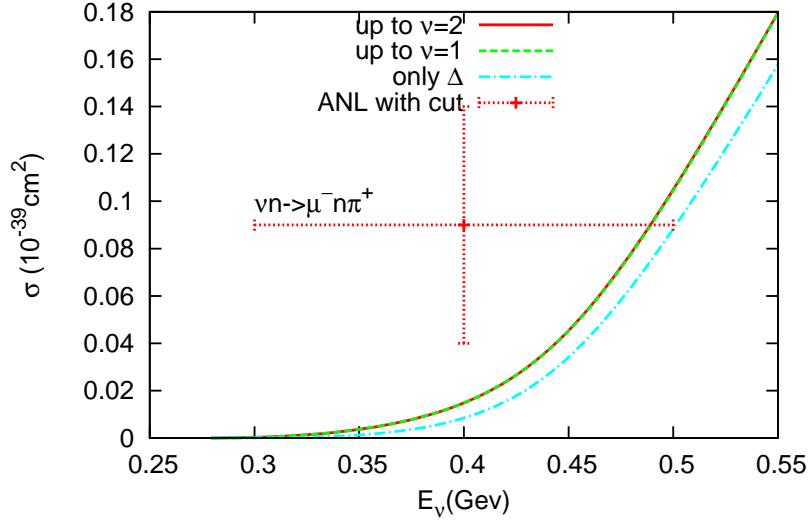}
\caption{Total cross section for $\nu_{\mu}+n\longrightarrow
\mu^{-}+n+\pi^{+}$. The curves are defined as in
Fig.~\ref{Fig:pppion+low}.} \label{Fig:nnpion+low}
\end{figure}

In Figs.~\ref{Fig:pppion+low}, \ref{Fig:nnpion+low},
\ref{Fig:nppion0low}, \ref{Fig:pppion-low}, \ref{Fig:nnpion-low},
and ~\ref{Fig:pnpion0low}, we begin to investigate the convergence
of our calculations in different channels in neutrino and
antineutrino scattering. We show the `mdff' calculations based on
our EFT lagrangian up to different orders. We see that the power
counting makes sense systematically in different channels: including
$N$ and contact terms up to $\nu=1$ changes the `only $\Delta$'
calculation non-negligibly. (Far away from resonance, we see that
the $\Delta$ contribution is not dominant compared to other
diagrams, and it begins to dominate around $ 0.4 \, \mathrm{GeV}$.
This is consistent with the power counting discussed in
Sec.~\ref{subsec:powercounting}). However, the $\nu=2$ terms do not
change the `up to $\nu=1$' results significantly. This partially
justifies the use of meson dominance, which automatically includes
higher-order terms. All the calculations of neutrino scattering are
consistent with the limited data from ANL. We can see that the cross
section for antineutrino scattering is generally smaller than that
of neutrino scattering, due to the relative sign chosen between
$V^{i\mu}$ and $A^{i\mu}$ in the Feynman diagrams with $\Delta$. The
signs between $V^{i\mu}$ and $A^{i\mu}$ in other diagrams is well
defined in our lagrangian. However, the relative sign between
currents due to the $\Delta$ and other diagrams is not well
constrained by the available data in our framework, as indicated in
Ref.~\cite{HERNANDEZ07}. Here we rely on the sign of $h_{A}$ fitted
in pion--nucleon scattering \cite{tang97simple} to set the sign
between the $\Delta$ current and the background contribution, as
shown in Eqs.~(\ref{eqn:axialtransitionff1}) to
(\ref{eqn:axialtransitionff}).

\begin{figure}[!ht]
\centering
\includegraphics[scale=0.7,angle=-90]{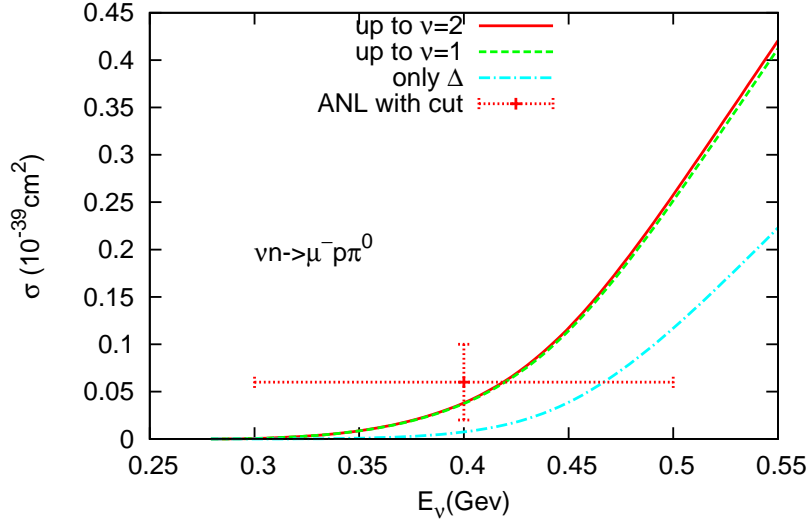}
\caption{Total cross section for $\nu_{\mu}+n\longrightarrow
\mu^{-}+p+\pi^{0}$. The curves are defined as in
Fig.~\ref{Fig:pppion+low}.}
\label{Fig:nppion0low}
\end{figure}

\begin{figure}[!ht]
\centering
\includegraphics[scale=0.7,angle=-90]{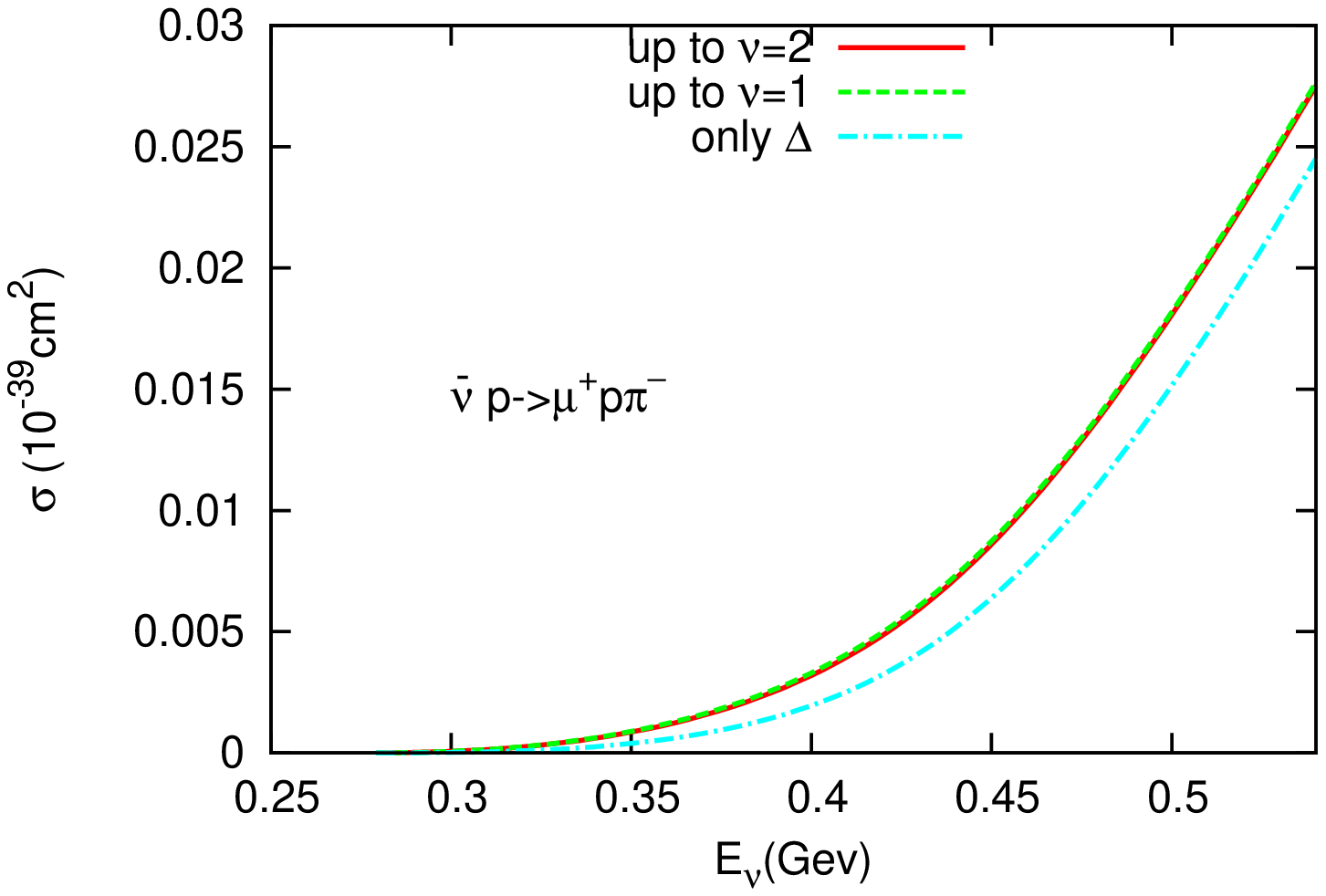}
\caption{Total cross section for $\bar{\nu}_{\mu}+p\longrightarrow
\mu^{+}+p+\pi^{-}$. The curves are defined as in
Fig.~\ref{Fig:pppion+low}.}
\label{Fig:pppion-low}
\end{figure}

\begin{figure}[!ht]
\centering
\includegraphics[scale=0.7,angle=-90]{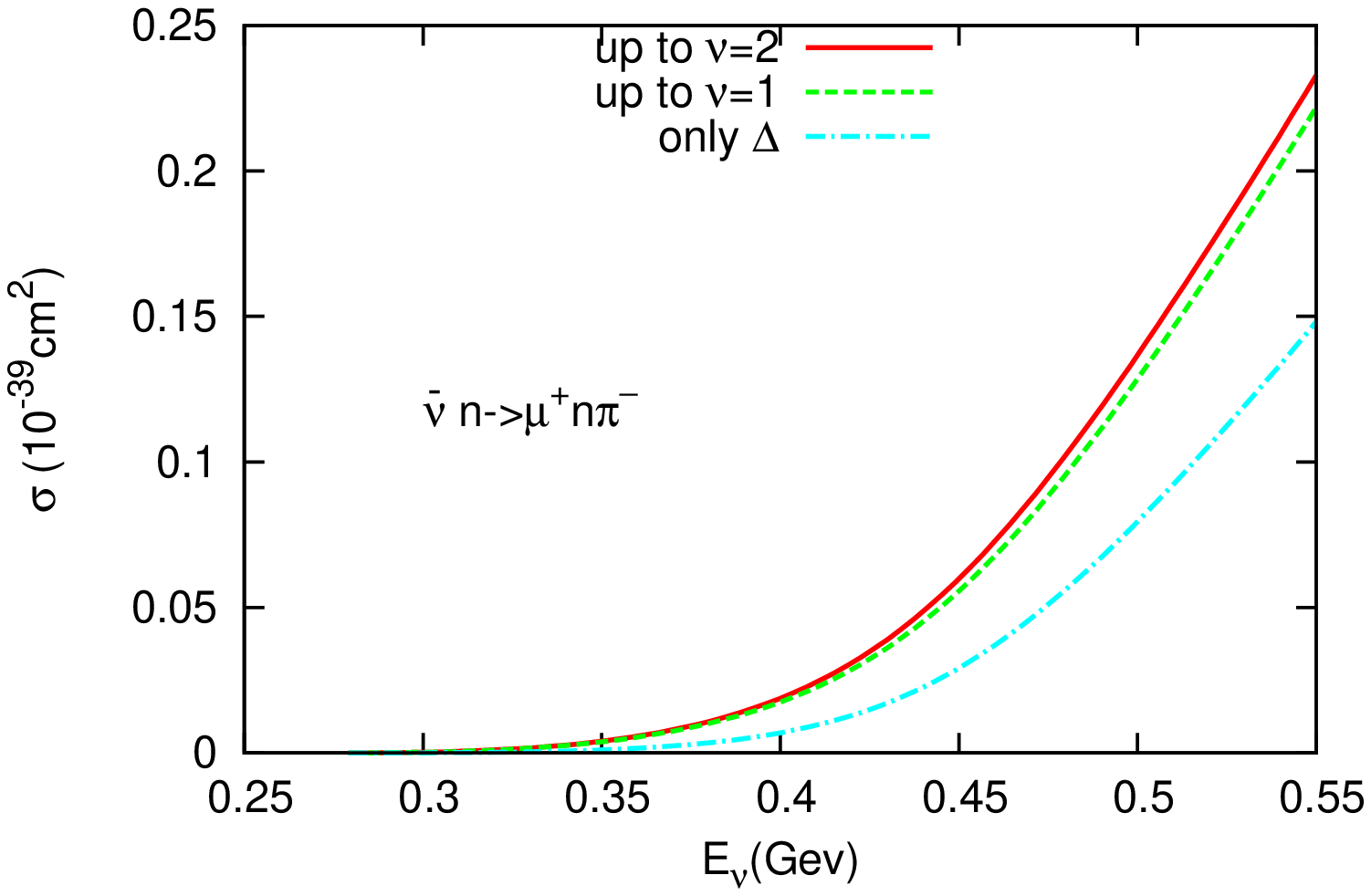}
\caption{Total cross section for $\bar{\nu}_{\mu}+n\longrightarrow
\mu^{+}+n+\pi^{-}$. The curves are defined as in
Fig.~\ref{Fig:pppion+low}.}
\label{Fig:nnpion-low}
\end{figure}

\begin{figure}[!ht]
\centering
\includegraphics[scale=0.7,angle=-90]{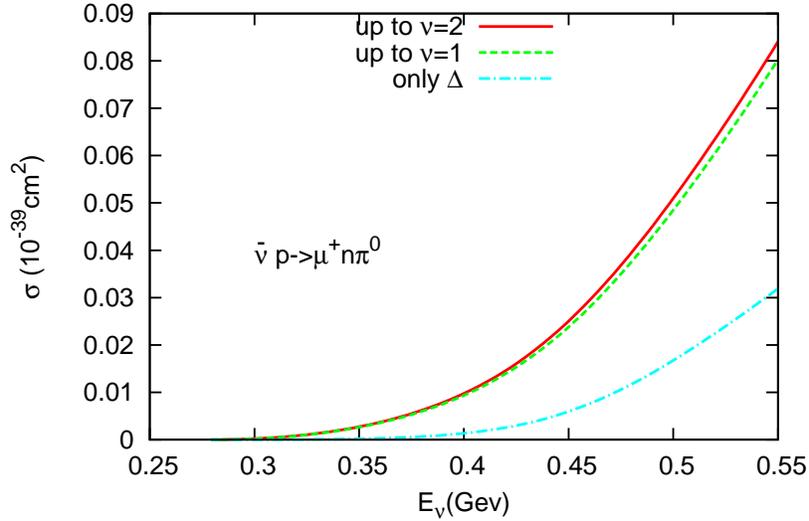}
\caption{Total cross section for $\bar{\nu}_{\mu}+p\longrightarrow
\mu^{+}+n+\pi^{0}$. The curves are defined as in
Fig.~\ref{Fig:pppion+low}.}
\label{Fig:pnpion0low}
\end{figure}

\subsection{NC pion production}

\begin{figure}[!ht]
\centering
\includegraphics[scale=0.7,angle=-90]{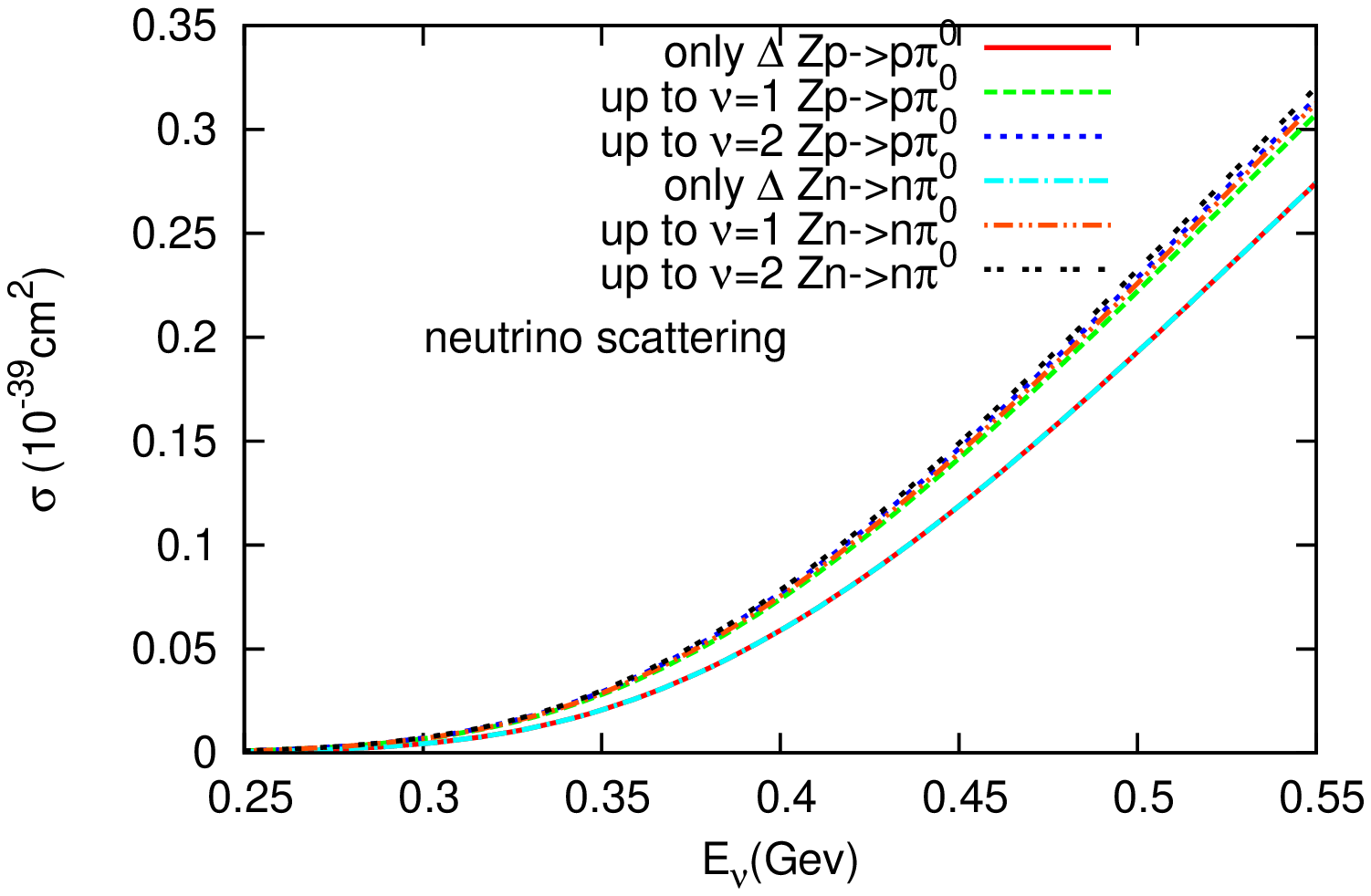}
\caption{Total cross section for NC $\pi^{0}$ production due to
neutrino scattering. The curves are defined as in
Fig.~\ref{Fig:pppion+low}, and the channels are also indicated. }
\label{Fig:ncpion0}
\end{figure}

\begin{figure}[!ht]
\centering
\includegraphics[scale=0.7,angle=-90]{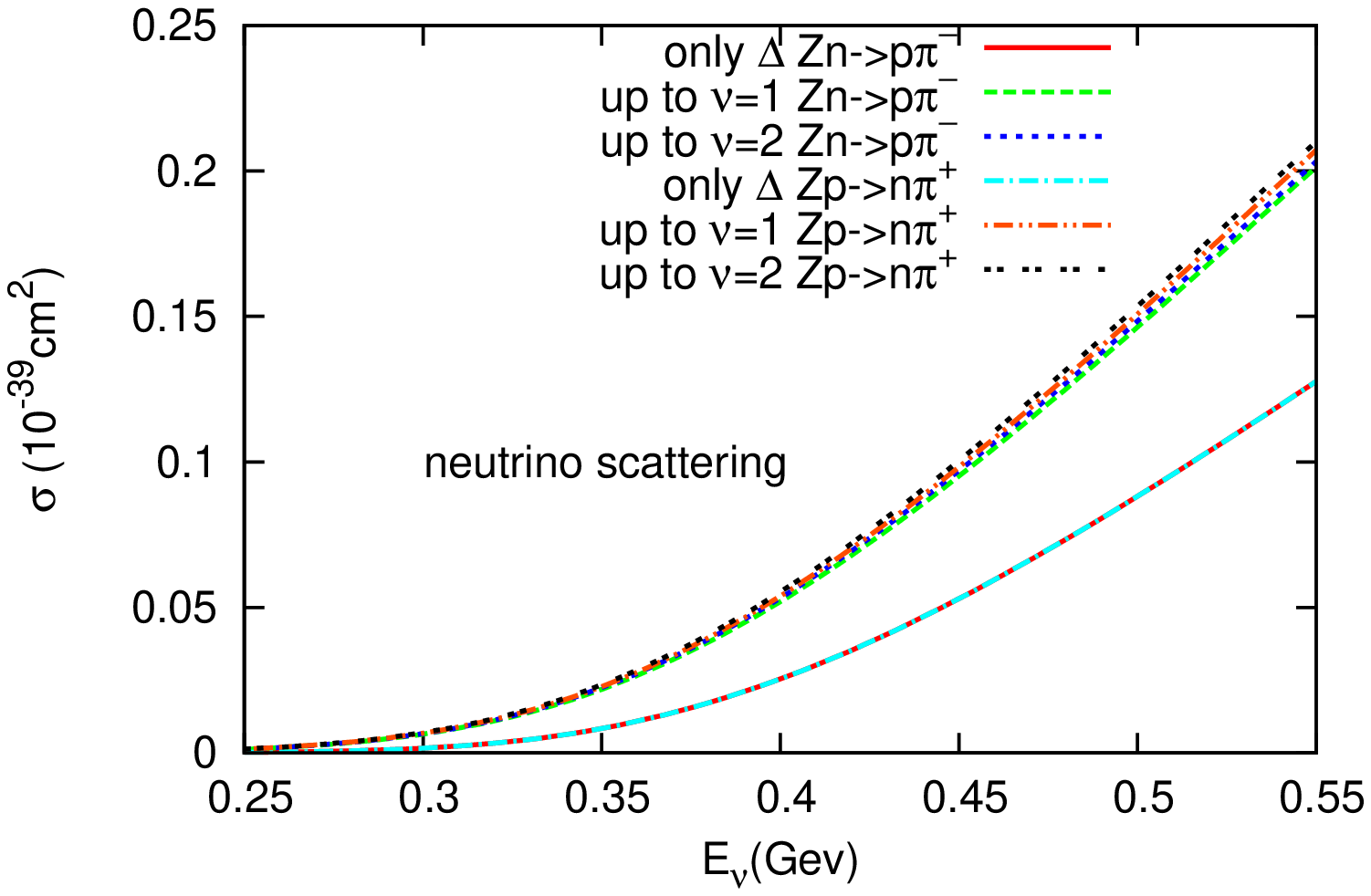}
\caption{Total cross section for NC $\pi^{\pm}$ production due to
neutrino scattering. The curves are defined as in
Fig.~\ref{Fig:pppion+low}, and the channels are also indicated.}
\label{Fig:ncpion+-}
\end{figure}

\begin{figure}[!ht]
\centering
\includegraphics[scale=0.7,angle=-90]{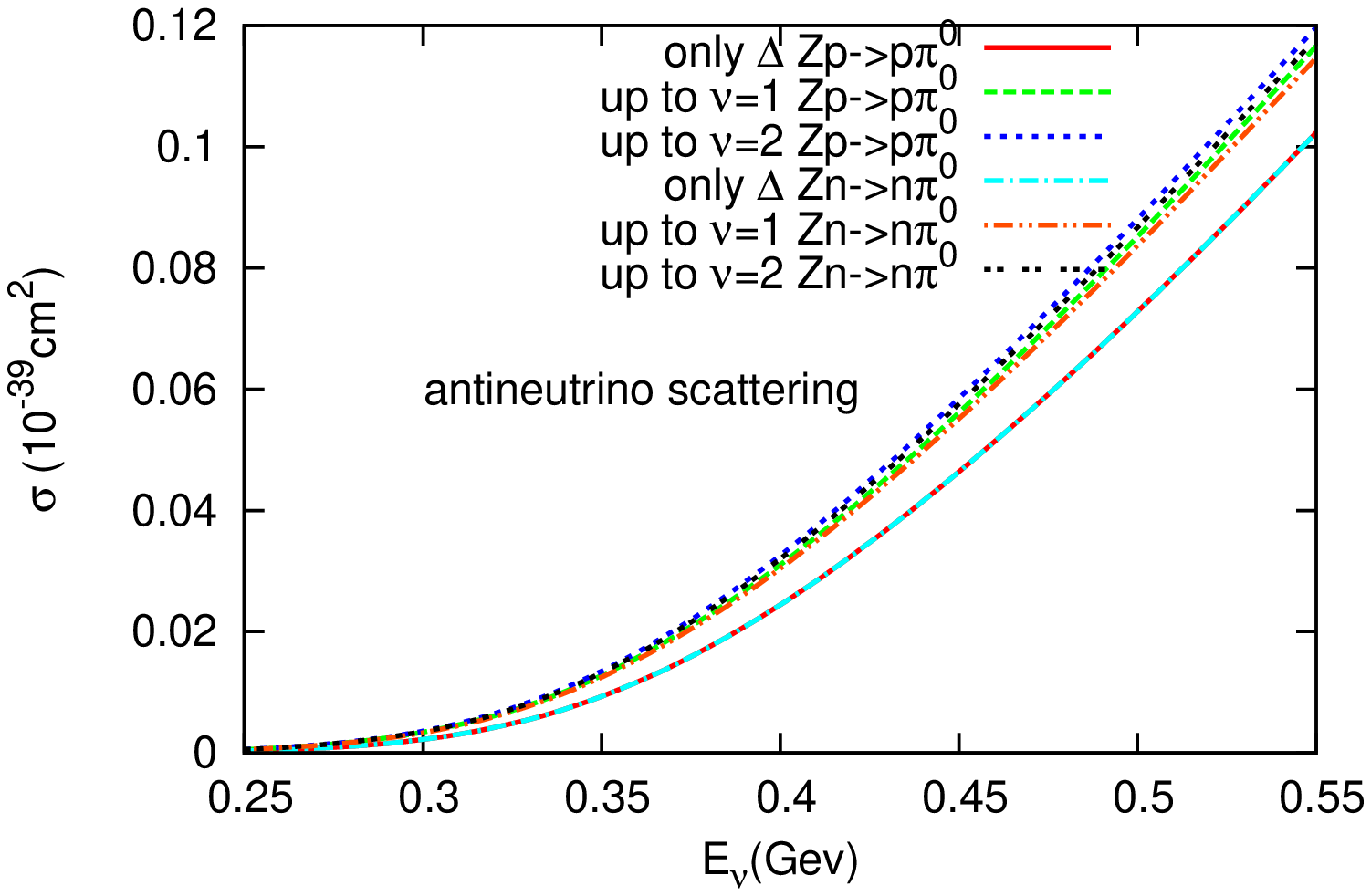}
\caption{Total cross section for NC $\pi^{0}$ production due to
antineutrino scattering. The curves are defined as in
Fig.~\ref{Fig:pppion+low}, and the channels are also indicated.}
\label{Fig:ncantipion0}
\end{figure}

\begin{figure}[!ht]
\centering
\includegraphics[scale=0.7,angle=-90]{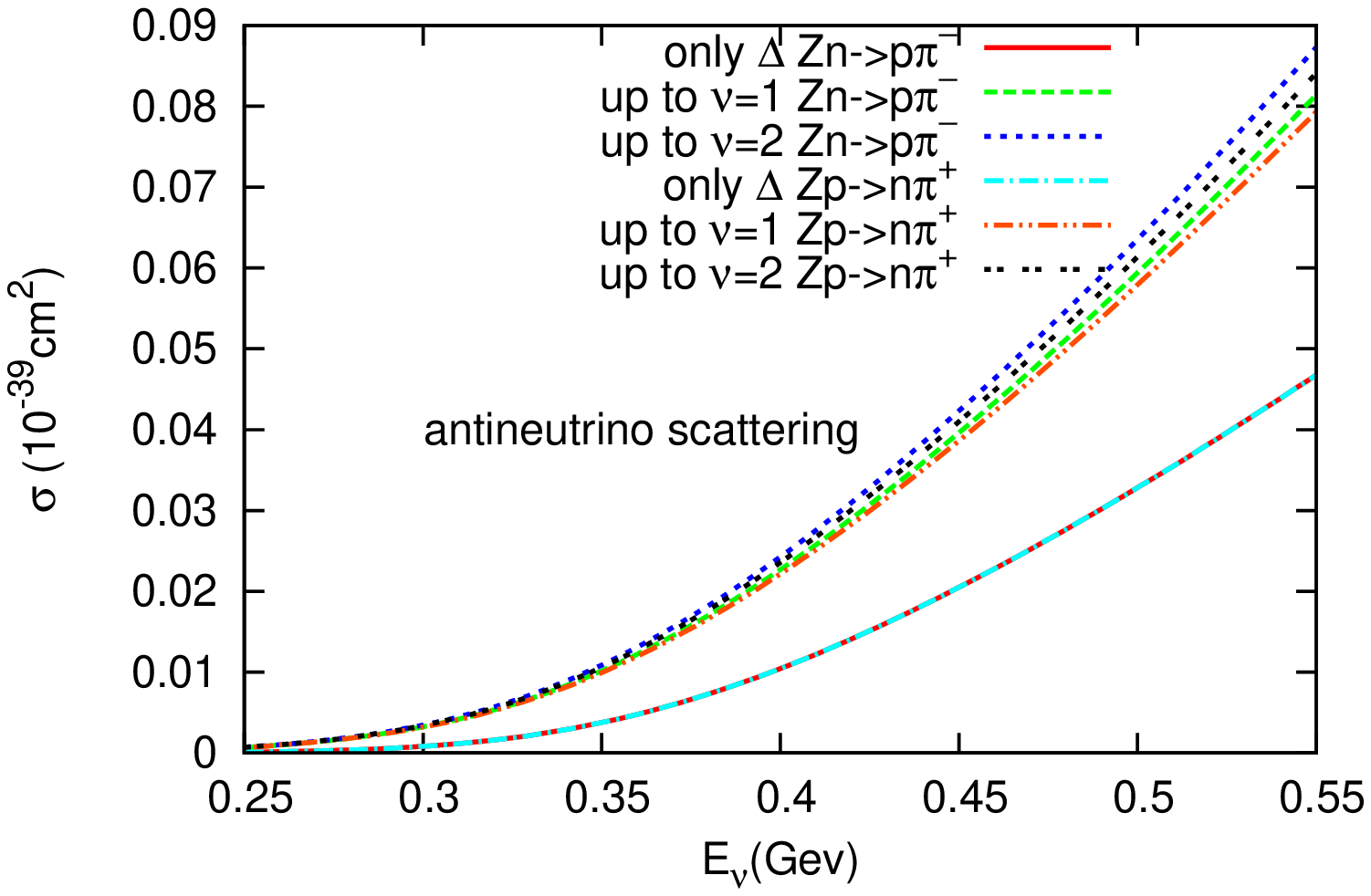}
\caption{Total cross section for NC $\pi^{\pm}$ production due to
antineutrino scattering. The curves are defined as in
Fig.~\ref{Fig:pppion+low}, and the channels are also indicated.}
\label{Fig:ncantipion+-}
\end{figure}

In this section, we discuss the results for NC pion production in
(anti)neutrino scattering. In Figs.~\ref{Fig:ncpion0}
and~\ref{Fig:ncpion+-}, the results in the `mdff' approach including
diagrams of different orders are shown for neutrino scattering,
while the results for antineutrino scattering are shown in
Figs.~\ref{Fig:ncantipion0} and~\ref{Fig:ncantipion+-}. The channels
are explained in each plot.

Since all of the available data for NC pion production are spectrum
averaged, and neutrinos with $E_{\nu} \leqslant 0.5\, \mathrm{GeV}$
have small weight in such spectrum integrated analyses, we will not
compare our results with data. In other words, current data does not
put strong constraints on the NC pion production in this energy
regime.

Nevertheless, we clearly see the convergence of our calculations;
introducing the $\nu=2$ terms does not change the total cross
section significantly. However, we also see the violation of isospin
symmetry in the `up to $\nu=1$' and `up to $\nu=2$' calculations in
each plot, if we compare each pair of channels in every plot,
namely, Figs.~\ref{Fig:ncpion0}, \ref{Fig:ncpion+-},
\ref{Fig:ncantipion0}, and~\ref{Fig:ncantipion+-}. In principle, if
there is no baryon current contribution in NC production, we should
see that the two channels yield the same results in each plot. For
example, isospin symmetry implies
\begin{eqnarray}
\bra{p,\pi^{0}} V^{0\mu}, A^{0\mu} \ket{p} &=&
          \bra{n,\pi^{0}} V^{0\mu}, A^{0\mu} \ket{n} \ ,
          \\[5pt]
\bra{p,\pi^{0}} J_{B}^{\mu} \ket{p} &=& -\bra{n,\pi^{0}}
          J_{B}^{\mu} \ket{n} \ .
\end{eqnarray}
So with `only $\Delta$', we will not see the difference between the
two cross sections, since the (isoscalar) baryon current cannot
induce transitions from $N$ to $\Delta$.  After introducing
background terms, which contain contributions from the baryon
current, we would expect the results for the two processes to be
different, as confirmed in Fig.~\ref{Fig:ncpion0}. This analysis is
applicable to the other plots.

\subsection{NC photon production}

\begin{figure}[!ht]
\centering
\includegraphics[scale=0.7,angle=-90]{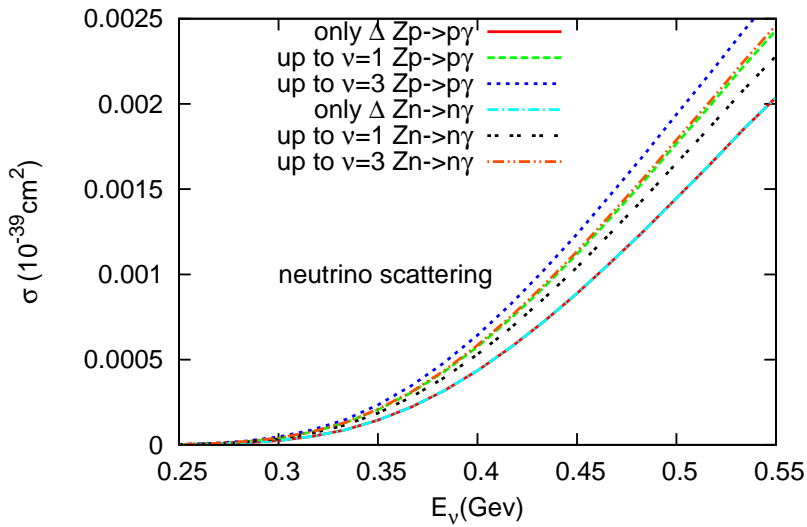}
\caption{Total cross section for NC photon production due to
neutrino scattering. `Only $\Delta$' indicates that only diagrams
with $\Delta$ (both $s$ and $u$ channels) are included. `Up to
$\nu=1$' includes all the diagrams at leading order. `Up to $\nu=3$'
includes higher-order diagrams. The $\nu=2$ terms are zero in these
channels. The next-to-leading order in this channel is $\nu=3$,
whose couplings are from Ref.~\cite{RHill09}. }
\label{Fig:ncphoton}
\end{figure}

\begin{figure}[!ht]
\centering
\includegraphics[scale=0.7,angle=-90]{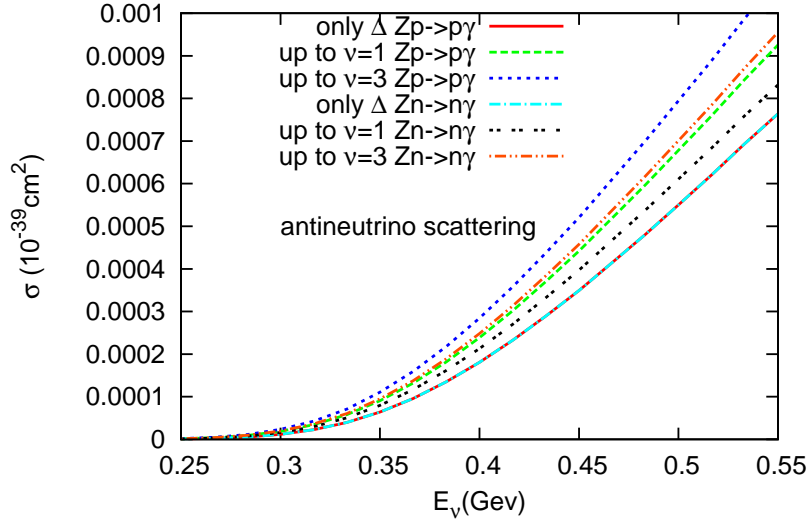}
\caption{Total cross section for NC photon productions due to
antineutrino scattering. The curves are defined as in
Fig.~\ref{Fig:ncphoton}, and the channels are also indicated. }
\label{Fig:ncantiphoton}
\end{figure}

In this section we focus on NC photon production. Besides NC
$\pi^{0}$ production, this process is another important background
in neutrino experiments. As we are ultimately concerned with the
excessive events in the MiniBooNE experiment \cite{MINIBOONE}, we
focus on $E_{\nu} \leqslant 0.5\, \mathrm{GeV}$, as mentioned in the
beginning of this paper. One important difference between NC photon
production and CC and NC pion production, is that all of the $\nu=2$
terms do not contribute in this process. Therefore, we include the
two $\nu=3$ terms in NC photon production, namely, the $e_{1}$ and
$c_{1}$ couplings in Eq.~(\ref{eqn:nccontact}). As mentioned in
Sec.~\ref{subsubsec:lagN}, there are many other interaction terms at
the same order as the $e_{1}$ and $c_{1}$ terms, \emph{but these are
the only two contributing in NC photon production.} Moreover, these
two couplings are singled out in Ref.~\cite{RHill09} as the
low-energy manifestations of anomalous $\rho$ and $\omega$ decay and
are believed to give important contributions in coherent photon
production from nuclei. Here we investigate the consequences of
these two couplings. We emphasize that from the EFT perspective, the
only way to determine these two couplings is by comparing the final
theoretical result with data, rather than by calculating them from
anomalous decay, which is not necessarily the only higher-energy
physics contributing to these two operators. For example, as we
discussed the off-shell couplings before, an off-shell coupling
between $N$, $\pi$, and $\Delta$ can introduce the same matrix
element as that induced by these two contact terms. Changing the
off-shell couplings would also change these two contact terms to
make the theory independent of the choice of off-shell couplings.
Nevertheless, to perform concrete calculations with these two terms
without precise information on the coupling strengths, we use the
values from Ref.~\cite{RHill09} in Figs.~\ref{Fig:ncphoton}
and~\ref{Fig:ncantiphoton}.

We can see the convergence of our calculations. The two couplings
introduced in the `up to $\nu=3$' calculations increase the total
cross section in both channels for both neutrino and antineutrino
scattering, although the change is quite small. This constructive
behavior is consistent with the results in Ref.~\cite{RHill09}.
However, as is easily seen, the contributions of this process are
negligible in scattering off a single nucleon.

Naive power counting, however, does not give an accurate comparison
between the $\Delta$ contributions and the $N$ contributions at low
energy. One reason is that the neutron does not have an electric
charge at low energy, so its current should appear at higher order
than the naive estimate. The second reason is that for the proton,
due to the cancelation between the baryon current and the vector
current, the neutral current is mainly composed of the axial-vector
current, which reduces the strength of the neutral current. Because
of these two factors, the contributions of $\Delta$ and $N$ are at
the same scale far away from resonance, but near the resonance, the
$\Delta$ dominates.

\section{summary} \label{sec:sum}

Weak pion and photon production from nucleons and nuclei produce
important backgrounds in neutrino-oscillation experiments and must
therefore be understood quantitatively. In this work, we studied
pion and photon neutrinoproduction in a Lorentz-covariant, chirally
invariant, meson--baryon EFT. For neutrino energies
$E^{\mathrm{Lab}}_\nu < 1 \,\mathrm{GeV}$, the resonant behavior of
the $\Delta$ is important. We therefore included the $\Delta$
degrees of freedom explicitly in our EFT lagrangian, in a manner
that is consistent with both Lorentz covariance and chiral symmetry.

It is well known that in a lagrangian with a finite number of
interaction terms, including the $\Delta$ as a Rarita--Schwinger
field leads to inconsistencies for strong couplings, strong fields,
or large field variations.  In a modern EFT with an infinite number
of interaction terms, however, these pathologies can be removed, if
we work at low energies with weak boson fields.  This is because the
problematic terms in the lagrangian produce local contact
interactions that can be absorbed into other contact terms in the
EFT lagrangian.  Ambiguous, so-called off-shell couplings have also
been shown to be redundant in the modern EFT framework.  Thus the
$\Delta$ resonance can be introduced into our EFT lagrangian in a
consistent way.  Moreover, we studied the structure of the dressed
$\Delta$ propagator and found that it has a pole only in the
spin-3/2 channel, so that we indeed have the correct number of
resonant degrees of freedom.

Because of the symmetries built into our lagrangian, the vector
currents are conserved and the axial-vector currents satisfy PCAC
automatically, which is not true in some of the other approaches to
this problem.  Needless to say, a conserved vector current is
crucial for computing photon production.  By using vector and
axial-vector transition currents that were calibrated at high
energies, we found results for pion production at lower energies
that are consistent with the (limited) data. This was also true when
vertices described by meson dominance were used.  We also studied
the convergence of our power-counting scheme at low energies and
found that next-to-leading-order tree-level corrections are very
small. Finally, we computed neutral-current photon production
including contact interactions consistent with anomalous $\rho$ and
$\omega$ decays and found that, at least for a nucleon target, the
resulting cross sections are unmeasurably small.

We are currently using this QHD/EFT framework to study the
electroweak response of the nuclear many-body system, so that we can
extend our results to pion and photon neutrinoproduction from
nuclei, which are the true targets in existing neutrino-oscillation
experiments.

\acknowledgments

This work was supported in part by the Department of Energy under
Contract No.\ DE--FG02--87ER40365.

\appendix
\section{isospin indices, $T$ matrices}\label{app:indices}
Suppose $\vec{t}$ are the generators of some (ir)reducible
representation of $SU(2)$; then it is easy to prove that, in matrix
form ($\underline{\widetilde\delta}\equiv-\underline{e}^{-i\pi
t^{y}}$),
\begin{eqnarray}
\underline{\widetilde\delta} \ \underline{\vec{t}} \
\underline{\widetilde\delta}^{\mkern4mu
-1}=-\underline{\vec{t}}^{\mkern6muT} \ ,
\end{eqnarray}
where the superscript $T$ denotes transpose. This equation justifies
the use of $\widetilde{\delta}$ as a metric linking the
representation and the complex conjugate representation. One easily
finds for $\mathcal{D}^{(3/2)}$, $\mathcal{D}^{(1)}$, and
$\mathcal{D}^{(1/2)}$:
\begin{eqnarray}
\widetilde\delta^{ab}=
                     \begin{pmatrix}
                     0&0&0&1 \\
                     0&0&-1&0 \\
                     0&1&0&0\\
                     -1&0&0&0
                     \end{pmatrix} \ ,&\qquad&
\widetilde\delta_{ab}=
                     \begin{pmatrix}
                     0&0&0&-1 \\
                     0&0&1&0 \\
                     0&-1&0&0\\
                     1&0&0&0
                     \end{pmatrix} \ ,  \\[5pt]
\widetilde\delta^{\mkern3mu ij}=\begin{pmatrix}
                     0&0&-1 \\
                     0&1&0 \\
                    -1&0&0
                      \end{pmatrix} \ , &\qquad&
\widetilde\delta_{\mkern3mu ij}=\begin{pmatrix}
                     0&0&-1 \\
                     0&1&0 \\
                    -1&0&0
                      \end{pmatrix}    \ ,  \\[5pt]
\widetilde\delta^{AB}=
                     \begin{pmatrix}
                     0&1 \\
                     -1&0
                     \end{pmatrix} \ , &\qquad&
\widetilde\delta_{AB}=
                     \begin{pmatrix}
                     0&-1 \\
                     1&0
                     \end{pmatrix} \ .  
\end{eqnarray}

We turn now to the $T$ matrices. As discussed in
Sec.~\ref{subsec:not},
\begin{eqnarray}
T^{\dagger \,\,\,iA}_{a}&=&\langle\frac{3}{2}; a \vert 1,\frac{1}{2};i,A
\rangle \ ,\\[5pt]
T^{a}_{\,\,iA}&=&\langle  1, \frac{1}{2};i,A \vert \frac{3}{2};
a\rangle \ . 
\end{eqnarray}
To be more specific:
\begin{eqnarray}
T^{\dagger \,\,\,+1\,A}_{a}&=&
                           \begin{pmatrix}
                            1&0\\
                            0&\sqrt{\frac{1}{3}} \\
                            0&0 \\
                            0&0
                           \end{pmatrix}_{aA} \ , \quad
T^{\dagger \,\,\,0\,A}_{a}=
                           \begin{pmatrix}
                            0&0 \\
                            \sqrt{\frac{2}{3}}&0 \\
                            0&\sqrt{\frac{2}{3}} \\
                            0&0
                           \end{pmatrix}_{aA} \ , \quad
T^{\dagger \,\,\,-1\,A}_{a}=
                           \begin{pmatrix}
                            0&0 \\
                            0&0 \\
                            \sqrt{\frac{1}{3}}&0 \\
                            0&1
                           \end{pmatrix}_{aA}  \ , \\[8pt]
T^{a}_{\,\,+1 \,A} &=&
                  \begin{pmatrix}
                  1&0&0&0 \\
                  0&\sqrt{\frac{1}{3}}&0&0
                  \end{pmatrix}_{Aa} \ , \quad
T^{a}_{\,\,0 \,A} =
                  \begin{pmatrix}
                  0&\sqrt{\frac{2}{3}}&0&0 \\
                  0&0&\sqrt{\frac{2}{3}}&0
                  \end{pmatrix}_{Aa} \ , \quad
T^{a}_{\,\,-1 \,A} =
                  \begin{pmatrix}
                  0&0&\sqrt{\frac{1}{3}}&0 \\
                  0&0&0&1
                  \end{pmatrix}_{Aa} \ . \notag\\[3pt]
&\null&
\end{eqnarray}
It is easy to prove the following relations (here $\tau^{i}$ is a
Pauli matrix):
\begin{eqnarray}
\tau^{i} \, \tau_{j} &=& \widetilde\delta^{\mkern3mu i}_{\mkern3mu
          j} + i\, \widetilde\epsilon^{\mkern5mu i}_{\;jk} \tau^{k} \ ,
          \\[5pt]
\left(P_{i}^{j}\right)_{A}^{\; B} &\equiv& T^{a}_{\;iA}\,
          T^{\dagger \;jB}_{a} =\widetilde\delta_{\mkern3mu i}^{j}
          \,\widetilde\delta_{A}^{\;B}-\frac{1}{3}(\tau_{i}\tau^{j})_{A}^{\;B} \ ,
          \\[5pt]
T^{\dagger \;iA}_{a}\;
          T^{b}_{\;iA}&=&\widetilde\delta_{a}^{\mkern3mu b} \ . 
\end{eqnarray}
Here $P_{i}^{j}$ is a projection operator that projects
$\mathcal{H}^{(\frac{1}{2})} \otimes \mathcal{H}^{(1)}$ onto
$\mathcal{H}^{(\frac{3}{2})}$.

A few words about $\widetilde\epsilon^{\mkern5mu i}_{\;jk}$ are in
order here. We have the following transformations of pion fields:
\begin{eqnarray}
\pi^{i}  &=& u^{i}_{I} \pi^{I}  \qquad \qquad
\text{here, $i=+1,0,-1$}\ ; \quad \text{ $I=x,y,z$ } \ ; \\[5pt]
\begin{pmatrix}
\pi^{+1} \\
\pi^{0} \\
\pi^{-1}
\end{pmatrix} &=&
\begin{pmatrix}
\displaystyle{\frac{-1}{\sqrt{2}}} & \displaystyle{\frac{-i}{\sqrt{2}}} & 0 \\
0 & 0 & 1 \\
\displaystyle{\frac{1}{\sqrt{2}}} &
\displaystyle{\frac{-i}{\sqrt{2}}} & 0
\end{pmatrix}
\begin{pmatrix}
\pi^{x} \\
\pi^{y} \\
\pi^{z}
\end{pmatrix}  \ ; 
\end{eqnarray}
and hence we have
\begin{eqnarray}
\pi \cdot \pi &=& \pi^{x}\pi^{x}+\pi^{y}\pi^{y}+\pi^{z}\pi^{z} \notag \\
&=& -\pi^{+1}\pi^{-1}-\pi^{-1}\pi^{+1}+\pi^{z}\pi^{z}  \notag \\
&=& \pi^{+1}\pi_{+1}+\pi^{-1}\pi_{-1}+\pi^{0}\pi_{0} \ . 
\end{eqnarray}
Meanwhile, under such transformations,
\begin{eqnarray}
\widetilde\epsilon^{\mkern5mu ijk}&\equiv& u^{i}_{I}\, u^{j}_{J}\, u^{k}_{K}
             \, \epsilon^{IJK}=\det(\underline{u^{i}_{I}})
             \epsilon^{ijk}=-i\, \epsilon^{ijk} \notag  \\
\Longrightarrow \quad\widetilde\epsilon^{\mkern5mu ijk}&=&
 \begin{cases}
  -i,            & \text{if $ijk=+1,0,-1$} \ ; \\
  -i \,\delta_{\mathcal{P}}, & \text{if $ijk=\mathcal{P}(+1,0,-1)$} \ .
 \end{cases}  
\end{eqnarray}

\section{$C$, $P$, and $T$ symmetry realized in QCD} \label{app:cptsymQCD}

The symmetries $C$, $P$, and $T$ are preserved in QCD. Based on the
lagrangian in Eq.~(\ref{eqn:qcdL}), we find the corresponding
transformation rules shown in Table~\ref{tab:bgfieldsundercpt}.
Inside the table,
$\mathcal{P}_{\nu}^{\mu}=\text{diag}(1,-1,-1,-1)_{\mn}$ and
$\mathcal{T}_{\nu}^{\mu}=\text{diag}(-1,1,1,1)_{\mn}$.
\begin{table}
 \begin{center}
   \begin{tabular}{|c|c|c|c|c|c|c|c|c|c|c|} \hline
           & $\vbg^{\mu}$
           & $\vbg_{(s)}^{\mu}$
           & $\abg^{\mu}$
           & $\sbg$
           & $\pbg$
           & $r^{\mu}$
           & $l^{\mu}$
           & $f_{R\mu\nu}$
           & $f_{L\mu\nu}$
           & $f_{s\mu\nu}$   \\  \hline
      $C$  &  $-\vbg^{T \mu}$
           & $-\vbg_{(s)}^{\mu}$
           & $\abg^{T \mu}$
           & $\sbg^{T}$
           & $\pbg^{T}$
           & $-l^{T\mu}$
           & $-r^{T\mu}$
           & $-f^{T}_{L\mu\nu}$
           &$-f^{T}_{R\mu\nu}$
           &$-f^{T}_{s\mu\nu}$ \\   \hline
      $P$  &  $\mathcal{P}^{\mu}_{\nu}\vbg^{\nu}$
           &  $\mathcal{P}^{\mu}_{\nu} \vbg_{(s)}^{\nu}$
           &  $-\mathcal{P}^{\mu}_{\nu}\abg^{\nu}$
           &  $\sbg$
           &  $-\pbg$
           &  $\mathcal{P}^{\mu}_{\nu}l^{\nu}$
           &  $\mathcal{P}^{\mu}_{\nu}r^{\nu}$
           &  $\mathcal{P}_{\mu}^{\lambda}\mathcal{P}_{\nu}^{\sigma}f_{L\lambda\sigma}$
           &  $\mathcal{P}_{\mu}^{\lambda}\mathcal{P}_{\nu}^{\sigma}f_{R\lambda\sigma}$
           &  $\mathcal{P}_{\mu}^{\lambda}\mathcal{P}_{\nu}^{\sigma}f_{s\lambda\sigma}$
           \\ \hline
      $T$  & $-\mathcal{T}^{\mu}_{\nu}\vbg^{\nu}$
           & $-\mathcal{T}^{\mu}_{\nu} \vbg_{(s)}^{\nu}$
           & $-\mathcal{T}^{\mu}_{\nu} \abg^{\nu}$
           & $\sbg $
           & $-\pbg$
           & $-\mathcal{T}^{\mu}_{\nu}r^{\nu}$
           & $-\mathcal{T}^{\mu}_{\nu}l^{\nu}$
           & $-\mathcal{T}_{\mu}^{\lambda}\mathcal{T}_{\nu}^{\sigma} f_{R\lambda\sigma}$
           & $-\mathcal{T}_{\mu}^{\lambda}\mathcal{T}_{\nu}^{\sigma} f_{L\lambda\sigma}$
           & $-\mathcal{T}_{\mu}^{\lambda}\mathcal{T}_{\nu}^{\sigma} f_{s\lambda\sigma}$
           \\ \hline
   \end{tabular}
   \caption{Transformations of background fields under $C$, $P$, and $T$ operations.
   The transformed spacetime arguments are not shown here.} \label{tab:bgfieldsundercpt}
 \end{center}
\end{table}

\section{$C$, $P$, and $T$ symmetry realized in QHD} \label{app:cptsymQHD}

The $C$, $P$, and $T$ transformation rules are summarized in
Table~\ref{tab:cpteft}. A plus sign means normal, while a minus sign
means abnormal, i.e., an extra minus sign exists in the
transformation. The convention for Dirac matrices sandwiched by
nucleon and/or $\Delta$ fields are
\begin{eqnarray}
C\psibar{N}\,\Gamma N C^{-1} &=& \begin{cases}
                               -N^{T}\, \Gamma^{T}\,  \psibar{N}^{T}\ ,
                               &\text{normal}\ ; \\
                               N^{T}\, \Gamma^{T} \, \psibar{N}^{T}\ ,
                               &\text{abnormal}\ .
                               \end{cases}  \label{eqn:cconjugdia}\\ [5pt]
C(\psibar{\Delta}\, \Gamma N + \psibar{N}\, \Gamma \Delta) C^{-1}
&=&
\begin{cases}
        - \Delta^{T}\,\Gamma^{T}\,\psibar{N}^{T}
        -N^{T}\,\Gamma^{T}\,\psibar{\Delta}^{T}\ ,  &\text{normal}\ ; \\
        +\Delta^{T}\,\Gamma^{T}\,\psibar{N}^{T}
        +N^{T}\,\Gamma^{T}\, \psibar{\Delta}^{T}\ ,
        &\text{abnormal}\ .
        \end{cases} \label{eqn:cconjugnondiagA}\\ [5pt]
Ci(\psibar{\Delta}\, \Gamma N - \psibar{N}\, \Gamma \Delta) C^{-1}
&=&
\begin{cases}
        +i \Delta^{T}\,\Gamma^{T}\,\psibar{N}^{T}
        -i N^{T}\,\Gamma^{T}\,\psibar{\Delta}^{T}\ , &\text{normal}\ ; \\
        -i\Delta^{T}\,\Gamma^{T}\,\psibar{N}^{T}
        +i N^{T}\,\Gamma^{T}\,\psibar{\Delta}^{T}\ ,
        &\text{abnormal}\ .
                                      \end{cases} \label{eqn:cconjugnondiagB}
\end{eqnarray}
\noindent Here, in
Eqs.~(\ref{eqn:cconjugdia}),~(\ref{eqn:cconjugnondiagA}),
and~(\ref{eqn:cconjugnondiagB}), the extra minus sign arises because
the fermion fields anticommute. The factor of $i$ in
Eq.~(\ref{eqn:cconjugnondiagB}) is due to the requirement of
hermiticity of the lagrangian. To make the analysis easier for
$\psibar{\Delta}\,\Gamma N + C.C.$, we can just attribute a minus
sign to an $i$ under the $C$ transformation. Whenever an $i$ exists,
the lagrangian takes the form $i(\psibar{\Delta}\,\Gamma
N-\psibar{N}\,\Gamma \Delta)$. When no $i$ exists, the lagrangian
will be like $\psibar{\Delta}\,\Gamma N+\psibar{N}\,\Gamma \Delta$.

For $P$ and $T$ transformations, the conventions are the same for
$N$ and $\Delta$ fields, except for an extra minus sign in the
parity assignment for each $\Delta$ field \cite{Weinb95}, so we list
only the $N$ case ($\mathcal{P}_{\nu}^{\mu}$ and
$\mathcal{T}_{\nu}^{\mu}$ can be found in
Appendix~\ref{app:cptsymQCD}):
\begin{eqnarray}
P\psibar{N}\,\Gamma_{\mu} N P^{-1} &=& \begin{cases}
          \psibar{N}\, \mathcal{P}_{\mu}^{\nu}\, \Gamma_{\nu}\, N\ ,
          &\text{normal}\ ; \\
          -\psibar{N}\, \mathcal{P}_{\mu}^{\nu}\, \Gamma_{\nu}\, N\ ,
          &\text{abnormal}\ .
                               \end{cases}  \\ [5pt]
T\psibar{N}\,\Gamma_{\mu} N T^{-1} &=& \begin{cases}
          \psibar{N}\, \mathcal{T}_{\mu}^{\nu}\, \Gamma_{\nu}\, N\ ,
          &\text{normal} \ ; \\
          -\psibar{N}\, \mathcal{T}_{\mu}^{\nu}\, \Gamma_{\nu}\, N\ ,
          &\text{abnormal}\ .
                               \end{cases}
\end{eqnarray}
It is easy to generalize these results to $\Gamma_{\mn}$, etc.

Suppose an isovector object is denoted as $O_{\mu}\equiv
O_{i\mu}t^{i}$, then the conventions are explained below:
\begin{eqnarray}
CO_{\mu}C^{-1}&=&\begin{cases}
           O_{\mu}^{T} \ ,    &\text{normal} \ ; \\
           -O_{\mu}^{T} \ ,   &\text{abnormal} \ .
    \end{cases} 
    \\[5pt]
PO_{\mu}P^{-1}&=&\begin{cases}
           \mathcal{P}_{\mu}^{\nu}O_{\nu} \ , &\text{normal} \ ; \\
           -\mathcal{P}_{\mu}^{\nu}O_{\nu} \ ,  &\text{abnormal} \ .
    \end{cases}    
    \\[5pt]
TO_{\mu}T^{-1}&=&\begin{cases}
           \mathcal{T}_{\mu}^{\nu}O_{\nu} \ , &\text{normal} \ ; \\
           -\mathcal{T}_{\mu}^{\nu}O_{\nu} \ , &\text{abnormal} \ .
    \end{cases}   
\end{eqnarray}
The same convention applies to the isovector (pseudo)tensors. For
isovector (pseudo)scalars, the $\mathcal{P}$ and $\mathcal{T}$
should be changed to $\mathbf{1}$. For the $C$ transformation,
$O^{T}$ means transposing both isospin and Dirac matrices in the
definition of $O$, if necessary.

\begin{table}
  \begin{center}
    \begin{tabular}{|c|c|c|c|c|c|c|c|c|c|c|c|c|c|c|c|c|c|c|c|c|c|} \hline
     & $\ugamma{\mu}$
     &$\sigma^{\mu\nu}$
     &$1$
     &$\ugamma{\mu}\ugammafive$
     &$i\ugammafive$
     &$i$
     &$i\overset{\leftrightarrow}{\partial}$
     &$\epsilon^{\mu\nu\alpha\beta}$
     &$\widetilde{a}_{\mu}$
     &$\widetilde{v}_{\mu}$
     &$\widetilde{v}_{\mu\nu}$
     &$\rho_{\mu}$
     &$\rho_{\mu\nu}$
     &$\psibar{\rho}_{\mu\nu}$
     &$V_{\mu}$
     &$V_{\mu\nu}$
     &$\psibar{V}_{\mu\nu}$
     &$F^{(\pm)}_{\mu\nu}$
     &$f_{s\mu\nu}$
     &$\psibar{F}^{\,(\pm)}_{\mu\nu}$
     &$\psibar{f}_{s\mu\nu}$  \\[2pt] \hline
$C$  & $-$ & $-$ & $+$ & $+$ & $+$ & $-$ & $-$ & $+$ & $+$ & $-$ &
     $-$ & $-$
     & $-$ &$-$ &$-$ &$-$ &$-$ & $\mp$   & $-$ & $\mp$     & $-$  \\[2pt] \hline
$P$  & $+$ & $+$ & $+$ & $-$ & $-$ & $+$ & $+$ & $-$ & $-$ & $+$ &
     $+$ & $+$
     & $+$ &$-$ &$+$ &$+$ &$-$ & $\pm$   & $+$ & $\mp$     & $-$  \\[2pt] \hline
$T$  & $-$ & $-$ & $+$ & $-$ & $-$ & $-$ & $-$ & $-$ & $-$ & $-$ &
     $-$ & $-$
     & $-$ &$+$ &$-$ &$-$ &$+$ & $-$     & $-$ & $+$       & $+$ \\[2pt] \hline
    \end{tabular}
    \caption{Transformation properties of objects under $C$, $P$, and $T$.}
    \label{tab:cpteft}
  \end{center}
\end{table}

\section{Form factors for currents} \label{app:ff}

Here we use matrix elements of the various currents to define the
form factors produced by the EFT lagrangian \cite{FST97}. Note that
$q^{\mu}$ is defined as the \emph{incoming} momentum transfer at the
vertex; in terms of initial and final nucleon momenta, $q^{\mu}
\equiv p^{\mu}_{nf} - p^{\mu}_{ni}$.  When a pion is emitted,
$p^{\mu}_{ni} + q^{\mu} = p^{\mu}_{nf} + k^{\mu}_{\pi}$.

To derive these expressions, it is necessary to expand terms in the
lagrangian to leading order in pion and external fields. Useful
results are given below:
\begin{eqnarray}
\Tr(\uhalftau{i}[U\ , \partial^{\mu}U^{\dagger}])&\approx&
        2i\epsilon^{ijk}\,\frac{\pi_{j}}{f_{\pi}}\,
        \frac{\partial_{\mu}\pi_{k}}{f_{\pi}}\ ,  \label{eqn:upartialu+c}\\
\Tr(\uhalftau{i}\{U\ , \partial^{\mu}U^{\dagger}\})&\approx& -2i\,
        \frac{\partial_{\mu}\pi^{i}}{f_{\pi}}\ , \label{eqn:upartialu+ac} \\
\xi^{\dagger} \uhalftau{i}\xi + \xi\uhalftau{i}\xi^{\dagger}&\approx&\tau^{i}\ , \\
\xi^{\dagger} \uhalftau{i}\xi - \xi\uhalftau{i}\xi^{\dagger}&\approx&
        -\epsilon^{ijk}\, \frac{\pi_{j}}{f_{\pi}}\, \tau_{k}\ , \\
\xi^{\dagger} \uhalftau{i}\xi &\approx& \uhalftau{i}-\epsilon^{ijk}\, \frac{\pi_{j}}{f_{\pi}}\,\dhalftau{k}\ , \\
\xi\uhalftau{i}\xi^{\dagger}&\approx&\uhalftau{i}+\epsilon^{ijk}\frac{\pi_{j}}{f_{\pi}}\dhalftau{k}\ , \\
\widetilde{v}_{\mu}&\approx& \frac{1}{2f^{2}_{\pi}}\,
        \epsilon^{ijk}\pi_{j}\partial_{\mu}\pi_{k}\dhalftau{i}
        -\vbg_{i\mu}\uhalftau{i}-\epsilon^{ijk}\, \frac{\pi_{j}}{f_{\pi}}\, \dhalftau{k}\, \abg_{i\mu}\ ,
        \label{eqn:vmu}\\
\widetilde{a}_{\mu}&\approx& \frac{1}{f_{\pi}}\,
        \partial_{\mu}\pi^{i}\, \dhalftau{i}
        +\abg_{i\mu}\uhalftau{i}+\epsilon^{ijk}\, \frac{\pi_{j}}{f_{\pi}}\, \dhalftau{k}\vbg_{i\mu}\ ,
        \label{eqn:amu} \\
\widetilde{v}_{\mn}&\approx&\frac{1}{f^{2}_{\pi}}\,
        \epsilon^{ijk}\partial_{\mu}\pi_{j}\partial_{\nu}\pi_{k}
        \, \dhalftau{i} - \left( i \left[\frac{1}{f_{\pi}}\, \partial_{\mu}\pi^{i}\,\dhalftau{i}\, , \,
        \abg_{\nu}+\epsilon^{ijk}\, \frac{\pi_{j}}{f_{\pi}}\, \dhalftau{k}\, \vbg_{i\nu} \right]
        -(\mu \leftrightarrow \nu) \right)  \notag \\[5pt]
&&{}+ \text{background\ interference\ terms,} \label{eqn:vmn}\\[5pt]
\rho_{\mn}&=&\partial_{[\mu}\rho_{\nu
        ]}+i\overline{g}_{\rho}[\rho_{\mu}\ , \ \rho_{\nu}]
        + i ([\widetilde{v}_{\mu}\ , \ \rho_{\nu}] - \mu \leftrightarrow \nu)\ ,
        \label{eqn:rhomn} \\[5pt]
f_{L\mn}+f_{R\mn}&=& 2\partial_{[\mu}\vbg_{\nu]} -2i[\vbg_{\mu}\ , \ \vbg_{\nu}]
        -2i[\abg_{\mu}\ , \ \abg_{\nu}]\ , \\[5pt]
f_{L\mn}-f_{R\mn}&=& -2\partial_{[\mu}\abg_{\nu]} +2i[\vbg_{\mu}\ , \ \abg_{\nu}]
        +2i[\abg_{\mu}\ , \ \vbg_{\nu}]\ , \\[5pt]
F^{(+)}_{\mn}&=&\xi^{\dagger} \uhalftau{i}\xi f_{Li\mn}+\xi\uhalftau{i}\xi^{\dagger}f_{Ri\mn}
        \notag \\[5pt]
&=&\uhalftau{i}(f_{Li\mn}+f_{Ri\mn}) -\epsilon^{ijk}\,
        \frac{\pi_{j}}{f_{\pi}}\, \tau_{k} (f_{Li\mn}-f_{Ri\mn})
        \notag \\
&\approx& 2\partial_{[\mu}\vbg_{\nu]} +2 \epsilon^{ijk}\,
        \frac{\pi_{j}}{f_{\pi}}\,\dhalftau{k}\,
        \partial_{[\mu}\abg_{i\nu]} + \text{background\ interference,} \label{eqn:F+mn}\\
F^{(-)}_{\mn}&=&\xi^{\dagger} \uhalftau{i}\xi f_{Li\mn}-\xi\uhalftau{i}\xi^{\dagger}f_{Ri\mn} \notag \\
&=&\uhalftau{i}(f_{Li\mn}-f_{Ri\mn}) -\epsilon^{ijk}\,
        \frac{\pi_{j}}{f_{\pi}}\, \tau_{k} (f_{Li\mn}+f_{Ri\mn})
        \notag \\
&\approx& -2\partial_{[\mu}\abg_{\nu]} -2
        \epsilon^{ijk}\, \frac{\pi_{j}}{f_{\pi}}\, \dhalftau{k}\, \partial_{[\mu}\vbg_{i\nu]}
        + \text{background\ interference.} \label{eqn:F-mn}
\end{eqnarray}
We now proceed to determine the matrix elements.

\begin{eqnarray}
\bra{N, B} V^{i}_{\mu} \ket{N, A}&=& \left[
\psibar{u}_{f}\dgamma{\mu}u_{i}
          + \frac{\beta^{(1)}}{M^{2}}\,\psibar{u}_{f}(q^{2}\dgamma{\mu}
          -\slashed{q} q_{\mu}) u_{i} \right.\notag \\[5pt]
&&\left.{}-\frac{g_{\rho}}{g_{\gamma}}
          \frac{q^{2}g_{\mn}-q_{\mu}q_{\nu}}{q^{2}-m^{2}_{\rho}} \,
          \psibar{u}_{f}\ugamma{\nu}u_{i} \right]\, \bra{B} \uhalftau{i}\ket{A} \notag
          \\[5pt]
&&{}+\left[
          2\lambda^{(1)}\,\psibar{u}_{f}\frac{\dsigma{\mn}iq^{\nu}}{2M}\, u_{i}
          -\frac{f_{\rho}g_{\rho}}{g_{\gamma}} \frac{q^{2}}{q^{2}-m^{2}_{\rho}}\,
          \psibar{u}_{f} \frac{\dsigma{\mn}iq^{\nu}}{2M}\, u_{i} \right]\,
          \bra{B} \uhalftau{i}\ket{A} \notag \\[5pt]
&\equiv& \bra{B} \uhalftau{i}\ket{A}\, \psibar{u}_{f}
          \left(\dgamma{\mu}
          +2\delta F_{1}^{V,md}\,\frac{q^{2}\dgamma{\mu}-\slashed{q} q_{\mu}}{q^{2}}
          +2F_{2}^{V,md}\,\frac{\dsigma{\mn}iq^{\nu}}{2M}\right)u_{i} \notag \\
&&\null   \\[5pt]
&\equiv& \bra{B} \uhalftau{i}\ket{A}\, \psibar{u}_{f} \Gamma_{V\mu}(q) u_{i}
          \label{eqnapp:NNvectorcurrentwithff} \\[5pt]
&\overset{\mathrm{on\ shell}}{\equiv}&  \bra{B} \uhalftau{i}\ket{A}\,
          \psibar{u}_{f} \left( 2F_{1}^{V,md}\dgamma{\mu}
          +2F_{2}^{V,md}\,\frac{\dsigma{\mn}iq^{\nu}}{2M} \right)u_{i}\ , 
          \\[5pt]
\bra{N, B} J^{B}_{\mu} \ket{N, A}&=&  \left[ \psibar{u}_{f}\dgamma{\mu}u_{i}
          + \frac{\beta^{(0)}}{M^{2}}\, \psibar{u}_{f}(q^{2}\dgamma{\mu}
          -\slashed{q} q_{\mu}) u_{i} \right. \notag \\[5pt]
&&\left. {}-\frac{2 g_{v}}{3 g_{\gamma}}
          \frac{q^{2}g_{\mn}-q_{\mu}q_{\nu}}{q^{2}-m^{2}_{v} }\,
          \psibar{u}_{f}\ugamma{\nu}u_{i} \right]\, \delta_{B}^{A} \notag
          \\[5pt]
&&+\left[ 2\lambda^{(0)}\, \psibar{u}_{f}\frac{\dsigma{\mn}iq^{\nu}}{2M}\,u_{i}
          -\frac{2 f_{v}g_{v}}{3 g_{\gamma}}
          \frac{q^{2}}{q^{2}-m^{2}_{v}}\,
          \psibar{u}_{f} \frac{\dsigma{\mn}iq^{\nu}}{2M}\, u_{i} \right]\,
          \delta_{B}^{A} \notag \\[5pt]
&\equiv&  \delta_{B}^{A}\, \psibar{u}_{f} \left( \dgamma{\mu}
          +2\delta F_{1}^{S,md}\, \frac{q^{2}\dgamma{\mu}-\slashed{q} q_{\mu}}{q^{2}}
          +2F_{2}^{S,md}\frac{\dsigma{\mn}iq^{\nu}}{2M} \right) u_{i} 
          \\[5pt]
&\equiv& \delta_{B}^{A}\, \psibar{u}_{f} \Gamma_{B\mu}(q) u_{i}
          \label{eqnapp:NNbaryoncurrentwithff} \\[5pt]
&\overset{\mathrm{on\ shell}}{\equiv}&  \delta_{B}^{A} \,
          \psibar{u}_{f}
          \left(2F_{1}^{S,md}\,\dgamma{\mu}+2F_{2}^{S,md}\,
          \frac{\dsigma{\mn}iq^{\nu}}{2M} \right) u_{i} \ ,
\end{eqnarray}

\begin{eqnarray}
&& \bra{N, B; \pi,j,k_{\pi}} A^{i}_{\mu} \ket{N, A}\notag \\[5pt]
&=&-\frac{\epsilon^{i}_{\,jk}}{f_{\pi}}\, \bra {B} \uhalftau{k}\ket{A}\,
          \psibar{u}_{f}\ugamma{\nu}u_{i}\, \left[g_{\mn}
          +\frac{\beta^{(1)}}{M^{2}}\, (q\cdot(q-k_{\pi}) g_{\mn}-(q-k_{\pi})_{\mu} q_{\nu})
          \right. \notag \\[5pt]
&&\left. {}-\frac{g_{\rho}}{g_{\gamma}}\,\frac{q\cdot(q-k_{\pi})g_{\mn}-(q-k_{\pi})_{\mu}q_{\nu}}
          {(q-k_{\pi})^{2}-m^{2}_{\rho}} \right]   \notag \\[5pt]
&&{}-\frac{\epsilon^{i}_{\,jk}}{f_{\pi}}\, \bra {B} \uhalftau{k}\ket{A}\,
          \psibar{u}_{f}\frac{\dsigma{\mn}iq^{\nu}}{2M}\, u_{i}
          \left[2\lambda^{(1)}-\frac{f_{\rho}g_{\rho}}{g_{\gamma}}\,
          \frac{q\cdot(q-k_{\pi})}{(q-k_{\pi})^{2}-m^{2}_{\rho}} \right] 
          \\[5pt]
&\equiv& -\frac{\epsilon^{i}_{\,jk}}{f_{\pi}}\, \bra {B} \uhalftau{k}\ket{A}\,
          \psibar{u}_{f}\ugamma{\nu}u_{i} \notag \\[5pt]
&&{}\times \left[g_{\mn}+2\delta F_{1}^{V,md}((q-k_{\pi})^{2})
          \frac{ q\cdot(q-k_{\pi})g_{\mn}-(q-k_{\pi})_{\mu}q_{\nu}}{(q-k_{\pi})^{2}}
          \right] \notag \\[5pt]
&&{}-\frac{\epsilon^{i}_{\,jk}}{f_{\pi}}\, \bra {B} \uhalftau{k}\ket{A}\,
          \psibar{u}_{f}\frac{\dsigma{\mn}iq^{\nu}}{2M}\, u_{i}
          \left[2\lambda^{(1)}+2\delta F^{V,md}_{2}((q-k_{\pi})^{2})\,
          \frac{q\cdot(q-k_{\pi})}{(q-k_{\pi})^{2}} \right] 
          \\[5pt]
&\equiv&  \frac{\epsilon^{i}_{\,jk}}{f_{\pi}}\, \bra {B}
          \uhalftau{k}\ket{A}\, \psibar{u}_{f} \Gamma_{A\pi \mu}(q,k_{\pi})
          u_{i} \ . \label{eqnapp:NNpionaxialcurrentwithff}
\end{eqnarray}

Now we consider $\bra{N, B} A^{i}_{\mu} \ket{N, A}$ and $\bra{N, B;
\pi,j} V^{i}_{\mu} \ket{N, A}$. In the \emph{chiral limit}, we find
\begin{eqnarray}
\bra{N, B} A^{i}_{\mu} \ket{N, A}&=& -\bra{B} \uhalftau{i}\ket{A}
          \psibar{u}_{f} \ugamma{\nu}\ugammafive u_{i}\left[
          g_{A}\left(g_{\mn}-\frac{q_{\mu}q_{\nu}}{q^{2}}\right)
          -\frac{\beta^{(1)}_{A}}{M^{2}}(q^{2}g_{\mn}
          -q_{\mu}q_{\nu}) \right.\notag \\[5pt]
&&\left.{}-2c_{a_{1}}g_{a_{1}}\,
          \frac{q^{2}g_{\mn}-q_{\mu}q_{\nu}}{q^{2}-m^{2}_{a_{1}}} \right]
          \ ,
\end{eqnarray}
\begin{eqnarray}
&& \bra{N, B; \pi,j,k_{\pi}} V^{i}_{\mu} \ket{N, A}\notag \\[5pt]
&=&\frac{\epsilon^{i}_{\,jk}}{f_{\pi}}\, \bra {B}
\uhalftau{k}\ket{A}\,
          \psibar{u}_{f}\ugamma{\nu}\ugammafive u_{i}\, \left[g_{A}g_{\mn}
          -\frac{\beta^{(1)}_{A}}{M^{2}}\, [q\cdot(q-k_{\pi})
          g_{\mn}-(q-k_{\pi})_{\mu} q_{\nu}]
          \right. \notag \\[5pt]
&&\left. {}-2c_{a_{1}}g_{a_{1}}
          \,\frac{q\cdot(q-k_{\pi})g_{\mn}-(q-k_{\pi})_{\mu}q_{\nu}}
          {(q-k_{\pi})^{2}-m^{2}_{a_{1}}} \right]\ .
\end{eqnarray}
Now suppose there is only one manifestly chiral-symmetry-breaking
term, i.e., the mass term for pions; then the pion-pole contribution
associated with the $g_{A}$ coupling in $\bra{N, B} A^{i}_{\mu}
\ket{N, A}$ will become
$g_{A}[g_{\mn}-q_{\mu}q_{\nu}/(q^{2}-m_{\pi}^{2})]$, while the other
parts in $\bra{N, B} A^{i}_{\mu} \ket{N, A}$, as well as the whole
$\bra{N, B; \pi,j} V^{i}_{\mu} \ket{N, A}$, will remain unchanged.
However, we must realize that there are other possible
chiral-symmetry-breaking terms contributing to $\bra{N, B}
A^{i}_{\mu} \ket{N, A}$. For example, $(m_{\pi}^{2}/M)\,
\psibar{N}i\ugammafive (U-U^{\dagger})N$ will contribute to $\bra{N,
B} A^{i}_{\mu} \ket{N, A}$ as
\begin{eqnarray}
-\frac{2m_{\pi}^{2}}{M^{2}} \frac{q_{\mu} \slashed{q}
\ugammafive}{q^{2}-m_{\pi}^{2}}\bra{B} \uhalftau{i}\ket{A}\ . \notag
\end{eqnarray}
To simplify the fitting procedures, we will use the following form
factors:
\begin{eqnarray}
\bra{N, B} A^{i}_{\mu} \ket{N, A}&=& -G_{A}^{md}(q^{2})\bra{B}
          \uhalftau{i}\ket{A} \psibar{u}_{f} \left(g_{\mn}
          -\frac{q_{\mu}q_{\nu}}{q^{2}-m_{\pi}^{2}} \right) \ugamma{\nu}\ugammafive u_{i}
          \ , \\[5pt]
\bra{N, B; \pi,j,k_{\pi}} V^{i}_{\mu} \ket{N,
          A}&=&\frac{\epsilon^{i}_{\,jk}}{f_{\pi}}\, \bra {B}
          \uhalftau{k}\ket{A}\,
          \psibar{u}_{f}\ugamma{\nu}\ugammafive u_{i}\,
          \bigg[g_{A}g_{\mn} \notag \\[5pt]
&&{}\left. {}+\delta G^{md}_{A}((q-k_{\pi})^{2}) \,
          \frac{q\cdot(q-k_{\pi})g_{\mn}-(q-k_{\pi})_{\mu}q_{\nu}}
          {(q-k_{\pi})^{2}} \right]\ .
\end{eqnarray}
The required definitions can be found in Eqs.~(\ref{eqn:defofGA1})
and (\ref{eqn:defofGA2}).

Finally, we calculate the pion form factor $\bra{\pi,k}
V^{i}_{\mu}\ket{\pi,j}$:
\begin{eqnarray}
\bra{\pi,k,k_{\pi}} V^{i}_{\mu}\ket{\pi,j,k_{\pi}-q} &=&i\epsilon^{ij}_{\;\; k}
          (2k_{\pi}-q)_{\mu} \notag \\[5pt]
&& {}+2i\, \frac{g_{\rho\pi\pi}}{g_{\gamma}}\,  \epsilon^{ij}_{\;\;k}\,
          \frac{q^{2}}{m^{2}_{\rho}} \frac{1}{q^{2}-m^{2}_{\rho}}\,
          (q\cdot k_{\pi} q_{\mu}-q^{2}k_{\pi\mu}) \notag \\[5pt]
q^{2}\to m_{\rho}^{2} \ \text{in numerator} \longrightarrow &&
          i\epsilon^{ij}_{\;\; k}\, (2k_{\pi}-q)_{\mu} \notag
          \\[5pt]
&& {}+2i\, \frac{g_{\rho\pi\pi}}{g_{\gamma}}\, \epsilon^{ij}_{\;\;k}\,
          \frac{1}{q^{2}-m^{2}_{\rho}}\, (q\cdot k_{\pi} q_{\mu}-q^{2}k_{\pi\mu}) 
          \\[5pt]
\text{pion on shell} \, &=& i\epsilon^{ij}_{\;\; k}\, (2k_{\pi}-q)_{\mu}
          \left( 1-\frac{g_{\rho\pi\pi}}{g_{\gamma}}\frac{q^{2}}{q^{2}-m^{2}_{\rho}}\right)
          \notag \\[5pt]
&\equiv&  i \epsilon^{ij}_{\;\; k}\, (2k_{\pi}-q)_{\mu} F_{\pi}^{md}(q^{2}) 
          \\
\Longrightarrow  && \notag \\
\bra{\pi,k,k_{\pi}} V^{i}_{\mu}\ket{\pi,j,k_{\pi}-q} &=& i\epsilon^{ij}_{\;\; k}\,
          \left[(2k_{\pi}-q)_{\mu} +2 \delta F_{\pi}^{md}(q^{2}) \left(k_{\pi \mu}
          -\frac{q\cdot k_{\pi}}{q^{2}}\, q_{\mu}\right)\right] \notag
          \\[5pt]
&\equiv&  i\epsilon^{ij}_{\;\; k}\, P_{V \mu}(q,k_{\pi}) \ .
          \label{eqnapp:pionvectorcurrentff}          \\[5pt]
\text{Here,} \qquad  \delta F_{\pi}^{md}(q^{2})
          &=&F_{\pi}^{md}(q^{2}) -F_{\pi}^{md}(0) \ . 
\end{eqnarray}

\section{Free $\Delta$ propagator and self-energy insertion}
\label{app:deltapropagator}

Normally, the free, spin-$3/2$ field's propagator can be decomposed
as \cite{tang98,PVN81,BDM89}
\begin{eqnarray}
S_{F}^{ 0 \mn}(p) &\equiv& -\frac{1}{\slashed{p}-m+i \epsilon}\,
          P^{(\frac{3}{2}) \mn}
          -\frac{1}{\sqrt{3}m}\, P^{(\half) \mn}_{12}
          -\frac{1}{\sqrt{3}m}\, P^{(\half) \mn}_{21} \notag \\[5pt]
&&{}+\frac{2}{3m^{2}}\, (\slashed{p}+m)\, P^{(\half) \mn}_{22} \ , 
          \\[5pt]
P^{(\frac{3}{2}) \mn}&=& g^{\mn}-\frac{1}{3}\, \ugamma{\mu}\ugamma{\nu}
          +\frac{1}{3p^{2}}\, \ugamma{[\mu}p^{\nu]} \slashed{p}
          -\frac{2}{3p^{2}}\, p^{\mu}p^{\nu} \ , 
          \\[5pt]
P^{(\frac{1}{2}) \mn}_{11}&=&\frac{1}{3}\, \ugamma{\mu}\ugamma{\nu}
          -\frac{1}{3p^{2}}\, \ugamma{[\mu} p^{\nu]} \slashed{p}
          -\frac{1}{3p^{2}}\, p^{\mu}p^{\nu} \ , 
          \\[5pt]
P^{(\frac{1}{2}) \mn}_{12}&=& \frac{1}{\sqrt{3}p^{2}}\,
          (-p^{\mu}p^{\nu}+\ugamma{\mu}p^{\nu}\slashed{p}) \ , 
          \\[5pt]
P^{(\frac{1}{2}) \mn}_{21}&=& \frac{1}{\sqrt{3}p^{2}}\,
          (p^{\mu}p^{\nu}-\ugamma{\nu}p^{\mu}\slashed{p}) \ , 
          \\[5pt]
P^{(\frac{1}{2}) \mn}_{22}&=&\frac{1}{p^{2}}\, p^{\mu} p^{\nu} \ .
\end{eqnarray}
It is easy to prove the following relations:
\begin{eqnarray}
(P^{(I) }_{ij})^{\mn} (P^{(J)}_{kl })_{\nu \lambda} &=&\delta_{IJ}\,
          \delta_{jk}\, (P^{(I)}_{il})^{\mu}_{\lambda} \ , 
          \\[5pt]
\ugamma{\mu} P^{\left(\frac{3}{2}\right)}_{\mn} &=&P^{\left(\frac{3}{2}\right)}_{\mn}
          \ugamma{\nu} =0 \ , \label{eqn:spinprojection2} \\[5pt]
p^{\mu}P^{\left(\frac{3}{2}\right)}_{\mn} &=&P^{\left(\frac{3}{2}\right)}_{\mn} p^{\nu}
          =0 \ , \label{eqn:spinprojection3} \\[5pt]
P^{(\threehalf)}+P^{(\half)}_{11}+P^{(\half)}_{22}&=& \bf{1} \ , 
          \\[5pt]
 P^{(\half)}_{11}+P^{(\half)}_{22}& \equiv& P^{(\threehalf \perp)} \ , 
          \\[5pt]
\left[P^{(\threehalf)} \, , \,  \slashed{p}\right]&=&0 \ , 
          \\[3pt]
\left[P^{(\half)}_{11} \, , \,  \slashed{p}\right]&=&0 \ , 
          \\[3pt]
\left[P^{(\half)}_{22} \, , \,  \slashed{p}\right]&=&0 \ . 
\end{eqnarray}
With the preceding properties of the projection operators, we easily
find
\begin{eqnarray}
S_{F}^{0}(p) &=& P^{(\frac{3}{2})}\, \frac{-1}{\slashed{p}-m+i \epsilon}\,
          P^{(\frac{3}{2})} -\frac{1}{\sqrt{3}m}\, P^{(\half)}_{12}
          -\frac{1}{\sqrt{3}m}\, P^{(\half) }_{21}
          +P^{(\half)}_{22}\frac{2}{3m^{2}}\, (\slashed{p}+m)\, P^{(\half)}_{22} \notag
          \\[5pt]
&=&P^{(\frac{3}{2})}\, \frac{-1}{\slashed{p}-m+i\epsilon}\,
          P^{(\frac{3}{2})} \notag
          \\[5pt]
&&{}+ P^{(\threehalf \perp)} \left[-\frac{1}{\sqrt{3}m}\, P^{(\half)}_{12}
          -\frac{1}{\sqrt{3}m}\, P^{(\half) }_{21}
          +P^{(\half)}_{22}\frac{2}{3m^{2}}\, (\slashed{p}+m)\, P^{(\half)}_{22}
          \right] P^{(\threehalf \perp)} 
          \\[5pt]
&\equiv& S_{F}^{0 (\threehalf)}(p) + S_{F}^{0 (\threehalf \perp)}(p)
          \ .
\end{eqnarray}

Following the analysis in Ref.~\cite{tang98}, the self-energy of the
$\Delta$ can be defined as $\Sigma_{\mn} \equiv \Sigma^{\Delta}
g_{\mn}+\delta \Sigma_{\mn}$. It follows that in
$\delta\Sigma_{\mn}$, the indices can only have structures like the
products $(\dgamma{\mu}, p_{\mu}) (\dgamma{\nu}, p_{\nu})$. Then we
find quite an interesting property of $\Sigma$ (the $\mn$ indices
are suppressed):
\begin{eqnarray}
\Sigma &=&\Sigma^{\Delta} g +\delta \Sigma \notag \\[3pt]
&=& (P^{(\frac{3}{2})}+P^{(\threehalf \perp)})(\Sigma^{\Delta} g+\delta \Sigma)
          (P^{(\frac{3}{2})}+P^{(\threehalf \perp)}) \notag \\[3pt]
&=&  P^{(\frac{3}{2})}\Sigma^{\Delta} g P^{(\frac{3}{2})}
          + P^{(\threehalf \perp)} \Sigma  P^{(\threehalf \perp)} \notag
          \\[3pt]
&&{}+  P^{(\frac{3}{2})} (\Sigma^{\Delta} g+\delta \Sigma) P^{(\threehalf \perp)}
          + P^{(\threehalf \perp)} (\Sigma^{\Delta} g+\delta \Sigma) P^{(\frac{3}{2})}
          \notag \\[3pt]
&=&  P^{(\frac{3}{2})}\Sigma^{\Delta} P^{(\frac{3}{2})}
          + P^{(\threehalf \perp)} \Sigma  P^{(\threehalf \perp)} \notag
          \\[3pt]
&\equiv& \Sigma^{(\frac{3}{2})} + \Sigma^{(\frac{3}{2} \perp)} \ .
\end{eqnarray}
In the proof, we make use of $\left[P^{(\threehalf)} \, , \,
\Sigma^{\Delta} \right]=0$ and $\left[P^{(\threehalf \perp)} \, , \,
\Sigma^{\Delta} \right]=0$, since the only possible spin structures
of $\Sigma^{\Delta}$ are $\bf{1}$, $\not\!\! p$,
and $\ugammafive$ (parity violation), which commute with the two
projection operators. This implies $P^{(\frac{3}{2})}\Sigma^{\Delta}
g P^{(\threehalf \perp)}=0$ and $P^{(\frac{3}{2}
\perp)}\Sigma^{\Delta} g P^{(\threehalf)}=0$. We also make use of
Eqs.~(\ref{eqn:spinprojection2}) and (\ref{eqn:spinprojection3}), so
we get $P^{(\frac{3}{2})} \delta \Sigma P^{(\threehalf \perp)}=0$
and $P^{(\threehalf \perp)} \delta \Sigma P^{(\frac{3}{2})}=0$.

\section{Construction of a Delta interaction term} \label{app:deltalagrangianexample}

We show an example of constructing a term in the lagrangian with
interactions between pions and the $\Delta$. Consider
$$
\frac{-i}{2}\, \psibar{\Delta}^{\mkern4mu
          a}_{\mu}\left\{\usigma{\mn}\, , \,
          \slashed{\widetilde{a}}_{a}^{\mkern3mu b} \ugammafive \right\}
          \Delta_{b\nu} \ .
$$
It can be shown that
\begin{eqnarray}
\frac{-i}{2}\, \psibar{\Delta}^{\mkern4mu
          a}_{\mu}\left\{\usigma{\mn}\, , \,
          \slashed{\widetilde{a}}_{a}^{\mkern3mu b} \ugammafive \right\}
          \Delta_{b\nu} && \notag
          \\[5pt]
&=&\psibar{\Delta}^{\mkern4mu
          a\mu}\slashed{\widetilde{a}}_{a}^{\mkern3mu b} \ugammafive
          \Delta_{b\mu}+\psibar{\Delta}^{\mkern4mu a}_{\mu}
          \left[-\ugamma{\mu}\,\widetilde{a}^{\mkern2mu \nu}\ugammafive
          -\ugamma{\nu}\,\widetilde{a}^{\mkern2mu \mu} \ugammafive
          +\ugamma{\mu}\slashed{\widetilde{a}}\,\ugamma{\nu}\ugammafive
          \right]_{a}^{\mkern3mu b}\Delta_{b\nu} \ . \notag
          \\
&&\null
\end{eqnarray}
According to the argument in Sec.~\ref{subsubsec:deltaprop}, the
original coupling and $\displaystyle{\psibar{\Delta}^{\; a\mu}
\slashed{\widetilde{a}}_{a}^{\mkern3mu b}\, \ugammafive
\Delta_{b\mu} }$ are equivalent in low-energy effective theory. We
use the second form in our lagrangian.

\section{kinematics} \label{app:kinmatics}

Following a standard calculation, we find the total cross section:
\begin{eqnarray}
\sigma &=& \int \frac{\overline{\vert M \vert}^{\mkern4mu 2}}{4
          \vert p_{li}^{L} \cdot p_{ni}^{L}\vert}\, (2\pi)^{4}
          \delta^{(4)}\left(\sum_{i}p_{i}^{L}\right)  \frac{d^{3}
          \vec{p}_{lf}^{\mkern3mu L}}{(2\pi)^{3}2E_{lf}^{L}}
          \frac{d^{3} \vec{p}_{\pi}^{\mkern3mu L}}{(2\pi)^{3}2E_{\pi}^{L}}
          \frac{d^{3} \vec{p}_{nf}^{\mkern3mu L}}{(2\pi)^{3}2E_{nf}^{L}} \notag
          \\[5pt]
&=& \int \frac{\overline{\vert M \vert}^{\mkern4mu 2}}{4 \vert
          p_{li}^{L} \cdot p_{ni}^{L}\vert}\, (2\pi)^{4}
          \delta^{(4)}(q+p_{ni}-p_{nf}-p_{\pi})\,  \frac{d^{3}
          \vec{p}_{lf}^{\mkern3mu L}}{(2\pi)^{3}2E_{lf}^{L}}
          \frac{d^{3} \vec{p}_{\pi}}{(2\pi)^{3}2E_{\pi}}
          \frac{d^{3} \vec{p}_{nf}}{(2\pi)^{3}2E_{nf}} \notag
          \\[5pt]
&=&\int \frac{\overline{\vert M \vert}^{\mkern4mu 2}}
          {4 \vert p_{li}^{L} \cdot p_{ni}^{L}\vert}\, (2\pi)^{4}
          \delta(q^{0}+p_{ni}^{0}-p_{nf}^{0}-p_{\pi}^{0})\,
          \frac{1}{(2\pi)^{3}2E_{nf}}
          \frac{d^{3} \vec{p}_{lf}^{\mkern3mu L}}{(2\pi)^{3}2E_{lf}^{L}}
          \frac{d^{3} \vec{p}_{\pi}}{(2\pi)^{3}2E_{\pi}} \notag
          \\[5pt]
&=&\int \frac{\overline{\vert M \vert}^{\mkern4mu 2}}{32 M_{n}}
          \frac{1}{(2\pi)^{5}}  \frac{\modular{p}{\pi}}{E_{\pi}+E_{nf}}
          \frac{\vert \vec{p}_{lf}^{\mkern3mu L}\vert }
          {\vert \vec{p}_{li}^{\mkern3mu L}\vert}\,
          d\Omega_{\pi}\, dE_{lf}^{L}\, d\Omega_{lf}^{L} \ .  
\end{eqnarray}

It is quite complicated to calculate the boundary of phase space in
terms of the integration variables in the preceding equations.
Later, we will work out the boundary of phase space in terms of the
invariant variables $Q^{2}$ and $M_{\pi n}$ in the cm frame of the
whole system, so we would like to have the following:
\begin{eqnarray}
Q^{2}&=& -M_{lf}^{2}+2E_{li}^{L}(E_{lf}^{L}-\vert
          \vec{p}_{lf}^{\mkern3mu L}\vert \cos{\theta_{lf}^{L}} )\ , 
          \\[5pt]
M_{\pi n}^{2}&=&(q^{L}+p_{ni}^{L})^{2}=-Q^{2}+M_{n}^{2}+2M_{n}(E_{li}^{L}-E_{lf}^{L})
          \ , 
\end{eqnarray}
from which it follows
\begin{eqnarray}
dQ^{2}dM_{\pi n}^{2} = 4M_{n} E_{li}^{L}\, \vert
          \vec{p}_{lf}^{\mkern3mu L}\vert dE_{lf}^{L}\, d\cos{\theta_{lf}^{L}}\ . 
\end{eqnarray}
By using the invariance of the cross section with respect to
rotations around the incoming lepton direction, we have $\int
d\Omega_{lf}^{L}=\int d\cos{\theta_{lf}^{L}\, 2\pi}$, and thus
\begin{eqnarray}
\sigma=\int \frac{\vert \psibar{M} \vert^{2}}{64 M_{n}^{2}}
          \frac{1}{(2\pi)^{5}}  \frac{\modular{p}{\pi}}{E_{\pi}+E_{nf}}
          \frac{\pi}{\vert \vec{p}_{li}^{\mkern3mu L}\vert E_{li}^{L}}\,
          d\Omega_{\pi}\,
          dM_{\pi n}^{2}\, d Q^{2} \ . 
\end{eqnarray}

In the isobaric frame, there is no preference in the direction of
the outgoing pion. Thus the boundary of $ \Omega_{\pi}$ is the whole
solid angle in the isobaric frame. Now let's work out the boundary
of phase space in the cm frame. We have
\begin{eqnarray}
M_{A}^{2}&\equiv&p_{A}^{2}=(p^{L}_{ni}+p^{L}_{li})^{2}
          =(M_{n}+E^{L}_{li})^{2}-(E^{L}_{li})^{2}=M_{n}^{2}+2M_{n}E^{L}_{li}\
          ,
          \\[5pt]
M_{\pi n}^{2}&\equiv& (p_{\pi}+p_{nf})^{2}
          =(p^{C}_{A}-p^{C}_{lf})^{2}=M_{A}^{2}+M_{lf}^{2}-2M_{A}E^{C}_{lf}\
          .
\label{eqn:MpinElf}
\end{eqnarray}
Here, $E^{C}_{lf}$ is the final lepton's energy in the cm frame.
From now on, all the quantities in the cm will be labeled in this
way. So, for given $E^{L}_{li}$, i.e., $M_{A}$, we can see that
\begin{eqnarray}
M_{n}+M_{\pi} \leqslant M_{\pi n} \leqslant M_{A}-M_{lf}  \ . 
\end{eqnarray}
By using Eq.~(\ref{eqn:MpinElf}), we find
\begin{eqnarray}
(E^{C}_{lf})_{ \mathrm{max}}&=&\frac{M_{A}^{2}+M_{lf}^{2}
          -(M_{\pi n }^{2})_{\mathrm{min}}}{2M_{A}}\ , 
          \\[5pt]
(E^{C}_{lf})_{ \mathrm{min}}&=&\frac{M_{A}^{2}+M_{lf}^{2}
          -(M_{\pi n}^{2})_{\mathrm{max}}}{2M_{A}}\ . 
\end{eqnarray}
Then, for given $E^{L}_{li}$ and $M_{\pi n}$ (or $ E^{C}_{lf})$,
using
$Q^{2}=-M_{lf}^{2}+2E^{C}_{li}E^{C}_{lf}-2E^{C}_{li}\modular{p}{lf}^{C}
\cos{\theta^{C}_{lf}}$ (where $\theta^{C}_{lf}$ is the angle between
the outgoing lepton's direction and the incoming lepton's direction
in the cm frame, and $E_{li}^{C}=(M_{A}^{2}-M_{n}^{2})/2M_{A}$ is
the initial lepton's energy in the cm frame), we finally arrive at
\begin{eqnarray}
[Q^{2}(E^{C}_{lf})]_{\mathrm{min}} &=&
          -M_{lf}^{2}+\frac{2E^{C}_{li}M_{lf}^{2}}
          {E^{C}_{lf}+\sqrt{(E^{C}_{lf})^{2}-M_{lf}^{2}}} \ , 
          \\[5pt]
[Q^{2}(E^{C}_{lf})]_{\mathrm{max}} &=& -M_{lf}^{2}+ 2E^{C}_{li}
\left(
          E^{C}_{lf}+\sqrt{(E^{C}_{lf})^{2}-M_{lf}^{2}}\, \right) \ . 
\end{eqnarray}
These equations give a description of the phase-space boundary in
terms of the invariants $M_{\pi n}$ and $Q^{2}$.

\end{document}